\RequirePackage{fix-cm}
\documentclass[smallcondensed]{svjour3}     
\smartqed  
\usepackage{graphicx,xcolor,natbib}
\usepackage{mathrsfs}
\usepackage{latexsym,amssymb}
\usepackage{amsmath}
\usepackage[normalem]{ulem}
\newcommand{\be}{\begin{equation}}
\newcommand{\ee}{\end{equation}}
\newcommand{\ud}{{\rm d}}
\newcommand{\h}{{\rm H}}
\newcommand{\hh}{{\rm H_{2}}}
\newcommand{\he}{{\rm He}}
\def\m{\ifmmode ^{-1} \else $^{-1}$ \fi}
\def\mm{\ifmmode ^{-2} \else $^{-2}$ \fi}
\def\mmm{\ifmmode ^{-3} \else $^{-3}$ \fi}
\def\HII{H{\sc ii} }

\usepackage{calc}
\newsavebox\CBox
\newcommand\hcancel[2][0.5pt]{%
  \ifmmode\sbox\CBox{$#2$}\else\sbox\CBox{#2}\fi%
  \makebox[0pt][l]{\usebox\CBox}%
  \rule[0.5\ht\CBox-#1/2]{\wd\CBox}{#1}}
  
\def\apjl{ApJ}%
\def\apjs{ApJS}%
\def\ao{Appl.~Opt.}%
\def\aap{A\&A}%

\begin{document}

\title{Impact of low-energy cosmic rays on star formation
}


\author{Marco~Padovani     \and
	Alexei~V.~Ivlev \and
	Daniele~Galli \and	
        Stella~S.~R.~Offner \and
        Nick~Indriolo \and 
        Donna~Rodgers-Lee \and
	Alexandre~Marcowith \and 
        Philipp~Girichidis \and
	Andrei~M.~Bykov \and
	J.~M.~Diederik~Kruijssen
}


%
\institute{M.~Padovani, D. Galli \at
              INAF-Osservatorio Astrofisico di Arcetri - Largo E. Fermi, 5 - 50125 Firenze, Italy\\
              \email{padovani@arcetri.astro.it}
        \and
           A.~V.~Ivlev \at
           Max-Planck-Institut f\"ur extraterrestrische Physik - Giessenbachstr. 1 - 85748 Garching, Germany
        \and
           S.~S.~R.~Offner \at
              Department of Astronomy, The University of Texas at Austin - 2500 Speedway - Austin TX 78712, USA
        \and
           N.~Indriolo \at
              ALMA Project, National Astronomical Observatory of Japan,
              National Institutes of Natural Sciences, 2-21-1 - Osawa, Mitaka,
              Tokyo 181-8588, Japan
        \and
           D.~Rodgers-Lee \at
              School of Physics, Trinity College Dublin, University of Dublin, College Green - 
              Dublin 2, Co. Dublin, D02 PN40, Ireland
        \and
          A.~Marcowith \at
          Laboratoire Univers et Particules de Montpellier, UMR 5299 du CNRS, Universit\'e de Montpellier - Place E. Bataillon, cc072 - 34095 Montpellier, France               
        \and
          P. Girichidis \at
          Leibniz-Institut f\"{u}r Astrophysik Potsdam (AIP) - An der Sternwarte 16 - 14482 Potsdam, Germany
     \and            
        A.~M.~Bykov  \at
              Ioffe Institute, 26 Politehnicheskaya
              - St. Petersburg, 194021, Russian Federation
        \and
          J.~M.~D.~Kruijssen \at
          Astronomisches Rechen-Institut, Zentrum f\"{u}r Astronomie der Universit\"{a}t Heidelberg - M\"onchhofstr. 12-14 - 69120 Heidelberg, Germany
        }

\date{Received: date / Accepted: date}

\maketitle
\begin{abstract}
In recent years, exciting developments have taken place in the identification  
of the role of cosmic rays in star-forming environments. 
Observations from radio to infrared wavelengths and theoretical modelling have shown that low-energy cosmic rays 
($< 1$~TeV) play a fundamental role in shaping the chemical richness of the interstellar medium,   
determining the dynamical evolution of molecular clouds. 
In this review we summarise in a coherent picture the main results obtained by observations and by theoretical models 
of propagation and generation of cosmic rays, from the smallest scales of protostars and circumstellar discs, to young stellar clusters, 
up to Galactic and extragalactic scales. We also discuss the new fields that will be explored in the near future thanks 
to new generation instruments, such as: CTA, for the $\gamma$-ray emission from high-mass protostars;
SKA and precursors, for the synchrotron emission at different scales; and ELT/HIRES, JWST, and ARIEL, 
for the impact of CRs on exoplanetary atmospheres and habitability.
\end{abstract}

\setcounter{tocdepth}{4}
\tableofcontents

\section{Introduction}
\label{intro}


Thanks to increasingly powerful observing facilities and sophisticated theoretical models, we are now able to address quantitatively several key aspects in process of star formation in the Milky Way and nearby galaxies, such as the formation of protostellar discs and bipolar jets in collapsing molecular clouds, the coalescence of dust grains and the formation of planetesimals and planets, and the origin of chemical complexity from molecular cloud to comets and planets.
A field where results have been particularly exciting in recent years is the study of the interaction of cosmic rays (CRs) with the interstellar matter.
For star formation to take place, gas and dust must be sufficiently cold such that gravity overcomes the 
thermal pressure and the ionisation fraction must be sufficiently low to allow significant decoupling
of the gas from the Galactic magnetic field. 
As soon as the visual extinction is of the order of 3--4 magnitudes, 
the UV photon flux of the interstellar radiation field is completely 
attenuated, thus the only source of ionisation and heating is represented by low-energy CRs\footnote{Low-energy CRs are defined as charged particles outside of the thermal distribution. The energy peak of the
Maxwellian distribution is at about 1~meV, 5~meV, and 1~eV for
typical temperatures of dense cores ($T=10$~K), diffuse clouds
($T=50$~K), and protostellar jets ($T=10^4$~K), respectively.}.
Since the cross sections of the most relevant processes for star formation peak 
at relatively low energies (e.g., $\sim10$~eV for dissociation of H$_2$ by
electrons and $\sim10$~keV for ionisation of H$_2$ by protons),
low-energy CRs turn out to be a fundamental ingredient for the dynamical
and chemical evolution of star-forming regions on a wide range of
physical scales.
%
On the largest global scale of star formation, CRs can contribute a significant amount of pressure to the Galaxy influencing Galactic dynamics. 

The two main components of low-energy CRs ($\lesssim 1$~TeV) considered in this review are those accelerated and confined in the Galactic disk (producing  what is known as the {\em Galactic CR spectrum}), and 
in shocks associated to the star formation process itself (hereafter {\em local CRs}).
The majority of low-energy CRs contributing to the observed Galactic spectrum are likely to originate from sources within our Galaxy, such as supernova remnants \citep{zwicky_1934,Blasi2013} and colliding stellar winds \citep{casse_1980,aharonian_2019}. Diffusive shock acceleration (DSA, also known as first-order Fermi acceleration) and magnetic reconnection are some of the possible acceleration mechanisms \citep[see e.g. reviews by][]{BELL2013,Pino_2014}. The local spectrum of Galactic CRs has been measured by the 
 Voyager 1 and 2 spacecrafts in the interstellar medium~\citep[ISM;][]{Cummings+2016,Stone+2019}, 
 and is thought to be unmodulated by the solar wind. The observed Galactic CR 
proton spectrum has a slope of $-2.78$ \citep[see e.g.][]{VosPotgieter2015}, steeper than expected from DSA which would result in a slope of $-2$. This may simply reflect the fact that the proton spectrum from the source has been modulated by some physical process en route to us, such as diffusive steepening, or a different acceleration mechanism. 
The Galactic CR spectrum may be specific to our location within the Galaxy and varies spatially (see Sect.~\ref{Observations}). It will certainly vary close to individual sources of CRs, for instance, and will be different in other galaxies (see Sect.~\ref{extra}).

In the context of star formation, there has been much interest in recent years in locally accelerated sources of low-energy CRs which will be discussed in some detail in this review. Even if these locally accelerated low-energy CRs do not contribute significantly to the Galactic CR spectrum, 
they represent the key to explain the high ionisation rate
and the synchrotron emission observed in protostellar environments.
The main accelerators are thought to be protostellar shocks.
More precisely, local CRs are accelerated at the surface of the shocks located along
the bipolar jets and on the protostellar surface, where the accretion disc matter channelled
by magnetic fields hits the protostellar surface itself.

This review is organised as follows:  
we focus on the
direct measurements and the different observational methods to constrain
the CR flux (Sect.~\ref{Observations}); 
we summarise the mechanisms of interaction of CRs with the interstellar
medium, providing accurate details on the regimes of CR transport and the
energy loss processes (Sect.~\ref{propagation_AI});
we examine in depth the role of CRs at cloud (Sect.~\ref{molecularclouds})
and circumstellar disc (Sect.~\ref{protostars}) scales;
we describe how CRs can be efficiently accelerated in protostellar
environments (Sect.~\ref{localcosmicrays});
we discuss the implications of
presence of CRs at different Galactic scales (Sect.~\ref{diffgalscales});
finally in Sect.~\ref{conclusions} we summarise the most important findings,
concluding with
an outlook on future work.



\section{Cosmic-ray observations and observables}
\label{Observations}

CRs are unlike radiation in that they do not travel directly to the Earth from their points of origin, instead diffusing through the Galaxy due to interactions with magnetic fields. As such, observations of CRs do not represent particles accelerated in any one particular object or region. Rather, these observations represent the population of particles accelerated throughout the Galaxy that happen now to be traversing our neighborhood. In order to investigate the CR spectrum elsewhere we must rely on observables produced by the various energy loss mechanisms described in Sect.~\ref{propagation_AI}. Here, we briefly discuss the direct measurement of CRs made in the solar system and local ISM and next consider different observables attributed to CR interactions and how they constrain aspects of the CR spectrum.

\subsection{In-situ measurements of cosmic rays}

As CRs diffuse throughout the Galaxy, some will inevitably pass through our solar system, where they can be measured by instruments aboard satellites in Earth-orbit and balloons flown at high altitude. Several experiments have been designed for observing CRs in different energy ranges \citep[for a recent summary, see Table 1 in][]{Grenier+2015}, and only by combining the observations from each is the full CR spectrum revealed. While CRs with energies above a few GeV can reach these experiments relatively unhindered, lower-energy particles are deflected from the inner solar system by the magnetic field coupled to the solar wind, an effect referred to as modulation. Direct measurement of the unmodulated CR spectrum below a few GeV requires an instrument that is outside the heliosphere, a condition now satisfied by the Cosmic Ray Subsystem instruments on both Voyager spacecrafts. 
The best measurements of the local interstellar CR proton and electron spectra down to energies of about 3~MeV are currently provided by observations from Voyager 1 \citep{Cummings+2016} and Voyager 2 \citep{Stone+2019}. 

\subsection{The observable influence of cosmic rays}

\subsubsection{Ionisation and interstellar chemistry}
\begin{figure}[!h]
\begin{center}
\resizebox{0.7\hsize}{!}{\includegraphics{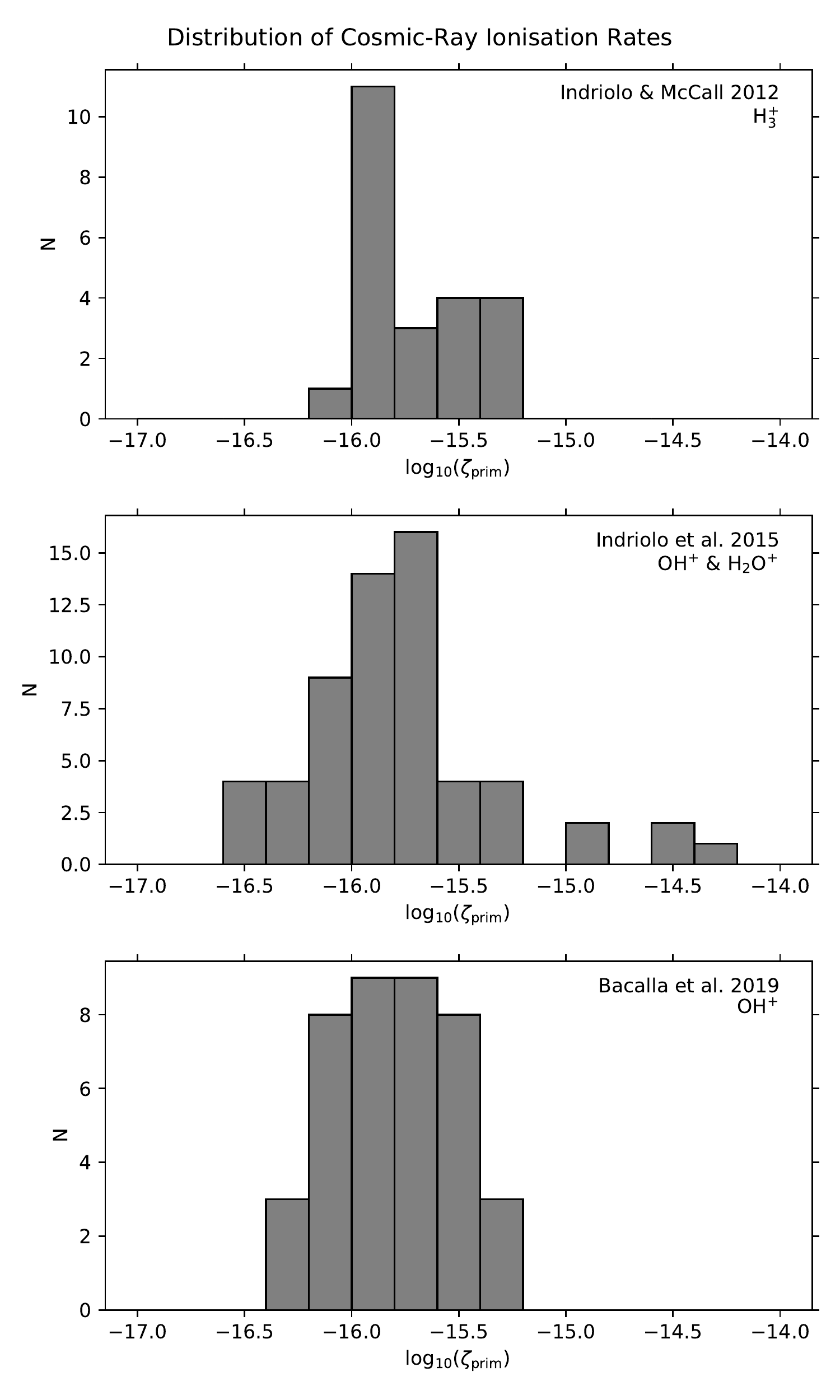}}
\caption{Primary CR ionisation rate ($\zeta_{\rm prim}$) distributions inferred from different molecular tracers in diffuse clouds. The histograms only account for clouds where absorption lines are detected, not upper limits inferred from non-detections. Note that all ionisation rates reported by \citet{Bacalla+2019} have been multiplied by 35/100. This was done to account for the different gas densities adopted by \citet{Indriolo+2015} and \citet{Bacalla+2019} (35~cm$^{-3}$ and 100~cm$^{-3}$, respectively).
}
\label{crir_dist}
\end{center}
\end{figure}

Ionisation of H and H$_2$ are the dominant energy loss mechanisms for low-energy CRs protons. While the resulting H$^+$ and H$_2^+$ are not directly observable, these ions react with other gas phase species in the ISM, initiating a network of ion-molecule reactions that drive interstellar chemistry. In diffuse molecular clouds the abundances of H$_3^+$, OH$^+$, and H$_2$O$^+$ depend strongly on the CR flux, and observations of these molecules are used to constrain the CR ionisation rate, $\zeta$. Note that there are multiple definitions of the CR ionisation rate, including the primary ionisation rate of H ($\zeta_{\rm prim}$; ionisation rate of atomic hydrogen due only to CR nuclei), the total ionisation rate of H ($\zeta_{\rm H}$; ionisation rate of atomic hydrogen due to both CR nuclei and secondary electrons produced by the primary ionisation events), and the total ionisation rate of H$_2$ ($\zeta_2$; ionisation rate of molecular hydrogen due to both CR nuclei and secondary electrons produced by the primary ionisation events). The exact relationship between these different ionisation rates depends on gas composition, but $\zeta_{\rm H}\approx1.5\zeta_{\rm prim}$ and $\zeta_{\rm H_2}\approx2.3\zeta_{\rm prim}$ are reasonable approximations \citep{GlassgoldLanger1974}. Here, we will use $\zeta_{\rm prim}$ when quoting results from various studies for the sake of consistency. Surveys of H$_3^+$, OH$^+$, and H$_2$O$^+$ absorption have now been used to place constraints on the CR ionisation rate in about 200 different diffuse cloud components within the Galactic disc \citep{Indriolo+2007,IndrioloMcCall2012,Porras+2014,Indriolo+2015,Zhao+2015,Bacalla+2019}. Figure \ref{crir_dist} shows the distribution of primary CR ionisation rates inferred from these observations. Results from the different surveys are consistent with each other within uncertainties, finding a mean ionisation rate of  $\zeta_{\rm prim}=(1.8^{+3.1}_{-1.1})\times10^{-16}$~s$^{-1}$.

The wide range of ionisation rates inferred from molecular abundances implies that the underlying CR spectrum is not uniform throughout the Galactic disc. These variations can be caused either by scaling the entire spectrum up or down in flux or by adding a varying component at low energies while maintaining the same underlying {\it floor} spectrum everywhere. An ionisation rate can also be calculated using the particle spectra measured by Voyager spacecrafts to determine how the particle flux in the local ISM compares to that elsewhere. Doing so results in a value of $\zeta_{\rm prim}\sim10^{-17}$~s$^{-1}$, over a factor of 10 times lower than the average value in diffuse molecular clouds \citep{Cummings+2016}, suggesting that the local CR flux may not be representative of the Galaxy as a whole. However, there are some upper limits on ionisation rates inferred from non-detections of H$_3^+$ and OH$^+$ that are consistent with this low value \citep{IndrioloMcCall2012,Bacalla+2019}, so the local ISM may not be unique in this regard. Either the local ISM represents the floor spectrum, with higher ionisation rates elsewhere due to an increased particle flux at low energies, or other regions in the Galactic disc experience an overall higher flux of CRs.

\subsubsection{Pion production and $\gamma$-ray emission}
For CR protons with $E\gtrsim1$~GeV energy losses due to pion production dominate those due to ionisation. The production of $\pi^0$ particles via elastic collisions with ambient atoms, and subsequent rapid decay into a pair of $\gamma$-ray photons, provides a potential observable. Indeed, this pionic $\gamma$-ray emission has been detected from supernova remnants interacting with molecular clouds \citep[e.g.][]{Ackermann+2013}, from the gas within the Galactic plane \citep{Ackermann+2012}, and even from nearby molecular clouds \citep[e.g.,][]{Yang+2014,Neronov+2017}. The $\gamma$-ray emission spectrum resulting from $\pi^0$ produced by CRs depends on the spectrum of CR particles and on the distribution of {\it target} ambient atoms. The total column density of ambient material along a given line-of-sight can be constrained by observations of H, CO and dust, and this can be used in concert with the observed $\gamma$-ray spectrum to infer the underlying CR spectrum at energies above about 1~GeV. $\gamma$-ray observations of several nearby molecular clouds with Fermi-LAT suggest a CR spectrum that is consistent with that measured in the local ISM \citep{Neronov+2017}, at least for the sampled energy range. 

\subsubsection{Inner shell ionisation and X-ray emission}
The ionisation of atoms is not limited to the removal of an electron from any specific energy level. When an electron is removed from an inner shell, the vacancy in that shell will rapidly be filled by the de-excitation of an electron from a more energetic shell, a process that generates a photon. A vacancy in the $n=1$ level is most often filled by an electron from the $n=2$ level. Emission arising from the $n=2\rightarrow1$ transition is referred to as K$\alpha$ and typically occurs in the X-ray regime. The Fe K$\alpha$ line at 6.4~keV has been observed in the Galactic centre and from molecular clouds near supernova remnants, but different studies of these regions have interpreted the emission as the product of ionisation by electrons \citep{Yusefzadeh+2007}, low-energy protons \citep{Tatischeff+2012,Nobukawa+2018} and X-ray photons \citep{Ponti+2010}. While Fe K$\alpha$ line emission indicates that energetic particles or photons are present, it can be difficult to distinguish which is the dominant ionising mechanism. As such, placing constraints on the underlying CR spectrum from Fe K$\alpha$ emission is highly uncertain.

 \subsubsection{Observational constraints on the ionisation rate in protoplanetary discs}
 
 The exact level of ionisation within protoplanetary discs has far-reaching consequences. In this section we discuss observational efforts to constrain the ionisation rate at disc scales. 
 In principle, the presence of CRs can be inferred from the products of interactions, including high-energy photons. However, stellar or Galactic CRs that interact with protoplanetary discs will not result in an observable level of $\gamma$ rays, since  the latter would likely be absorbed by the disc. Instead, very sensitive  observations carried out by the Atacama Large Millimetre Array (ALMA) can measure the ratio of different molecules that directly constrain the (total) ionisation rate  \citep[][for instance]{Cleeves+2015}. Chemical models can also indirectly constrain the ionisation rate from CO observations \citep[such as][]{Schwarz+2018}.

\citet{Cleeves+2015}, using observations from the Submillimeter Array of HCO$^+$ and ALMA observations of N$_2$H$^+$, constrained the total ionisation rate (i.e., from all sources of ionisation including FUV, X-rays, CRs etc.) for TW Hya's disc to be $\zeta_{\rm H_2}\lesssim10^{-19}\mathrm{s^{-1}}$, which is relatively low. TW Hya is a prime observational target due to its close proximity. However, it is a relatively old disc and the total disc mass of $0.02M_\odot$ derived from dust observations (assuming a particular dust to gas mass ratio) is quite large in comparison to the scatter observed in young star-forming regions \citep{Pascucci+2016}. Thus, the observed low ionisation rate in TW Hya's disc may not be representative of most discs in star-forming regions.
On the other hand \citet{Schwarz+2018} present results from CO chemical models for a relatively large sample of observed protoplanetary discs and conclude an unknown physical process must be operating in these discs. These results may provide indirect evidence for the presence of stellar cosmic rays (SCRs). 

More direct observations are needed to constrain the ionisation rate in a bigger sample of protoplanetary discs. Modelling the influence of SCRs in a wide range of discs is critical to identify discs likely to display high ionisation rates. Such observations will in turn constrain some of the free parameters adopted in propagation models. Using the empirical relationship between the X-ray and SCR luminosity obtained through combined (sub-)millimetre  and X-ray observations will also help to constrain the X-ray contribution to the ionisation rate and the luminosity of SCRs appropriate to model specific discs.

\section{Overview of CR transport and interactions in the ISM}
\label{propagation_AI}

In this section we focus on the interaction of Galactic CRs with dense astrophysical objects, such as molecular clouds, cores, and discs, and describe approaches to calculate the CR penetration into these objects and the resulting ionisation/heating.
In the ISM the details of CR transport are governed by the dominant role of their interaction with gas. The energy
losses are proportional to the gas density and, hence, the adiabatic losses, which play an important role on Galactic
scales, can be completely neglected. Consequently, one can also neglect CR diffusion in momentum space, leading to CR re-acceleration in the Galaxy. These processes will be discussed in Sect.~\ref{galaxy}.

\subsection{Transport and penetration of CRs into dense astrophysical objects}
\label{transport}

The gyroradii of low-energy CRs, whose energy is relevant for ionisation and heating in molecular clouds and discs, are much smaller than the spatial scales characterising inhomogeneities in the magnetic field and gas density within these objects. Thus, we adopt the concept of the guiding centre for CRs and consider their propagation along the local magnetic field, with coordinate $s$ measured along the field line. We introduce the {\it effective} column density of gas traversed by a CR
particle before it reaches a given point \citep{Padovani+2018a},
\begin{equation}\label{N_eff}
N=\int_0^sn_{\rm H}(s)\:ds,
\end{equation}
where $n_{\rm H}=n({\rm H})+2n({\rm H}_2)$ is the total number density of hydrogen atoms. In the inner regions of clouds and
discs, where the field lines are likely strongly twisted and entangled, the effective column density may be much larger than the line-of-sight value. Therefore, Eq.~(\ref{N_eff}) is the proper measure of attenuation of Galactic
CRs penetrating such regions.

The CR kinetics in dense objects is conveniently described in terms of their distribution function in energy space,
$f_E=f(E,s,\mu)$. It is related to the distribution function in momentum space via $vf_E=p^2f_p$, where $v(E)$ is the
physical velocity of a particle and $\mu$ is the cosine of pitch angle. The number of CRs per unit volume and energy is then given by $\mathcal{N}(E,s)= 2\pi\int_{-1}^{1}f_E\:d\mu$.

The transport equation for $f_E$, which describes the continuity of the distribution in phase space, takes into account the energy losses by CRs due to their collisional interactions with gas. The equation can be expressed generally as:
\citep[e.g.,][]{Morfill+1976,CesarskyVolk1978,Berezinskii1990book}:
\begin{equation}\label{KE}
\frac{\partial S}{\partial s}+\frac{\partial}{\partial E}\left(\dot E_{\rm con}f_E\right)+\nu_{\rm cat}f_E=0,
\end{equation}
where $S(f_E)$ is the differential flux of CRs (per unit area, energy, and solid angle), whose dependence on $f_E$ is
determined by the dominant CR transport regime, as discussed below. CR losses are divided into two categories:
$\nu_{\rm cat}(E)$ is the rate of all {\it catastrophic} collisions leading to an instantaneous loss of energy by a particle,
and $\dot E_{\rm con}(E)$ is the {\it continuous} energy decrease per unit time due to collisions accompanied by loss of a
small portion of energy. Both $\nu_{\rm cat}$ and $\dot E_{\rm con}$ scale with $n_{\rm H}$. In Sect.~\ref{energylosses}
we discuss individual processes contributing to continuous or catastrophic losses. The latter affect the CR transport
only at high column densities (corresponding to inner regions of circumstellar discs, see Sect.~\ref{propagation}), and, therefore, below we assume continuous losses. 

The distribution function is generally a function of the pitch angle, which depends on the local strength of the regular magnetic field, $B(s)$ (see Sect.~\ref{mirr-foc}),  and CR scattering that occurs due to
resonant CR interactions with magnetohydrodynamic (MHD) turbulence and gas nuclei. Depending on the degree of pitch angle scattering, there are two CR transport regimes  -- {\it free streaming} and {\it diffusive} -- as
discussed in the following Sect.~\ref{section_FS} and \ref{section_D}.

\subsubsection{Free streaming (continuous slowing-down approximation)}
\label{section_FS}

The free-streaming approximation \citep[also known as the continuous slowing-down approximation, CSDA,
see][]{Takayanagi1973,Padovani+2009} is the most common approach used to calculate the propagation of CRs in molecular
clouds. Scattering processes are inefficient in this regime, so that the resulting mean squared deviation of the pitch
angle along a CR track is small. Hence, $\mu$ is conserved and the differential CR flux, $S$, is given by
\begin{equation}\label{S_FS}
S(E,s,\mu)=\mu vf_E,
\end{equation}
i.e., $S$ is simply proportional to the CR spectrum, defined as
\begin{equation}
j=vf_E.
\end{equation}
The dependence on $\mu$ is completely determined by the boundary conditions for the initial (interstellar, Galactic)
spectrum $j^{\rm IS}(E,\mu)$ of CRs entering a cloud or disc. The kinetic energy $E$ of a particle after it has traversed the
effective column density $N$ is related to its initial energy $E_i$ via
\begin{equation}\label{N}
N=\mu\int_{E}^{E_i}\frac{dE}{L(E)}\:,
\end{equation}
where $L(E)$ is the energy loss function (see Sect.~\ref{energylosses}),\footnote{By setting the lower integration limit
to zero and the upper one to $E$ in Eq.~(\ref{N}), we obtain the stopping range $R(E)$ for particles with zero pitch
angles ($\mu=1$), plotted for different CR species in Fig.~\ref{fig1bis} (Sect.~\ref{energylosses}).} defined as
$L(E)=-\dot E_{\rm con}/(n_{\rm H}v)$. 
This yields the following solution of Eq.~(\ref{KE}):
\begin{equation}\label{solution_FS}
j(E,N,\mu)L(E)=j^{\rm IS}(E_i,\mu)L(E_i),
\end{equation}
which relates the initial (interstellar) and local (propagated) spectra.

To illustrate the effect of energy losses, consider the penetration of
a CR particle into a molecular cloud. For gas column
densities up to $N\sim10^{25}$~cm$^{-2}$, attenuation is completely determined by the ionisation losses
by both CR protons and electrons affecting subrelativistic particles (see Figs.~\ref{Lfunc} and \ref{fig1bis} in Sect.~\ref{energylosses}). For a
power-law initial CR spectrum,
\begin{equation}\label{ini}
j^{\rm IS}(E)=j_0\left(\frac{E}{E_0}\right)^{-a},
\end{equation}
it is possible to obtain simple analytical expressions for the local spectrum and corresponding ionisation rate as a function of column density. 
The loss function in the energy range relevant for ionisation in 
molecular clouds can be well approximated by
\citep{Padovani+2018a,SilsbeeIvlev2019}
\begin{equation}\label{L_i}
L(E)=L_0\left(\frac{E}{E_0}\right)^{-b}\,.
\end{equation}
For CR protons in the energy range where ionisation losses
dominate\footnote{\label{fn1} In \citet{Padovani+2018a}, the total column density of {\it all} gaseous
species was used (assuming all the hydrogen to be in molecular form), whereas $N$ in Eq.~(\ref{N_eff}) is the total
column density of hydrogen atoms. Therefore, $N$ in \citet{SilsbeeIvlev2019} is a factor of 1.67 higher than that in \citet{Padovani+2018a}, which is
taken into account in the value of $L_0$.}
(from $\sim10^5$~eV to
$\sim10^9$~eV), 
$L_0=1.27\times10^{-15}$~eV~cm$^2$, $E_0=1$~MeV, and $b=0.82$. Then Eqs.~(\ref{N}) and (\ref{solution_FS}) give
\begin{equation}\label{j_FS}
j(E,N,\mu)=j^{\rm IS}(E)\left[1+\frac{N}{\mu R(E)}\right]^{-\frac{a+b}{1+b}},
\end{equation}
where
\begin{equation}\label{R_FS}
R(E)=\frac{E_0}{(1+b)L_0}\left(\frac{E}{E_0}\right)^{1+b},
\end{equation}
is the stopping range (for the above range of energies). This result intuitively shows
that only CRs with a stopping range $R(E)\lesssim N$ are attenuated (and their local spectrum changes to $j\propto E^b$),
while the spectrum at higher energies coincides with the initial spectrum.

If the local CR proton spectrum is known, it is straightforward to calculate the local ionisation rate of H$_2$ due to CR protons. Since only particles
with $\mu>0$ penetrate into a cloud in the free-streaming regime, 
\begin{equation}\label{zeta}
\zeta_{p,\rm H_2}=\int_0^1d\mu\int_0^\infty j_{p}(E,N,\mu)\sigma_{p,\rm H_2}^{\rm ion}(E)\:dE,
\end{equation}
where $\sigma_{p,\rm H_2}^{\rm ion}(E)$ is the ionisation cross section for molecular hydrogen. For protons with energies between
$10^5$~eV and $5\times10^8$~eV, the ratio between $L_{p}(E)$ and $\sigma_{p,\rm H_2}^{\rm ion}(E)$ is nearly constant and equal to
$\epsilon\sim37$~eV \citep{CravensDalgarno1978,Padovani+2018a}. This fact allows us to replace the cross section in Eq.~(\ref{zeta}) with
$L_{p}(E)/\epsilon$ determined by Eqs.~(\ref{L_i}). For an initial power-law spectrum, we also substitute
Eq.~(\ref{j_FS}) and obtain the dependence of the ionisation rate on the column density in the free-streaming regime
\citep{SilsbeeIvlev2019},
\begin{equation}\label{zeta_FS}
\zeta_{p,\rm H_2}(N)=\zeta_0\left(\frac{N}{R_0}\right)^{-\frac{a+b-1}{1+b}},
\end{equation}
where $R_0=R_{p}(E_0)\sim 10^{21}$~cm$^{-2}$ and $\zeta_0\approx j_0E_0L_0/\epsilon$ with a reasonable accuracy. Note
that Eq.~(\ref{zeta_FS}) is derived for sufficiently steep initial spectra, with $a>1-b\sim0.2$. Otherwise, e.g.,
for the spectrum of subrelativistic protons measured by the Voyager
spacecrafts \citep{Cummings+2016,Stone+2019}, the main contribution to the integral in
Eq.~(\ref{zeta}) is provided by CRs at several hundreds of MeV; the resulting $\zeta_{p,\rm H_2}(N)$
remains practically constant as long as $N\lesssim10^{25}$~cm$^{-2}$.

\subsubsection{Spatial diffusion}
\label{section_D}

The dominant regime of CR transport in regions surrounding dense cores embedded within molecular clouds is debated.  MHD turbulence in these regions can resonantly scatter the pitch angles of penetrating CRs \citep{KulsrudPearce1969},
leading to spatial diffusion. 
The spectrum of MHD turbulence determines the magnitude of the CR diffusion coefficient and its dependence on the particle energy \citep{Schlickeiser2016,SilsbeeIvlev2019}. However, MHD turbulence can also be driven by anisotropy in the CR distribution function \citep{SkillingStrong1976,MorlinoGabici2015,Ivlev+2018}, which arises in response to CR absorption in dense cores.

We begin by briefly summarising the basic principles of diffusive CR transport. 
This regime is realised when the mean free path of
a CR particle due to pitch-angle scattering is much smaller than the relevant spatial scale of the problem. Therefore, the
local distribution function is isotropised, and the 
number of CRs per unit volume and energy is related to the distribution function via
$\mathcal{N}(E,s)\approx 4\pi f_E$. The corresponding flux $S$ does not depend on $\mu$ and generally contains two terms
\citep[e.g.,][]{Berezinskii1990book},
\begin{equation}\label{S_D}
S(E,s)=-D\frac{\partial f_E}{\partial s}+uf_E,
\end{equation}
where $D(E)$ is the CR diffusion coefficient. The second term, which is proportional to the advection velocity $u$, arises due to the
anisotropy of MHD turbulence. This typically occurs when MHD waves are generated by CRs themselves (see
Sect.~\ref{section_self-gen}). Preexisting turbulence is usually assumed to be isotropic on small spatial scales (responsible
for scattering of low-energy CRs), and then $u=0$.

In order to calculate $D(E)$, one can employ a simplified expression proposed by \citet{Skilling1975},
\begin{equation}\label{D}
D(E)=\frac{vB^2}{6\pi^2\mu_*k_{\rm res}^2W(k_{\rm res})}\:,
\end{equation}
relating the diffusion coefficient to the spectral energy density of weak magnetic disturbances, $W(k)$, with $B$ being the
strength of the regular magnetic field. The relation is determined by $k_{\rm res}(E)$, the wavenumber of longitudinal MHD
waves satisfying the condition of the cyclotron resonance with CR particles of energy $E$,
\begin{equation}\label{k_res}
k_{\rm res}(E)=\frac{m\Omega}{\mu_*p(E)}\:,
\end{equation}
where $\mu_*$ is the {\it effective} cosine of the resonant pitch angle (usually set to a constant of order unity),
$p(E)$ is the momentum of a CR particle with rest mass $m$, and $\Omega=eB/mc$ is the CR gyrofrequency.

Further analysis of diffusive CR transport depends critically on the origin of MHD turbulence. If the turbulence is
driven by CRs, its spectrum $W(k)$ is determined by the CR spectrum $f_E$ and vice versa. This {\it essentially nonlinear}
regime is discussed separately in Sect.~\ref{section_self-gen}. The case of preexisting turbulence can be  analysed generally by equating $W(k)$ and the turbulent kinetic energy of ions,
\begin{equation}\label{EqP}
kW(k)\approx\frac12\rho_iv_{\rm turb}^2(k),
\end{equation}
where $\rho_i$ is the ion mass density and $v_{\rm turb}^2$ is their mean squared turbulent velocity. Then, for a
power-law turbulent spectrum with $v_{\rm turb}(k)\propto k^{-\lambda}$, the diffusion coefficient scales as
\citep{SilsbeeIvlev2019}
\begin{equation}\label{D1}
D\propto E^{1-\lambda}\rho_i^{-1},
\end{equation}
where $\lambda=1/3$ for a Kolmogorov spectrum and $\lambda=1/4$ for a Kraichnan spectrum. Assuming that both the
ionisation fraction ($\propto \rho_i/n_{\rm H}$) and turbulent velocity are independent of $N$,
Eq.~(\ref{KE}) is reduced to a linear diffusion equation for the product $jL$, with the coordinate $N$ and (pseudo)
time
\begin{equation}
T(E)=-\int_0^E\frac{dE}{X(E)}\:,
\end{equation}
where $X(E)=vL/(n_{\rm H}D)$ is a function of $E$ only. This allows solution in a general form; for a power-law initial CR
spectrum (\ref{ini}), the local spectrum is \citep{SilsbeeIvlev2019}
\begin{equation}\label{j_diff}
j(E,N)=j^{\rm IS}(E)\int_0^1{\rm erfc}\left[\frac{N/N_0}{\sqrt{(E/E_0)^\alpha(x^{-\frac{\alpha}{a+b}}-1)}}\right]dx,
\end{equation}
where $\alpha=3/2+b-\lambda$ and
\begin{equation}
N_0=\sqrt{\frac{4n_{\rm H}D_0E_0}{\alpha v_0L_0}}
\end{equation}
is the characteristic attenuation range (column density) in the diffusive regime, with $D_0=D(E_0)$ and $v_0=\sqrt{2E_0/m}$.
Depending on the turbulent spectrum, \citet{SilsbeeIvlev2019} estimated the magnitude of $n_{\rm H}D_0$ to be
$\sim 10^{28}$~cm$^{-1}$s$^{-1}$ for molecular cloud envelopes (where the ionisation fraction is set by carbon photoionisation). The qualitative behavior of Eq.~(\ref{j_diff}) is the same as of Eq.~(\ref{j_FS}): for a given $N$, the
local spectrum scales as $j\propto E^b$ at lower energies and tends to $j^{\rm IS}(E)$ at higher energies.

Substituting Eq.~(\ref{j_diff}) in Eq.~(\ref{zeta}) yields the local ionisation rate \citep{SilsbeeIvlev2019}
\begin{equation}\label{zeta_diff}
\zeta_{\rm H_2}(N)=\zeta_0\left(\frac{N}{N_0}\right)^{-\frac{4(a+b-1)}{3+2b-2\lambda}},
\end{equation}
where $\zeta_0$ is about that in Eq.~(\ref{zeta_FS}). Comparing Eqs.~(\ref{zeta_FS}) and (\ref{zeta_diff}) shows that a characteristic value of the diffusive attenuation range, $N_0\sim10^{20}$~cm$^{-2}$, is about an order of
magnitude smaller than that of the free-streaming range, $R_0$. Furthermore, the ratio of the attenuation exponents in the
diffusive and free-streaming regimes, $4(1+b)/(3+2b-2\lambda)$, is larger than unity. This indicates that the attenuation is
stronger in the diffusive regime -- which is also intuitive, since in this case a CR particle must cross a larger gas
column, $N$, before reaching a given location.

In Sect.~\ref{molecularclouds} we present a detailed discussion of the CR ionisation rate predicted in the envelopes of
molecular clouds and compare the theoretical results with observations in diffuse clouds (see Sect.~\ref{Observations}).

\subsection{Energy losses and attenuation of CRs}\label{energylosses}

In this section we summarise the basic energy loss processes that affect the penetration of primary CRs and secondary particles in interstellar clouds {(see, e.g.,~\citealt{Hayakawa1969book,Takayanagi1973,UmebayashiNakano1981})}. 
Several of these processes  result in the production of showers of secondary species (photons,
electrons, and positrons) through processes such as pion decay, bremsstrahlung (BS) and pair production. In addition to primary 
CRs, these secondary particles contribute to the total ionisation rate of the medium.

We consider CR species $k$ (including secondaries) and derive their spectra $j_{k}(E,N)$ as function of
the column density, $N$, along the direction of propagation, i.e., along the local magnetic field. 
We adopt the ISM composition by
\cite{Wilms+2000}, which is summarised in Table~\ref{tabwilms}. 
%
%
The column density is related to the surface density, $\Sigma=\bar A m_{p}N$, where $\bar A$ is the mean molecular weight ($\bar A=2.35$ for a molecular cloud and $\bar A=1.28$ for an atomic cloud) and $m_{p}$ is the proton mass.
%

\begin{table}[!h]
\caption{Assumed composition of the ISM~\citep{Wilms+2000}. Number of electrons ($Z$), mass number ($A_{Z}$), 
ratio of gas abundance to total IS medium abundance 
($1-\beta_Z$),
and abundance with
respect to the total number of particles ($f_{Z}$). 
}
\begin{center}
\begin{tabular}{ccccc}
\hline\hline
species & $Z$ & $A_{Z}$ & $1-\beta_Z$ & $f_{Z}^{(a)}$ \\
\hline
H     & 1 & 1 & 1&$9.10\times10^{-1}$ \\
He    & 2 & 4 & 1&$8.89\times10^{-2}$ \\
C     & 6 & 12& 0.5&$2.18\times10^{-4}$ \\
N     & 7 & 14& 1&$6.91\times10^{-5}$ \\
O     & 8 & 16& 0.6&$4.46\times10^{-4}$ \\
Ne    & 10& 20& 1&$7.93\times10^{-5}$ \\
Na    & 11& 23& 0.25&$1.31\times10^{-6}$ \\
Mg    & 12& 24& 0.2&$2.28\times10^{-5}$ \\
Al    & 13& 27& 0.02&$1.95\times10^{-6}$ \\
Si    & 14& 28& 0.1&$1.69\times10^{-5}$ \\
P     & 15& 31& 0.6&$2.39\times10^{-7}$ \\
S     & 16& 32& 0.6&$1.12\times10^{-5}$ \\
Cl    & 17& 35& 0.5&$1.20\times10^{-7}$ \\
Ar    & 18& 40& 1&$2.34\times10^{-6}$ \\
Ca    & 20& 40& 0.003&$1.44\times10^{-6}$ \\
Ti    & 22& 48& 0.002&$5.88\times10^{-8}$ \\
Cr    & 24& 52& 0.03&$2.95\times10^{-7}$ \\
Mn    & 25& 55& 0.07&$1.99\times10^{-7}$ \\
Fe    & 26& 56& 0.3&$2.45\times10^{-5}$ \\
Co    & 27& 59& 0.05&$2.57\times10^{-8}$ \\
Ni    & 28& 59& 0.04&$1.02\times10^{-6}$ \\
\hline
\end{tabular}
\end{center}
\footnotesize{$(a)$ computed assuming no depletion 
($\beta_Z=0$).}
\label{tabwilms}
\end{table}%

The expression for the energy loss function, $L_k^{l}$, for a particle of species $k$ depends on the
type of process $l$, as anticipated in Sect.~\ref{transport}: if only a small fraction of the particle kinetic energy is lost in each particle collision, the process can be considered as {\em continuous} and described by the loss function
\be\label{loss_cont} L_k^l(E) = \int_{0}^{E^{\rm max}}E'\frac{\ud\sigma_k^l(E,E^{\prime})}{\ud E^{\prime}}\ud E^{\prime}\,,
\ee
where $\ud\sigma_k^l/\ud E^{\prime}$ is the differential cross section of the process and $E^{\rm max}$ is the maximum
energy lost in a collision. In the opposite case, where a large fraction of the kinetic energy is lost in
a single collision or the CR particle ceases to exist after the collision, the process is called  
{\em catastrophic} and the loss function is
\be\label{loss_cat} L_k^l(E) = E\sigma_k^l(E)\,, \ee
where $\sigma_k^l$ is the cross section of the process.
In the following, we express the total energy loss function $L_k=\sum_lL_k^l$
in terms of the loss functions for collisions with atomic hydrogen ($L_{k,\h}$) or helium ($L_{k,\he}$). The loss function for 
collisions with molecular hydrogen can be approximated as
$L_{k,{\rm H}_2}\approx 2L_{k,{\rm H}}$, except for elastic and excitation losses.

\subsubsection{Protons}

The proton energy loss function, $L_{p}(E)$, is dominated by 
ionisation losses ($L_{p}^{\rm ion}$) at low energies,
and losses due to pion production ($L_{p}^{\pi}$) above the threshold 
energy $E^{\pi}$=280~MeV :
\be\label{lossprotons} 
L_{p}(E)=\varepsilon^{\rm ion}L_{p,\h}^{\rm ion}(E)
+\varepsilon^{\pi}L_{p,\h}^{\pi}(E)\,.
\ee
%
Ionisation losses are proportional to the atomic number,
$Z$, of the target species {(Bethe-Bloch formula, see~\citealt{Hayakawa1969book})}, so that $L_{p,Z}^{\rm ion}=ZL_{p,\h}^{\rm ion}$.
The total ionisation loss function is
\be\label{Lpion}
L_{p}^{\rm ion}(E)=\left(
\sum_{Z\geq1}f_{Z}Z\right)L_{p,\h}^{\rm ion}(E)=%
\varepsilon^{\rm ion}L_{p,\h}^{\rm ion}(E)\,,
\ee
where $\varepsilon^{\rm ion}=1.10$. 
At higher energies, above a threshold
$E^{\pi}=280$~MeV, we add energy losses due to pion production, as given by ~\cite{Schlickeiser2002book} and~\cite{KrakauSchlickeiser2015},
\be\label{pionlosses}
L^{\pi}_{p,Z}(E)\approx 2.57\times10^{-17}\frac{A_{Z}^{0.79}}{\beta}\left(\frac{E}{\rm GeV}\right)^{1.28}
\left(\frac{E+E^{\rm as}}{\rm GeV}\right)^{-0.2}~{\rm eV~cm^{2}},
\ee
where $\beta=v/c$ is the ratio between the proton speed and the speed of light, and the asymptotic
energy $E^{\rm as}=200$~GeV. The factor $A_{Z}^{0.79}$ is a phenomenological correction to the pion production cross
section for heavier target species~\citep{Geist1991}.
Pion losses become dominant for $E\gtrsim 1$~GeV, fully determining the propagation of high-energy CRs at high column
densities.
The total pion production loss function is then
\be\label{Lpi} L_{p}^{\pi}(E)=\left(
\sum_{Z\geq1}f_{Z}A_{Z}^{0.79}\right)L_{p,\h}^{\pi}(E)=
\varepsilon^{\pi}L_{p,\h}^{\pi}(E)\,, \ee
where $\varepsilon^{\pi}=1.18$.

\subsubsection{Electrons and positrons}

The loss function for electrons and positrons, hereafter simply $L_{e}$, is dominated by ionisation losses ($L_{e}^{\rm ion}$) at low
energies, by BS losses ($L_{e}^{\rm BS}$) above $\sim100$~MeV  and by synchrotron losses ($L_{e}^{\rm syn}$) above $E^{\rm syn}\sim1$~TeV :
\be\label{losselectrons} L_{e}(E)=\varepsilon^{\rm ion}L_{e,\h}^{\rm ion}(E)+\varepsilon^{\rm BS}L_{e,\h}^{\rm BS}(E)+L_{e}^{\rm syn}(E)\,.
\ee
%
Ionisation losses for electrons have the same correction factor as protons, $L_{e}^{\rm ion}(E)=\varepsilon^{\rm
ion}L_{e,\h}^{\rm ion}(E)$, see Eq.~(\ref{Lpion}).
Taking into account that $L_{e,\hh}^{\rm BS}=2L_{e,\h}^{\rm BS}$
and that the differential BS cross section is proportional to $Z(Z+1)$, this yields
\be\label{LBS}
L_{e}^{\rm BS}(E)=\left(
\sum_{Z\geq1}f_{Z}\frac{Z(Z+1)}{2}\right)L_{e,\h}^{\rm BS}(E)=%
\varepsilon^{\rm BS}L_{e,\h}^{\rm BS}(E)\,,
\ee
where $\varepsilon^{\rm BS}=1.22$.
Synchrotron losses dominate at energies above $E^{\rm syn}\sim 1$~TeV and do not depend on the composition. Following~\cite{Schlickeiser2002book}, $L^{\rm syn}_e(E)$ is
\be\label{Lsyn} L^{\rm syn}_e(E)\approx 5.0\times10^{-14}\left(\frac{E}{\rm TeV}\right)^{2}\, \mathrm{eV~cm^{2}}\,, \ee
which is independent of the gas density, $n$, if the magnetic field strength 
varies as $B\propto n_{\rm H}^{1/2}$.
Inverse Compton energy losses may be as important as synchrotron losses at high energies in diffuse gas, especially in the vicinity of an intense radiation source. However, in the deep interior of circumstellar discs stellar radiation is completely attenuated and inverse Compton losses are negligible.
%
%

\subsubsection{Photons}

Photons are generated through BS by electrons and positrons and through decay of neutral pions (produced
by CR protons). 
The
photon energy loss function, $L_{\gamma}$, is a sum of three contributions: photoionisation ($L^{\rm PI}_{\gamma}$), Compton
scattering ($L_{\gamma}^{\rm C}$), and pair production ($L_{\gamma}^{\rm pair}$),
\be\label{lossphotons}
L_{\gamma}(E)=L^{\rm PI}_{\gamma}(E)+\varepsilon^{\rm C}L_{\gamma,\h}^{\rm C}+\varepsilon^{\rm pair}L_{\gamma,\h}^{\rm pair}\,.
\ee
%
Photoionisation and pair production are catastrophic processes. Their loss functions are proportional to the
corresponding cross sections $\sigma^{\rm PI}$ \citep{YehLindau1985,Draine2011book} and $\sigma^{\rm pair}$ \citep{BetheMaximon1954}, see eq.~(\ref{loss_cat}). 
Conversely, the Compton effect is a continuous loss process; 
its associated energy loss function is described by \cite{Hayakawa1969book} and \cite{Padovani+2018a}.
Since Compton losses are proportional to $Z$, the correction due to heavy elements is the same as
for ionisation losses, $\varepsilon^{\rm C}=1.10$, while
$\varepsilon^{\rm pair}=\varepsilon^{\rm BS}$, because
the pair production and BS are symmetric processes.
The dominant contribution to $L^{\rm
PI}_{\gamma}$ is provided by the $K$-shell photoionisation of heavy species \citep[see e.g.][]{Draine2011book}.
At energies above $\sim100$~TeV, photon-photon scattering on the cosmic microwave background and/or the stellar radiation field can become important. However, this process is negligible in dense clouds and circumstellar discs.

\begin{figure}[!ht]
\begin{center}
\resizebox{.75\hsize}{!}{\includegraphics{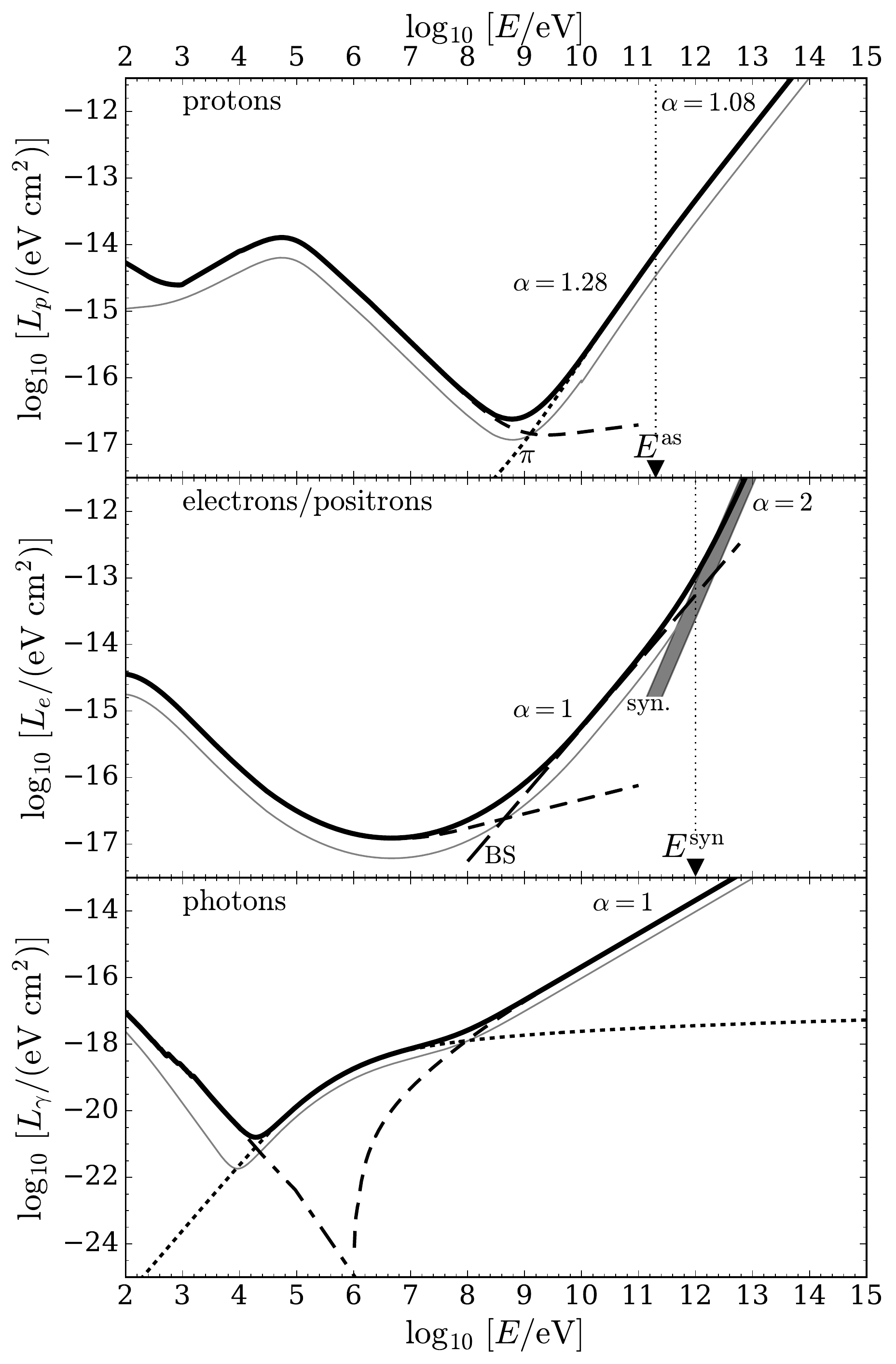}}
\caption{Total energy loss functions
of primary and secondary CR particles $k$ (protons, electrons and positrons, and photons), computed for a medium composition given in 
Table~\ref{tabwilms} assuming H in molecular form 
($L_k$, {\em thick black lines}) and for a medium
of pure atomic hydrogen
($L_{k,\h}$, {\em thin grey lines}). {\it Protons} (upper panel): ionisation losses ({\em short dashed lines})  
and pion production ({\em dotted lines});
the vertical dotted line shows the energy, $E^{\rm as}$, at which 
the proton loss function changes its asymptotic behaviour from $\alpha=1.28$ to $\alpha=1.08$.
{\it Electrons and positrons} (middle panel): ionisation losses ({\em short dashed line}), BS ({\em long-dashed line}), and synchrotron losses with the uncertainty on the magnetic field strength ({\em
shaded area});
the vertical dotted line shows the transition energy, $E^{\rm syn}$, between
BS ($\alpha=1$) and synchrotron ($\alpha=2$) losses. 
{\it Photons} (lower panel): photoionisation losses ({\em dash-dotted line}),
Compton scattering ({\em dotted line}), and pair production ({\em short-dashed line}).
Data for the ionisation by protons are taken from the 
Stopping and Range of Ions in Matter package 
~\citep{Ziegler+2010}; for the ionisation by electrons we adopt \cite{Dalgarno+1999} and the
National Institute of Standards and Technology database ({\tt
http://physics.nist.gov/PhysRefData/Star/Text}). Figure from \citet{Padovani+2018a}.}
 \label{Lfunc}
\end{center}
\end{figure}

\subsubsection{Loss functions}

\begin{figure}[!htb]
\begin{center}
\resizebox{.8\hsize}{!}{\includegraphics{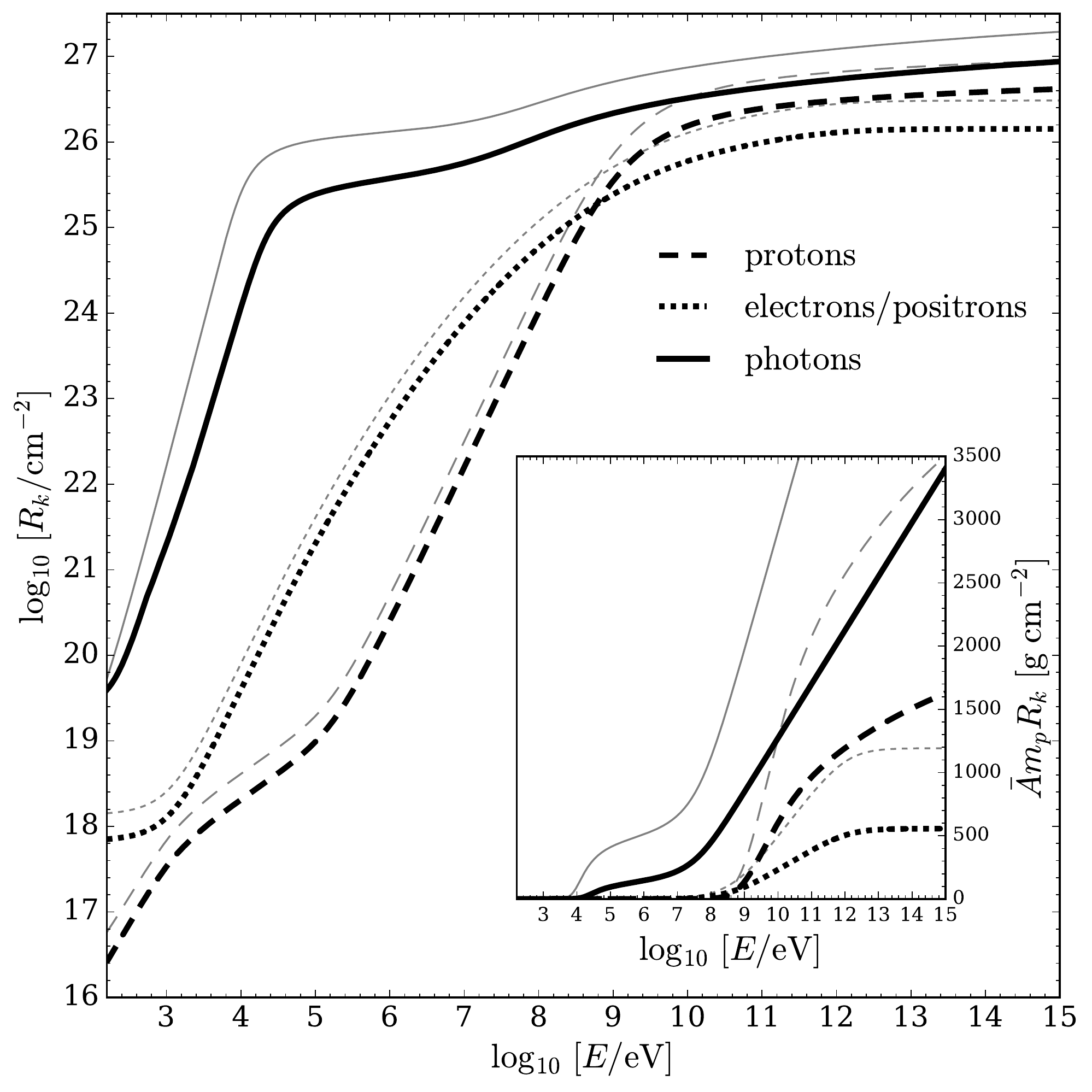}}
\caption{Total range functions, $R_k(E)$, of primary and secondary CR particles ({\em thick black lines}),  Eq.~(\ref{N}).
The inset shows the total range functions multiplied by $\bar A m_{p}$, to highlight the behaviour at large surface densities.
For comparison, the range functions for atomic hydrogen are also plotted ({\em thin grey lines}). Figure from \citet{Padovani+2018a}.} 
\label{fig1bis}
\end{center}
\end{figure}

Figure~\ref{Lfunc} shows the total proton, electron, positron and photon energy loss functions $L_k(E)$ calculated for the ISM composition given in Table~\ref{tabwilms} assuming that 
all H is in molecular form. In Fig.~\ref{fig1bis} the corresponding ranges are  calculated from Eq.~(\ref{N}).
In this case the correction factors in 
Eqs.~(\ref{lossprotons}), (\ref{Lpion}), (\ref{Lpi}),
(\ref{losselectrons}), (\ref{LBS}),
and (\ref{lossphotons}) are:
$\varepsilon^{\rm ion}=\varepsilon^{\rm C}=2.01$,
$\varepsilon^{\pi}=2.17$, and
$\varepsilon^{\rm BS}=\varepsilon^{\rm pair}=2.24$.
The loss functions in a medium of pure H atoms,
$L_{k,\h}(E)$, are also plotted for comparison.
Notice a change in the asymptotic behaviour $L_p\propto E^{\alpha}$ of the proton loss function, from $\alpha=1.28$
to $\alpha=1.08$, occurring at energy $E^{\rm as}$ (see Eq.~\ref{pionlosses}). On the other hand, the asymptotic
behaviour of the electron/positron loss function changes from $\alpha=1$ to $\alpha=2$ due to the transition from BS to
synchrotron (catastrophic) losses at energy $E^{\rm syn}$. The asymptotic behaviour of the photon loss
function is determined by the pair production (catastrophic) losses with $\alpha=1$; at low energies, where photoionisation
dominates, one can see small peaks (around 1~keV) due to $K$-shell ionisation of heavy species.

\subsection{Propagation of CRs at high column densities} 
\label{propagation}
At the high column
densities typical of circumstellar discs,
only the propagation of CR electrons
can be approximately described by CSDA, because protons and
secondary photons suffer catastrophic losses and
undergo diffusive processes. 
In fact, above $E^{\rm BS}\sim500$~MeV (see the middle panel
of Fig.~\ref{Lfunc}) the energy loss of electrons is dominated 
by BS, which is a quasi-catastrophic process, because the energy of a photon created by BS is generally of the order of the energy 
of the electron that generated it~\citep{GinzburgSyrovatskii1964}.
The effective cross section of
BS corresponds to a column density of $N^{\rm BS}\sim1.5\times10^{25}$~cm$^{-2}$, i.e.,
$\sim 58$~g~cm$^{-2}$.
Consequently, CSDA slightly overestimates the electron population at $E\gtrsim E^{\rm BS}$. However, electrons at these
energies have only a minor contribution to the ionisation rate:
smaller than 2\% at $\Sigma\sim 30$~g~cm$^{-2}$ and vanishing at higher $\Sigma$~\citep{Padovani+2018a}.


\subsubsection{Cosmic-ray protons}
\label{propagationp}

At energies larger than $E^{\pi}$ the interaction between CR protons and the medium leads to the production of
pions. Since the pion rest mass is significant, CR protons lose a non-negligible fraction of their energy in each collision \citep{Schlickeiser2002book}. 
%
\cite{Padovani+2018a} showed that CSDA works for $E\lesssim25$~MeV;
according to Fig.~\ref{fig1bis}, this corresponds to column
densities $N\lesssim7\times10^{22}$~cm$^{-2}$.
The scattering of CR protons becomes increasingly important at higher
energies, and the diffusive regime operates above $\sim 1$~GeV, corresponding to
$N\gtrsim2\times10^{25}$~cm$^{-2}$.
Thus, a CSDA solution obtained for the local CR proton spectrum at low energies should be combined with the diffusion
solution at high energies \citep[see Appendix C in][for details]{Padovani+2018a}.

The solution for the diffusive regime assuming continuous losses is governed by the following equation for ${\mathcal N}_{p}(E,\ell)$, the number of protons per unit volume and energy~\citep{GinzburgSyrovatskii1964}
as function of the coordinate $\ell$ along the local magnetic field (CR path),
\begin{equation}\label{ginz}
D_p\frac{\partial^{2}{\mathcal N}_{p}}{\partial \ell^{2}}=\frac{\partial}{\partial E}\left(\frac{{\rm d}E_p}{{\rm d}t}{\mathcal
N}_{p}\right)\,,
\end{equation}
%
%
where
\begin{equation}
D_p(E)\approx\frac{\beta c}{3n\sigma_{\rm mt}(E)}\,,
\qquad
\frac{{\rm d}E_p}{{\rm d} t}= -n\beta c L_{p}(E) \,,
\label{defloss}
\end{equation}
$\sigma_{\rm mt}$ is the total momentum transfer cross section,
$\beta c$ is the proton velocity,
%
and $n$ is the total particle number density, summed over all the medium species. The factor of 3 accounts for the fact that diffusion occurs in three dimensions.
The solution of Eq.~(\ref{ginz}) is given in \cite{Padovani+2018a}.
%
%

In principle, elastic scattering of CR protons leads to new source and sink terms in the general transport equation associated with efficient energy exchange with hydrogen nuclei, but for realistic conditions these new terms can be safely neglected.

\subsection{Generation and propagation of secondary particles}
\label{generation}

\begin{figure}[!h]
\begin{center}
\resizebox{.8\hsize}{!}{\includegraphics{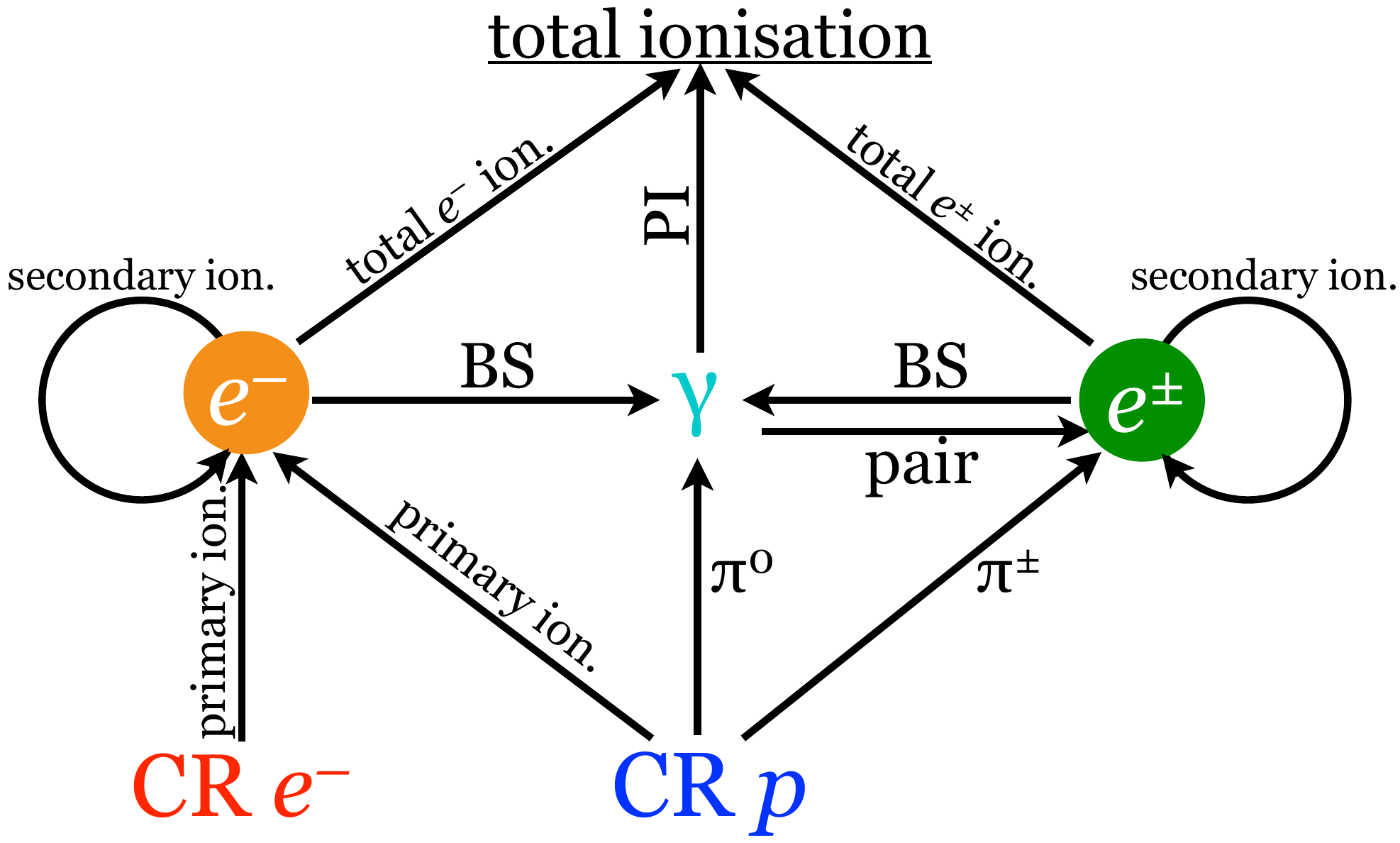}}
\caption{Ionisation diagram, explaining the effect of secondary particles that are generated (directly or indirectly) by
CR protons and electrons through ionisation, pion decay ($\pi^{0},\pi^{\pm}$),  BS, and pair production
(pair). The secondary particles include electrons ($e^-$, due to primary and secondary ionisation), positrons and electrons
($e^{\pm}$, due to pair production and $\pi^{\pm}$ decay; also electrons produced in secondary ionisation are
included), and photons ($\gamma$, due to BS and $\pi^0$ decay), all contributing to the respective ionisation routes. Figure from \citet{Padovani+2018a}.}
\label{diagram}
\end{center}
\end{figure}

Figure~\ref{diagram} shows the main ionisation routes associated with secondary particles produced
by CRs at high column densities such as in circumstellar discs. Assuming that at these high densities all H is in molecular form, CR protons and electrons generate secondary electrons by primary ionisation ($p_{\rm CR}+{\rm
H_{2}}\rightarrow p_{\rm CR}'+{\rm H_{2}^{+}}+e^{-}$). Secondary electrons with energy larger than the ionisation
potential of H$_{2}$ produce further generations of ambient electrons. CR protons colliding with  protons create charged pions; through muon decay, they become electrons and positrons ($\pi^{\pm}\rightarrow\mu^{\pm}\rightarrow e^{\pm}$), producing
secondary ionisations. In addition, $p$--$p$ collisions create neutral pions decaying into photon pairs
($\pi^{0}\rightarrow2\gamma$). The second important source of photons is BS by electrons and positrons. One
should also account for $e^{\pm}$ pair production by photons ($\gamma\rightarrow e^++e^-$). In the following sections we summarise the basic elements needed to compute the secondary particle flux.

\subsubsection{Photons}
\label{photons_sec}

The photon propagation regime depends on their energy $E_\gamma$. Photoionisation and pair production dominate below $\sim 5$~keV and above $\sim 50$~MeV, respectively.
Both processes are catastrophic, i.e., photons disappear after the interaction with nuclei.
As for Compton scattering,
the relative average energy lost by a photon in each interaction with electrons depends on the photon's energy. 
For $E_\gamma\gg m_{e}c^{2}$ losses are catastrophic, whereas in the 
opposite limit photons transfer a small amount of their energy 
to the electrons, and losses can be treated as continuous.
However, below $E\sim1$~keV photoionisation is the dominant process, and losses become catastrophic again. The solution of the transfer equation for photons in either regime is given in \cite{Padovani+2018a}.

\subsubsection{Electrons and positrons}
\label{elepos_sec}

Electrons and positrons have two different sources. First, pairs are produced by photons with energy above
$2m_ec^2$, so that the electron and positron energy is $0 \le E \le E_\gamma-2m_ec^2$.  Second,
electrons and positrons are created through decay of charged pions, generated in proton-nucleus collisions at energies
above $E^{\pi}=280$~MeV.
As pointed out at the beginning of 
Sect.~\ref{propagation}, the use of CSDA
to describe the propagation of electrons and positrons with energies larger than the
BS threshold, $E^{\rm BS}\sim500$~MeV,
leads to a slight overestimation of the 
flux. However, since the contribution of
$e^\pm$ of such energies to the ionisation
rate is negligible, CSDA can be employed.
Thus, when the stopping range is comparable
to (or larger than) the local column density
$N$, the resulting spectrum of electrons and
positrons is given by
\begin{equation}
\label{convolution}
j_{e^{\pm}}(E,N)=\frac{1}{2L_{e}(E)}\int_N^\infty S_{e^{\pm}}(E_0,N_0)L_{e}(E_0)\,\ud N_0,
\end{equation}
where  $S_{e^\pm}$ is the total source
function for electrons and positrons
(including pair production and
charged pion decay) and
the factor $1/2$ accounts for electrons and positrons propagating in two directions. The initial energy $E_0>E$ at $N_0$ is related to $N$ by
\begin{equation}
|N_0-N|=\int_{E}^{E_0}\frac{\ud E'}{L_{e}(E^\prime)}\,.
\end{equation}
If the range is small, $|N_{0}-N|\ll N$, the spectrum is localised,
\begin{equation}\label{loc}
j_{e^{\pm}}(E,N)=\frac{E}{L_{e}(E)} S_{e^{\pm}}(E,N).
\end{equation}
An a posteriori check of the energy spectra calculated with Eq.~(\ref{convolution}) for $N\gtrsim10^{24}$~cm$^{-2}$ shows
that they are accurately reproduced by Eq.~(\ref{loc}).




\subsection{Role of magnetic field geometry in CR propagation and ionisation}  
\label{mirr-foc}

The magnetic configuration in dense cores and discs can be very complicated \citep[e.g.,][]{Joos+2012,Li+2013,Padovani+2013,Hull+2017,Hull+2018},  
and the field strength is usually much larger than the interstellar value \citep{Crutcher2012}. Dust polarisation observations of dense cores often display an hourglass shape, which can be created by gravitational contraction  {\it pinching} the magnetic field \citep{Girart+2006,Girart+2009,Hull+2014}.
Simulations of magnetised collapsing cores reproduce the hourglass field morphology and show a significant toroidal component near the centre that is produced by rotation \citep{Padovani+2013,Lee+2017}. 
The magnetic field configuration in turn influences the details of CR propagation and local ionisation.
The field strength increases along the converging field lines, which leads to efficient mirroring of the
penetrating CRs -- their pitch angles increase in response to the growing field until reaching $90^\circ$, and thus more and
more particles are reflected back. On the other hand, field line convergence results in CR focusing. Below we
explore the net effect of these two competing processes.

First, we consider a situation without losses. In the absence of scattering processes, the pitch angle satisfies the
relation
\begin{equation}
\frac{1-\mu^2}{B(s)}=\frac{1-\mu_i^2}{B_i},
\end{equation}
which follows from the adiabatic invariance of the magnetic moment of a particle \citep{Chen1984book}. The Liouville theorem
yields conservation of the distribution function along the field lines, and therefore \citep{CesarskyVolk1978,Silsbee+2018}
\begin{equation}\label{Liou}
j(E,s,\mu)=j^{\rm IS}(E,\mu_i).
\end{equation}
We are interested in the propagation of CRs in the direction of increasing $B(s)$. Hence, if the initial distribution is
isotropic, which is assumed for low-energy Galactic CRs, the local distribution of penetrating CRs remains
isotropic, too. We immediately infer from Eq.~(\ref{zeta}) that the ionisation rate does not depend on $N$ in this case
-- an expected result, as Eq.~(\ref{Liou}) also implies conservation of the differential density of penetrating
particles.

Of course, introducing energy losses breaks this symmetry. Among all CRs being able to reach a given position before they are mirrored at deeper positions, particles with a larger initial pitch angle must cross a larger gas column ($N/\mu$)
and therefore experience stronger losses. This effect can be taken into account in the framework of free-streaming transport
discussed in Sect.~\ref{section_FS}. Rigorous analysis shows that for all possible forms of monotonically increasing
$B(s)$ the resulting dependencies $\zeta_{\rm H_2}(N)$ are bound between well-defined lower and upper limits
\citep{Silsbee+2018},
\begin{equation}\label{bounds}
\zeta_{\rm min}\leq\zeta_{\rm H_2}\leq \zeta_{\rm max}.
\end{equation}
The lower bound, $\zeta_{\rm min}(N)$, is realised in the case of a constant magnetic field strength and is equal to its initial
value up to $N$. Interestingly, the ratio of the upper and lower bounds {\it does not depend} on $N$; for a
power-law initial CR spectrum, it is equal to
\begin{equation}\label{zeta_ratio}
\frac{\zeta_{\rm max}}{\zeta_{\rm min}}=\frac{a+2b}{1+b}\:.
\end{equation}
We conclude that for any density profile and monotonically increasing magnetic field the combined impact of mirroring
and focusing only slightly increases the ionisation rate (relative to the constant-field value, $\zeta_{\rm
min}$). For realistic values of the spectral index $a\leq1$, $\zeta_{\rm max}/\zeta_{\rm min}<1.5$; this ratio
naturally tends to unity when $a=1-b\sim0.2$.

Thus, the effects of magnetic mirroring and focusing on the local CR ionisation (or the CR density) practically cancel each
other out when CRs propagate along the lines of monotonically increasing magnetic field. Therefore, irrespective of the field structure complexity, $\zeta_{\rm H_2}(N)$ can be calculated by simply assuming a constant field strength
\citep{Silsbee+2018}. 

\begin{figure}[!h]
\begin{center}
\resizebox{10cm}{!}{\includegraphics{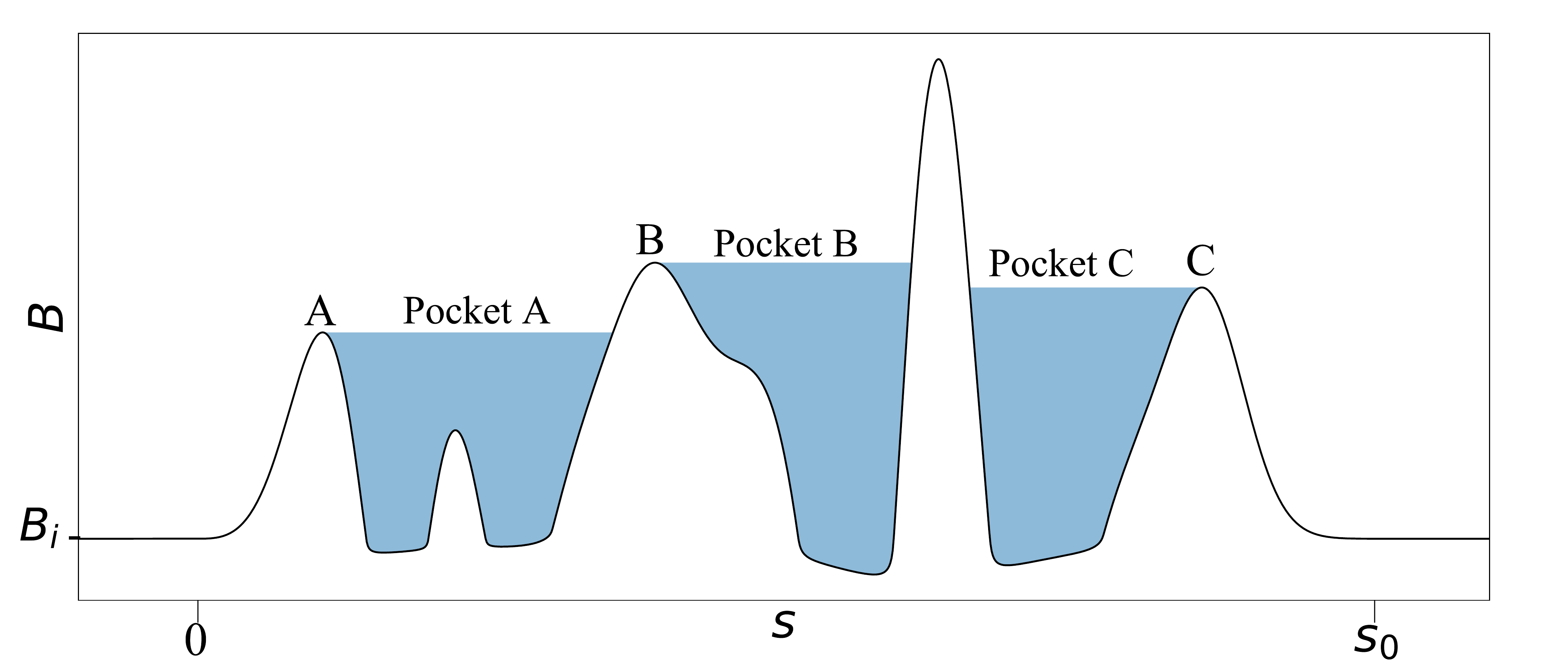}}
\caption{Sketch illustrating a situation where the magnetic field has multiple local minima along the field line. This
results in multiple magnetic pockets, indicated by the shading.
Figure from \citet{Silsbee+2018}.}
\label{pockets}
\end{center}
\end{figure}

The situation may change dramatically if the magnetic field strength has local minima along the field lines. The magnetic
mirroring and focusing are no longer able to balance each other in such {\it magnetic pockets} sketched in
Fig.~\ref{pockets}. Indeed, the pitch angles of particles moving within a pocket can only decrease with respect to the values
at the pocket border. This removes particles with pitch angles around $90^\circ$ from the distribution
function and reduces the CR density.

Calculations performed in the absence of losses give the following expression for the number of CRs per unit volume
and energy in a magnetic
pocket \citep{KaplanPikelner1970,Silsbee+2018}:
\begin{equation}\label{N_pocket}
\frac{\mathcal{N}(E,s)}{\mathcal{N}^{\rm IS}(E)}=1-\sqrt{1-\frac{B(s)}{B_{\rm low}}}\:,
\end{equation}
where $\mathcal{N}^{\rm IS}(E)$ is the CR density per unit energy outside pockets (equal to the interstellar density) and $B_{\rm low}$ is the field
strength at the lower peak bounding a pocket (indicated by A, B, and C in Fig.~\ref{pockets}).  Eq.~(\ref{N_pocket}) remains applicable even if $B(s)$ has small peaks inside a pocket (as, e.g., for pocket A). The
local ionisation rate naturally follows the same dependence. Inclusion of losses does not practically modify the effect:
 the ionisation rate is bound between the lower and
upper limits with the same relative range given by Eq.~(\ref{zeta_ratio}), similar to the case of a monotonically increasing field.

Magnetic pockets are expected to be omnipresent in the inner regions of dense molecular cloud cores. Repetitive stretching
and compression of the field lines due to large-scale turbulence in collapsing cores results in the formation of
deep pockets. This phenomenon becomes particularly prominent where the magnetic field has a strong toroidal component
produced by rotation \citep{Padovani+2013}. Consequently, in dense cores with $n_{\rm H}\gtrsim10^4$~cm$^{-3}$ numerical models show
the CR ionisation rate can be reduced by more than an order of magnitude. This has clear implications for gas chemistry and
angular momentum transport, which are important ingredients in the star-formation process.

In conclusion, we point out that the above analysis applies to the free-streaming regime of CR propagation. In the presence
of weak, small-scale MHD turbulence, resonant pitch-angle scattering can lead to the efficient isotropisation of the
distribution function, completely eliminating the effect of magnetic pockets. In this diffusive regime the results presented
in Sect.~\ref{section_D} are applicable for any magnetic configuration, irrespective of the behavior of $B(s)$. Note,
however, that very strong turbulence may further increase mirroring, and thus reduce the overall density of CRs in cores \citep{FatuzzoAdams2014}.

\subsection{Effect of CR-generated turbulence}
\label{section_self-gen}

It has long been known \citep[see][]{Lerche1967,KulsrudPearce1969} that a CR flux propagating through a plasma can excite MHD
waves and thus create magnetic disturbances. However, the effect of these disturbances on the penetration of CRs into
molecular clouds and on the resulting ionisation is still debated \citep{SkillingStrong1976,CesarskyVolk1978, MorlinoGabici2015,Ivlev+2018,
Phan+2018,Dogiel+2018}.

\citet{SkillingStrong1976} predicted from a qualitative analysis of the problem a depletion of CR density inside the clouds. They
concluded that Alfv\'en waves should be generated by streaming CRs in the cloud envelopes where the gas density is low enough
to sustain the waves, which suppresses penetration by sub-GeV CRs. This statement was confirmed by analytical and numerical
calculations carried out by \citet{Ivlev+2018}. It was also shown that the flux penetrating into {\it very dense} clouds
(where CRs are fully absorbed) has a universal energy dependence -- it is exclusively determined by the densities of ionised
and neutral components of the cloud envelopes. Furthermore, \citet{Ivlev+2018} and \citet{Dogiel+2018} demonstrated that the
turbulence leading to the universal flux can be generated either by CR absorption in the vicinity of a dense cloud or by CR energy losses in the outer part of the envelope. In both cases, turbulence makes the spectrum of low-energy
CRs harder and independent from that in the ISM.

We now consider the condition to generate MHD turbulence by CRs penetrating into a molecular cloud and obtain the resulting modulated CR flux \citep{Ivlev+2018,Dogiel+2018}. If the CR kinetic energy, $E$, exceeds a certain {\it excitation
threshold}, $E_{\rm ex}$ (discussed below), their flux is not able to excite MHD turbulence in the cloud envelope, i.e., the propagation of such CRs through the envelope is free-streaming, as described in
Sect.~\ref{section_FS}. The excitation occurs for $E\lesssim E_{\rm ex}$, where the magnitude of the turbulent magnetic
disturbances becomes large enough to ensure diffusive propagation. The CR flux, $S$, is then given by Eq.~(\ref{S_D})
with the advection velocity, $u$, equal to
\begin{equation}
v_{\rm A}=\frac{B}{\sqrt{4\pi \rho_i}}\:,
\end{equation}
the Alfv\'en velocity of the excited MHD waves propagating in the direction of the flux
. The diffusion coefficient, $D$, can be obtained from Eq.~(\ref{D}), where $W(k)$ is no
longer an independent quantity but is determined by the CR flux itself.

The turbulent spectrum, $W(k)$, is generally governed by a wave equation, including the dominant processes of excitation,
damping and transport, as well as nonlinear wave interaction \citep{LagageCesarsky1983,NormanFerrara1996, Ptuskin+2006}. Similar to the
simplified expression~(Eq.~\ref{D}) for the diffusion coefficient, it is possible to derive an approximate expression for the growth rate (amplitude) of MHD waves excited by streaming CRs \citep{Skilling1975},
\begin{equation}\label{gamma_CR}
\gamma_{\rm CR}(k,z) \approx\pi^2 \frac{e^2v_{\rm A}}{mc^2\Omega} pv\left(4\pi\langle S\rangle-v_{\rm A}\mathcal{N}\right),
\end{equation}
where the wavenumber, $k$, is related to the particle energy (momentum) by the resonance condition~(Eq.~\ref{k_res}),
$\langle\ldots \rangle=\frac12\int_{-1}^1\ldots d\mu$ is the pitch-angle average for penetrating CRs, and $\mathcal{N}(E,s)=
2\pi\int_{-1}^{1}f_E\:d\mu$ is the CR density per unit energy. Eq.~(\ref{gamma_CR}) can be applied both in the free-streaming
and diffusive regimes by substituting the respective expressions for the flux, $S$.

The excitation threshold, $E_{\rm ex}$, is determined by balancing the growth rate and the damping rate, $\nu_{in}$,
due to ion collisions with gas \citep{KulsrudPearce1969},
\begin{equation}
\nu_{in}\approx\frac12\frac{m_n}{m_i}\langle\sigma v\rangle_{in} n_n\,,
\end{equation}
which depends on the (velocity-averaged) momentum-transfer rate for ion-gas collisions, $\langle\sigma v\rangle_{in}$,
the ratio $m_n/m_i$ of the mean gas particle mass to the mean ion mass, and the density, $n_n$, of gas
particles. The value of $E_{\rm ex}$ is obtained by substituting the {\it net} average flux, $\langle S\rangle$, of
free-streaming CRs penetrating the cloud. If all penetrating CRs are absorbed in the cloud, the net free-streaming flux is given
by Eq.~(\ref{S_FS}) for $\mu>0$; then $4\pi\langle S\rangle=\frac14v\mathcal{N}^{\rm IS}$ for an isotropic interstellar
distribution, i.e., the net flux velocity is $u_{\rm net}=\frac14v$ as expected. If  absorption is incomplete, the net
velocity $u_{\rm net}=\langle S\rangle/f^{\rm IS}$ naturally decreases to the value determined by dominant energy losses in a
cloud. One can generally present the net velocity as \citep{Dogiel+2018}
\begin{equation}\label{u_net}
u_{\rm net}(E) = \frac14v\times\min\left\{N \left[\sigma(E)+\ell\frac{L(E)}{E} \right],1\right\} \,,
\end{equation}
where 
$\sigma(E)$ is the total cross section of catastrophic
(spallation) collisions, and $L(E)$ is the continuous (ionisation) loss function; the factor $\ell\sim1$ is determined by
the form of the interstellar CR spectrum at the cloud edge. The excitation-damping balance leads to the following equation
for $E_{\rm ex}$:
\begin{equation}\label{E_ex}
\left.\tilde p\tilde j^{\rm IS}(E)\left(\frac{u_{\rm net}(E)}{v_{\rm A}}-1\right)\right|_{E=E_{\rm ex}}=\tilde\nu,
\end{equation}
where $\tilde p(E)=p(E)/mc$ is the dimensionless momentum of a CR particle, $\tilde j^{\rm IS}(E)=j^{\rm IS}(E)/j_*$ is the dimensionless
interstellar spectrum normalised by its characteristic value $j_*=j^{\rm IS}(E=mc^2)$, and $\tilde\nu$ is the dimensionless damping
rate. The scaling dependence of $\tilde\nu$ on the physical evelope parameters is given by the following general
expression \citep{Ivlev+2018,Dogiel+2018}:\footnote{For convenience, the value of $\tilde\nu$ used here is twice the value of
$\nu$ in \citet{Ivlev+2018} and \citet{Dogiel+2018}.}
\begin{eqnarray}
\tilde\nu = 17.4 \left(\frac{m_n/m_p}{2.3}\right)
    \left(\frac{j_* m_p c^2}{0.57~\mbox{cm}^{-2}\mbox{s}^{-1}\mbox{sr}^{-1}}\right)^{-1} \hspace{2.6cm}\label{scale_nu}\\
	\times\left(\frac{n_i/n_n}{3\times 10^{-4}}\right)\left(\frac{n_n}{100~\mbox{cm}^{-3}}\right)^2
    \left(\frac{B}{0.1~\mbox{mG}}\right)^{-1}.\nonumber
\end{eqnarray}

To summarise, we conclude that CRs with $E>E_{\rm ex}$ penetrate a cloud freely, and their (pitch-angle averaged) flux is
simply equal to $u_{\rm net}(E)f^{\rm IS}(E)$. For $E\lesssim E_{\rm ex}$, the flux is self-modulated: CRs propagate diffusively,
with the diffusion coefficient determined by the flux itself. The flux penetrating a cloud in this case is given by
\citep{Ivlev+2018,Dogiel+2018}
\begin{equation}
\label{flux1}
S(E)=\frac{v_{\rm A}f^{\rm IS}(E)}{1-[1-v_{\rm A}/u_{\rm net}(E)]e^{-\eta_0(E)}}\,,
\end{equation}
where $\eta_0$ is a measure of the relative importance of diffusion and advection in the modulated flux. For $v_{\rm
A}/u_{\rm net}\lesssim\eta_0\lesssim1$ the flux is dominated by the diffusion term, viz. by the first term in
Eq.~(\ref{S_D}), while for $\eta_0\gtrsim1$ interstellar CRs are advected with the Alfv\'en velocity, $v_{\rm A}$.
The value of $\eta_0$ can be well approximated by
\begin{equation}\label{zeta_0}
\eta_0(E)\approx\sqrt{\tilde E(\tilde E+2)}\;\frac{\tilde j^{\rm IS}(E)}{\tilde\nu}\,,
\end{equation}
which is valid as long as $\eta_0\lesssim1$ (otherwise, its value is unimportant for calculating the flux). Thus, in the {\it
diffusion-dominated} regime $v_{\rm A}/u_{\rm net}\lesssim\eta_0\lesssim1$ and Eqs.~(\ref{flux1}) and 
(\ref{zeta_0}) give the flux as
\begin{equation}\label{S_DD}
S_{\rm DD}(E)=\frac{\tilde\nu}{\tilde p\tilde v}\frac{v_{\rm A}}{c}j_*,
\end{equation}
while in the {\it advection-dominated} regime it is simply $S_{\rm AD}(E)=v_{\rm A}f^{\rm IS}(E)$. The CR flux
penetrating into a cloud has a universal energy dependence at $E\lesssim E_{\rm ex}$, scaling as $S_{\rm DD}(E)\propto
E^{-1}$ in both the non-relativistic and ultra-relativistic limits. Note that this flux is solely determined by the physical
parameters of the envelope, i.e., it does not depend on the interstellar CR spectrum.

The importance of CR self-modulation for typical conditions of molecular clouds is discussed in Sect.~\ref{section_self-mod}.

\section{Galactic cosmic rays in 
molecular clouds}
\label{molecularclouds}


CRs are a crucial source of ionisation within molecular clouds, shaping gas chemistry, temperature and dynamics via the degree of magnetic field-gas coupling. The ionisation rate is a function of the energy spectrum of CRs penetrating into the cloud, and thus is controlled by the interplay of ionisation energy losses and the dominant transport regime of CRs (see Sect.~\ref{transport}).
Therefore, the CR ionisation rate likely varies strongly within molecular clouds, as we show in Sect.~\ref{section_FS_vs_D} and \ref{section_self-mod}. Finally, we discuss the impact of CR ionisation variation on gas chemistry and dust physics and the resulting implications for cloud properties and observed emission in Sect.~\ref{MC:chemistry}.

\subsection{Free-streaming versus diffusive penetration of CRs}
\label{section_FS_vs_D}

The CR ionisation rate depends on both the propagation model for CRs and on their spectrum at the edge of the cloud.
Ionisation at column densities in the range of $N=10^{20}$--$10^{23}$~cm$^{-2}$ is dominated by CR protons with energies
from 1~MeV to 1~GeV. Unfortunately, the spectrum of such particles cannot be measured accurately from near Earth because
they are largely excluded by the solar wind \citep{Potgieter2013}. The Voyager probes have measured the spectrum of Galactic
CRs down to 3~MeV \citep{Cummings+2016,Stone+2019}. However, the magnetic field direction measured by the probes has not changed, as would be
expected if they 
were beyond the influence of solar modulation \citep{GloecklerFisk2015}.
Furthermore, \citet{Padovani+2018a} and \citet{Phan+2018} noted that the proton and electron spectra from Voyager spacecrafts were too low
by about a factor of 10 to explain the values of the ionisation rate observed in nearby molecular clouds. For these reasons,
there is still considerable uncertainty in the density of low-energy CRs impinging on molecular clouds.

In this section we discuss the possible effect of preexisting MHD turbulence on CR ionisation (Sect.~\ref{section_D})
and compare the results with predictions derived from the free-streaming approximation (Sect.~\ref{section_FS}).
Molecular cloud envelopes are thought to be turbulent. However, it is not known whether magnetic
disturbances associated with the turbulence would have sufficient energy at small enough scales to ensure diffusive
transport for the CRs responsible for the majority of the ionisation. Radio scintillation observations
\citep[e.g.,][]{Armstrong+1995} suggest that ISM turbulence extends to scales at least as small as $10^{10}$~cm, less
than the gyroradius of a 100~KeV proton, but it is still not clear that this result is relevant for molecular clouds.

The diffusion approximation is applicable to describe CR penetration into a cloud at depths greater  
than the mean free path $\approx3D(E)/v$ due to pitch-angle scattering. This requires the column density
$N\gtrsim 3n_{\rm H}D(E)/v$. The particles dominating the ionisation at column $N$ are those whose stopping range $R(E)$,
Eq.~(\ref{R_FS}), is comparable to $N$. Using Eq.~(\ref{D1}), we derive the following condition
\citep{SilsbeeIvlev2019}:
\begin{equation}\label{N_min}
N\gtrsim\frac{3n_{\rm H}D_0}{v_0}\left(\frac{3n_{\rm H}D_0}{v_0R_0}\right)^{\frac{1-2\lambda}{1+2b+2\lambda}}.
\end{equation}
As pointed out in Sect.~\ref{section_D}, the value of $n_{\rm H}D_0$ does not depend on $N$ for a constant ionisation
fraction and is determined by the velocity exponent $\lambda$ of the turbulent spectrum. \citet{SilsbeeIvlev2019} showed that
condition~(\ref{N_min}) is expected to hold for $N\gtrsim8\times10^{19}$~cm$^{-2}$ for a Kolmogorov spectrum and for
$N\gtrsim3\times10^{18}$~cm$^{-2}$ for a Kraichnan spectrum.

At higher column densities, the ion density is expected to drop dramatically when there are insufficient UV
photons to keep carbon ionised. This depends on the strength of the UV field near the cloud, as well as on the assumed
properties of the medium \citep{HollenbachTielens1999}. Based on the work of \citet{KetoCaselli2008}, we assume the transition to take
place at $N_{\rm tran}\sim2\times10^{21}$~cm$^{-2}$, though we note that \citet{NeufeldWolfire2017} find a very sharp decline in
C$^+$ abundance near a column density of $6\times10^{20}$~cm$^{-2}$. For $N\gtrsim N_{\rm tran}$, the ion density is
expected to drop by a factor of $\sim100$, depending on the value of $\zeta_{\rm H_2}$ \citep{NeufeldWolfire2017}, leading to a
proportional increase in $D_0$. It appears unlikely that the turbulence would then be strong enough to greatly influence CR propagation. In environments with larger CR fluxes or/and UV radiation, this boundary may move to higher column
density. Specifically, \citet{NeufeldWolfire2017} find that if $\zeta_{\rm H_2}/n_{\rm H}>1.2\times10^{-17}$~cm$^3$s$^{-1}$, then
the ionisation fraction remains greater than $10^{-4}$ to $N\sim10^{22}$~cm$^{-2}$. Such conditions may be found near the
Galactic centre \citep{LePetit+2016}.

\begin{figure}[!h]
\begin{center}
\resizebox{8cm}{!}{\includegraphics{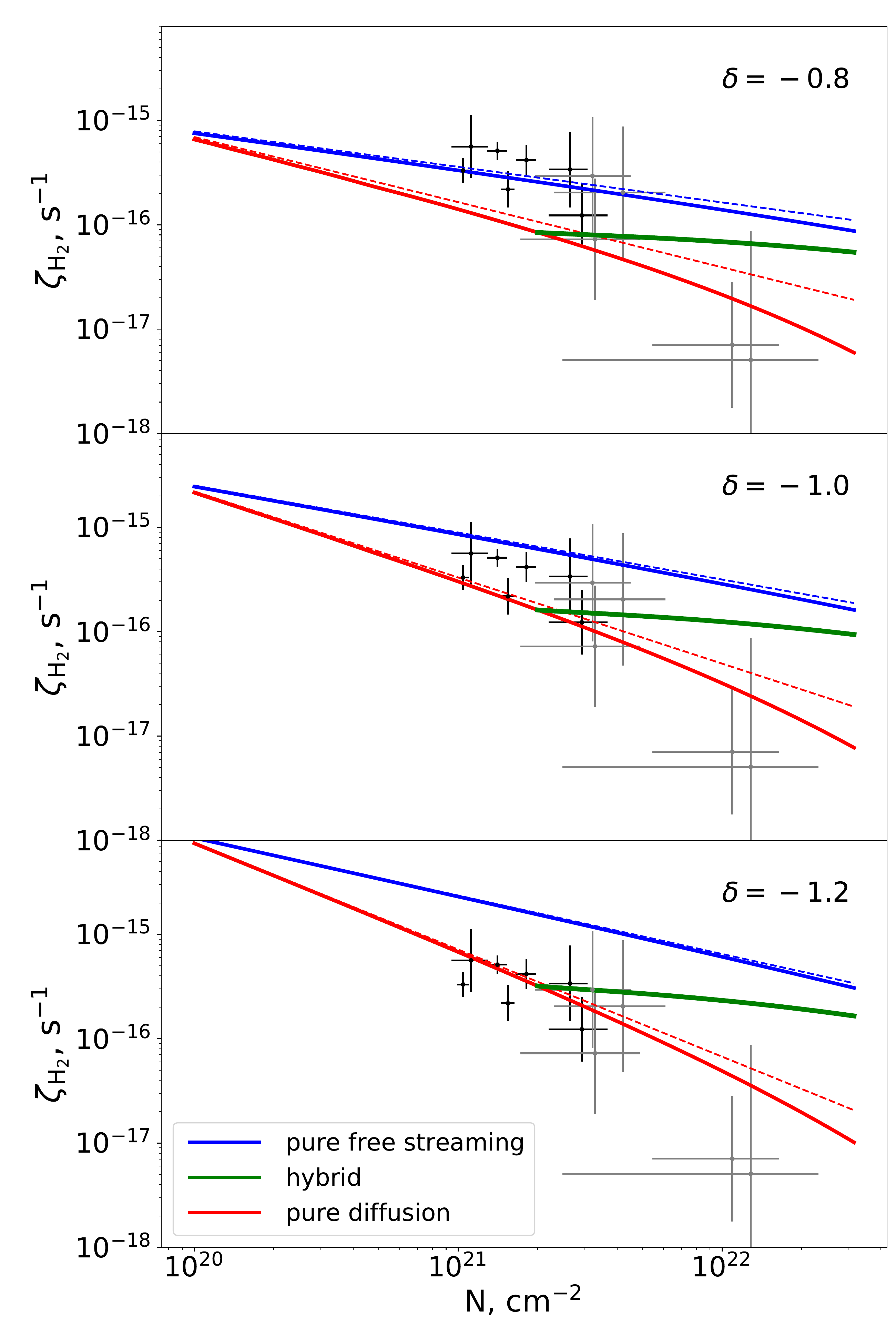}}
\caption{Ionisation rate of molecular hydrogen (not including secondaries), $\zeta_{\rm H_2}$, as a function of the effective column density of
hydrogen atoms, $N$, for three different CR propagation models (see the legend). Different panels correspond to different
values of $\delta$ in the assumed initial spectrum, Eq.~(\ref{model_j}). The points and error bars are estimated from
Fig.~6 of \citet{NeufeldWolfire2017}; black crosses are those for which the H$_2$ column density has been measured
directly. The solid lines represent the results for Eq.~(\ref{model_j}), the dashed lines are for the power-law
spectrum in Eq.~(\ref{ini}). Note that the column density displayed here is a factor of 1.67 larger than that in
\citet{Padovani+2018a} (see Footnote~\ref{fn1}).
Figure from \citet{SilsbeeIvlev2019}.} \label{FS_vs_D}
\end{center}
\end{figure}

Padovani et al. (2018) propose an interstellar proton spectrum of the following model form:
\begin{equation}\label{model_j}
j^{\rm IS}(E)=C\frac{E^\delta}{(E+E_t)^\beta}~\mathrm{eV\m cm\mm s\m sr\m}\,,
\end{equation}
where $C=2.4\times10^{15}$ and $E_t=650$~MeV.
The high-energy slope of this function, $\delta-\beta=2.7$, is well determined \citep[e.g.,][]{Aguilar+2014,Aguilar+2015}, while
the low-end slope, $\delta$, is highly uncertain.
A simple extrapolation of the Voyager data at lower energies fails to
reproduce the CR ionisation rate measured in diffuse clouds from H$_3^+$ emission (see Sect.~\ref{Observations}). For this
reason, 
it is appropriate to consider two different models for the CR proton spectrum: a ``low'' spectrum, ${\mathscr L}$,
obtained by extrapolating the Voyager data ($\delta=0.1$), and a ``high'' spectrum, ${\mathscr H}$ ($\delta=-0.8$). The
two spectra must be considered as lower and upper bounds, respectively, to the actual average Galactic spectrum of CR protons \citep{Ivlev+2015,Padovani+2018a}.


Figure~\ref{FS_vs_D} shows the ionisation rate calculated for three different propagation models \citep{SilsbeeIvlev2019},
assuming the interstellar spectrum of Eq.~(\ref{model_j}). The blue curve describes pure free-streaming
propagation. The green curve represents the hybrid model, which assumes that CRs propagate diffusively until $N_{\rm
tran}\sim 2\times10^{21}$~cm$^{-2}$ (such that carbon is no longer ionised) and then stream freely; at $N\gtrsim N_{\rm
tran}$, this results in a region where $\zeta_{\rm H_2}(N)$ is nearly flat in the beginning, as the ionisation is then
dominated by protons with $R(E)\gg N_{\rm tran}$. The red curve assumes pure diffusive propagation for the entire column,
ignoring the expected sharp decrease in the ion density. This represents a lower bound on $\zeta_{\rm H_2}(N)$, but such a
curve is probably unrealistic unless there are some processes, such as an anomalously high interstellar CR flux or/and a high UV
field, that enhance the ionisation fraction deeper within the cloud. Finally, the dashed lines are the analytic
approximations assuming a power-law initial spectrum. They are described by Eqs.~(\ref{zeta_FS}) and (\ref{zeta_diff})
with $a=-\delta$ and $j_0=CE_0^\delta/E_t^\beta$, which corresponds to the interstellar spectrum~(\ref{model_j}) at lower
energies.

The data points and error bars in Fig.~\ref{FS_vs_D} are taken from Fig.~6 of
\citet{NeufeldWolfire2017},\footnote{\citet{NeufeldWolfire2017} plot $\zeta_{\rm prim}$, the primary ionisation rate per hydrogen, which they
assume to be 1/2.3 times the total ionisation rate $\zeta_{\rm H_2}$ (including secondary ionisations) per H$_2$. Taking the fraction of secondary ionisation equal to 0.7~\citep{Glassgold+2012}, we shift the points from \citet{NeufeldWolfire2017} upwards
by a factor of 1.4.} assuming one magnitude of visual extinction to be equivalent to a column density of
$1.9\times10^{21}$~cm$^{-2}$. The spectrum in the top panel, corresponding to $\delta=-0.8$ in Eq.~(\ref{model_j}), was
constructed by \citet{Padovani+2018a} so that the free-streaming model passes through the points. For the other models, this
spectrum yields curves that are too low. The middle panel shows the resulting ionisation rate if $\delta = -1.0$. The low-energy slope of the resulting spectrum corresponds to the spectrum of particles produced in strong shocks
\citep{Drury1983}. Finally, in the bottom panel we consider a steeper low-energy slope of $\delta=-1.2$. In this case, the
diffusive model provides the best fit. We also point out that the slope for $\zeta_{\rm H_2}(N)$ obtained from
\citet{NeufeldWolfire2017} ($1.05\pm0.36$) is fit better by the diffusive model with $\delta=-1.2$, which appears to favor this
model. However, as illustrated by the magnitude of the error bars, the slope suggested by \citet{NeufeldWolfire2017} is rather
uncertain.

Thus, if a certain degree of preexisting turbulence is present in a cloud, CRs propagate diffusively. A substantially
steeper ionisation rate slope as a function of the effective column density is predicted in this case compared with
the free-streaming model. Under conditions appropriate for local molecular clouds, this mechanism would likely only operate
up to column densities of $\sim10^{21}$~cm$^{-2}$. However, the assumption of diffusive propagation makes a significant
difference to the behavior of $\zeta_{\rm H_2}(N)$. Equations~(\ref{zeta_FS}) and (\ref{zeta_diff}) provide convenient
analytic solutions for $\zeta_{\rm H_2}(N)$ in the two regimes, applicable for a variety of environments.

More detailed observations and analyses must be performed to distinguish between the two propagation modes. In particular, a dedicated analysis of $\zeta_{\rm H_2}(N)$ measured in molecular clouds is necessary to determine the
most probable slope. Furthermore, the analysis of \citet{NeufeldWolfire2017} should be done assuming that $\zeta_{\rm
H_2}$ varies within the cloud rather than assuming a uniform value.

\subsection{Self-modulation of CRs}
\label{section_self-mod}

In Sect.~\ref{section_self-gen} we obtained a general condition for the generation of MHD turbulence by CRs penetrating into a
molecular cloud, Eq.~(\ref{E_ex}). The CR flux, $S(E)$, modulated by this turbulence, Eq.~(\ref{flux1}), is
controlled by the dimensionless damping rate, $\tilde\nu$, Eq.~(\ref{scale_nu}), and the net velocity of CRs in a
free-streaming regime, Eq.~(\ref{u_net}), which depends on the effective cloud column density. 

\begin{figure}[!h]
\begin{center}
\resizebox{10cm}{!}{\includegraphics{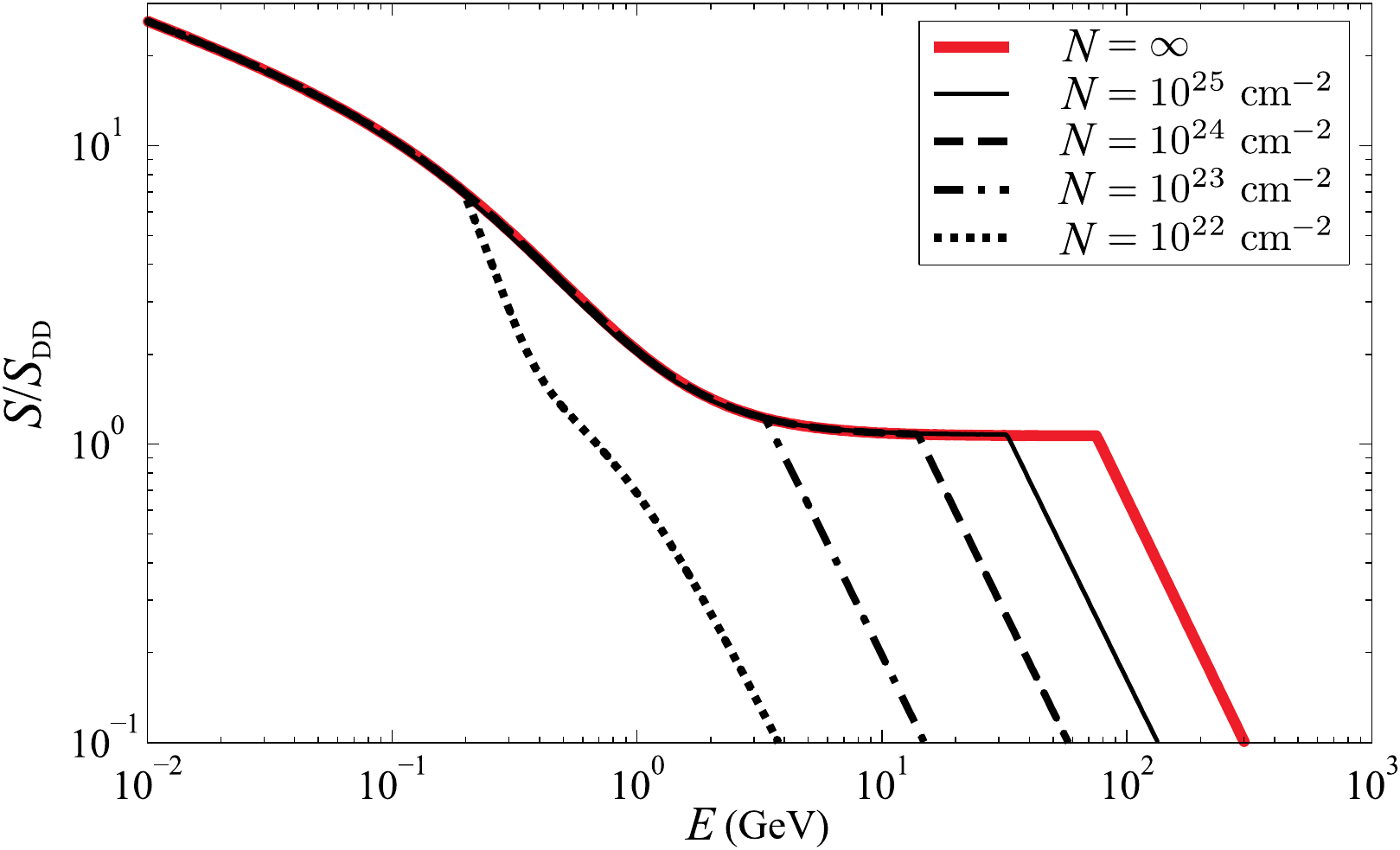}}
\caption{The ratio of the total CR flux $S(E)$ to the diffusion-dominated component
$S_{\rm DD}(E)$. Above the kink, $E>E_{\rm ex}$, the curves represent the free-streaming flux $u_{\rm
net}f^{\rm IS}$ for different values of $N$. The modulated flux below the kink is described by Eq.~(\ref{flux1}).
The shown results are for $\tilde\nu=0.78$ and $j^{\rm IS}(E)=j_*[2/(1+\tilde E)]^{2.8}$. Figure from \citet{Dogiel+2018}.} \label{SDD}
\end{center}
\end{figure}

Figure~\ref{SDD} illustrates the effect of self-modulation and shows the ratio of the modulated flux penetrating into a
cloud, $S(E)$, to its universal component, $S_{\rm DD}(E)$ \citep{Dogiel+2018}. The solid line represents the case of a
perfectly absorbing cloud, the other lines show cases of finite-size clouds. One can see that for $N>
10^{23}$~cm$^{-2}$ the flux is universal in a rather broad energy range below the excitation threshold (where the curves are
practically horizontal), as expected from the excitation-damping balance. In this case the modulation can be very strong,
reducing the penetrating flux by an order of magnitude or even more for sufficiently small values of $\tilde\nu$, i.e., for
small $n_n$ and/or large $j_*$, where $j_*=j^{\rm IS}(E=mc^2)$,
see Sect.~\ref{section_self-gen}.

Molecular clouds in the local ISM typically have column densities of $N\lesssim 10^{22}$~cm$^{-2}$. Therefore,
one might expect that the CR modulation in such clouds and the resulting effect on the ionisation are not strong -- though a
detailed analysis of this problem is still outstanding. In contrast, the situation could be very different for molecular clouds in the Central Molecular Zone
(CMZ). The effective CMZ column density is substantially higher, while the
CR density is expected to be several times larger than the local interstellar spectrum
\citep[e.g.,][]{Yang+2015,Acero+2016}.

\begin{figure}[!h]
\begin{center}
\resizebox{10cm}{!}{\includegraphics{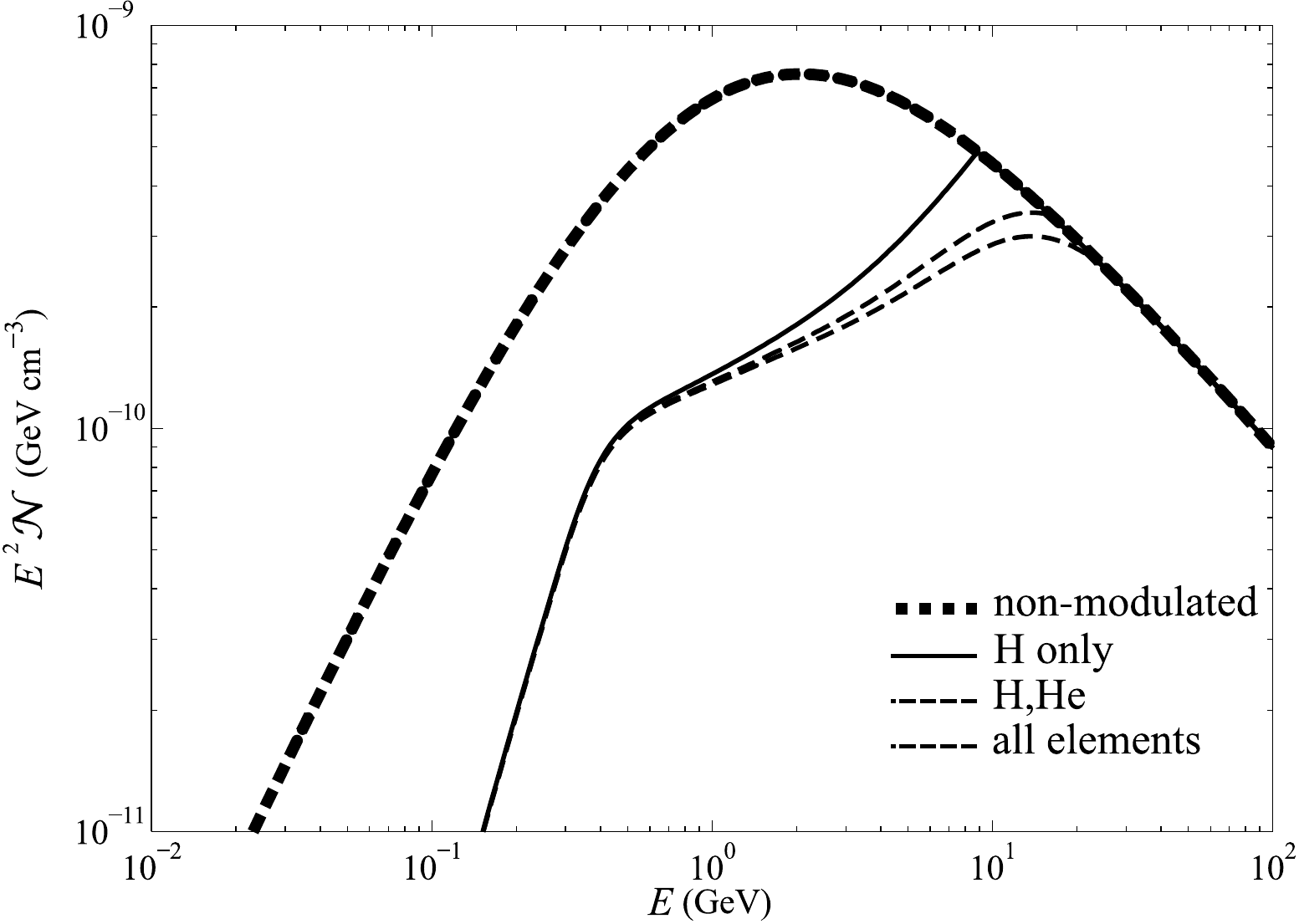}}
\caption{Energy density of CR protons in the CMZ, $E^2\mathcal{N}(E)$. The thick dotted line shows the interstellar spectrum,
the solid line represents the modulated spectrum where no effects of heavier CRs are included, the other curves are for
additional 10\% of He nuclei and for all elements. The hydrogen density in the envelope is $10$~cm$^{-3}$, the magnetic
field strength is $10~\mu$G, the effective gas column density of the CMZ is $10^{23}$~cm$^{-2}$. Figure from \citet{Dogiel+2018}.} \label{CMZ}
\end{center}
\end{figure}

To demonstrate the effect of CR self-modulation in dense molecular clouds, Fig.~\ref{CMZ} presents the results for
the CMZ, assuming that the proton spectrum is five times the local interstellar spectrum \citep{Dogiel+2018}. The solid line
shows the results for the case neglecting the effect of heavier CR species, Eq.~(\ref{flux1}). The contribution of
heavier CRs is  illustrated by the behavior of the peak at 
the excitation threshold, $E_{\rm ex}$ (see Sect.~\ref{section_self-gen}), which becomes smoother, while the excitation threshold position  shifts to higher energies. This strong modulation leads to a substantial suppression of the expected
$\gamma$-ray flux from the CMZ, reducing it by a factor of $\sim3$ for $\gamma$-ray photons with energies below 2~GeV
\citep{Dogiel+2018}. We refer to Sect.~\ref{cmz} for a discussion of observational constraints on the CR ionisation rate in the CMZ and its relation to the local ISM, star formation, and outflows.

\subsection{Impact of cosmic rays on gas chemistry and dust in Molecular Clouds}
\label{MC:chemistry}




\subsubsection{Molecular chemistry and gas temperature}
Astrochemistry calculations of molecular clouds have typically neglected CR attenuation and instead adopt a constant CR ionisation rate throughout the cloud \citep{GloverClark2012,Offner+2013,Bisbas+2017}. 
However, the net result of the CR processes described above is that --  even in the absence of local, internal CR sources (see Sect.~\ref{localcosmicrays}) --  molecular clouds likely experience a spatially varying ionisation rate, which in turn impacts gas chemistry and temperature \citep{Gaches+2019a}.
The inner,  high-extinction regions of molecular clouds are well-shielded from interstellar X-ray and FUV photons, allowing the gas to cool and carbon and nitrogen species to freeze onto dust grains. 
Due to the cold ($\sim$10\,K) conditions typical of starless cores, once species stick to grain surfaces, they cannot thermally evaporate \citep{CaselliCeccarelli2012}. However, CRs can regulate the gas phase abundances of molecules in these regions via non-thermal desorption caused by CRs hitting dust grains or photodesorption due to the FUV photons produced by CR interactions with H$_2$ \citep{Oberg+2009,Oberg+2009b}.  At moderate fluxes, these CR processes also promote the formation of organic species in dense regions, such as H$_2$CO, CH$_3$OH and HCOOCH$_3$ \citep{BennettKaiser2007,Oberg+2009c,Oberg+2010}.

Molecular clouds in environments experiencing high CR fluxes, such as within the CMZ or starburst galaxies, may have significantly altered carbon chemistry. The effects of high CR ionisation rates on CO is particularly important for observations of molecular clouds, since the CO emission to H$_2$ gas mass ratio, or $X_{\rm CO}$, is ubiquitously used to study molecular clouds and infer their masses \citep{Bolatto+2013}. However, chemical models demonstrate that clouds in environments with 10 times the typical Galactic CR ionisation rate exhibit significantly reduced CO/H$_2$ ratios \citep{Bisbas+2017,Gaches+2019b}. The fraction of atomic and ionised carbon rises steadily for CR ionisation rates $10^2-10^3$ times the Galactic mean value, and eventually C$^+$ becomes the dominant gas tracer \citep{Bisbas+2017}. 

The CR flux and its impact on gas chemistry also shape the cloud temperature. CR heating, produced by the energetic and excited products of CR interactions, can be an important heating source in regions with high CR fluxes, since nearly half of the CR energy goes into this heating mechanism \citep{Glassgold+2012}. Clouds with significant CR ionisation are also correspondingly warmer \citep{Bisbas+2017,GachesOffner2018}. The CR pressure is dynamically negligible \citep{GachesOffner2018}; however, the resulting elevated temperatures of $30-50$\,K increase the thermal pressure, which in turn may make it more difficult to form stars. However, it is worth noting that molecular gas is not necessarily a prerequisite for star formation \citep{Krumholz2012}.
Understanding the relationship between the cloud environment, gas chemistry and temperature is ultimately critical for models of star formation.


\subsubsection{Production of atomic hydrogen}
\label{atomichydrogen}
A wealth of studies has been carried out to characterise the origin of the atomic
hydrogen component in dense environments \citep[e.g.,][]{McCutcheon+1978, vanderWerf+1988, Montgomery+1995,LiGoldsmith2003,
GoldsmithLi2005}. An accurate description of H$_2$ dissociation is essential because many chemical processes, such as CO
hydrogenation and its depletion degree onto dust grains and the formation of complex organic molecules, critically depend on the atomic hydrogen abundance \citep[e.g.,][]{TielensHagen1982}.

The fractional abundance of hydrogen atoms, $f_{\rm H}$, measured in the densest regions of molecular clouds can only be
explained by the dissociation of H$_2$ by CRs. Recently, \citet{Padovani+2018b} carried our detailed studies of how CRs
control the value of $f_{\rm H}$ in dark clouds. They showed that the CR dissociation rate is primarily determined by
secondary electrons produced during the primary CR ionisation process. These secondary electrons represent the only source
of atomic hydrogen at column densities above $10^{21}$~cm$^{-2}$. Furthermore, they found that the ratio of H$_2$
dissociation and ionisation rates remains practically constant and equal to $\sim0.7$ in a broad range of column
densities between $\sim10^{21}$~cm$^{-2}$ and $\sim10^{25}$~cm$^{-2}$, irrespective of the form of the interstellar CR
spectrum. Finally, it was pointed out that only a CR proton spectrum enhanced at low energies (spectrum ${\mathscr H}$, see
Eq.~(\ref{model_j})) is capable of reproducing the upper values of measured $f_{\rm
H}$ in molecular clouds.

\subsubsection{Dust charging}
\label{dustcharging}
Interstellar dust grains in dense molecular clouds are subject to several electric charging processes
\citep[e.g.,][]{DraineSutin1987,WeingartnerDraine2001}. The resulting net charge carried by micron or sub-micron grains has important
consequences for the chemical and dynamical evolution of molecular clouds: it affects the process of dust coagulation
\citep{Okuzumi2009}, the rate of grain-catalysed electron-ion recombination \citep{Mestel1956}, the amount of gas-phase
elemental depletion \citep{Spitzer1941}, and the electrical resistivity of the gas \citep{Elmegreen1979}. Collection of
electrons and ions from the surrounding gas and photoemission induced by the interstellar UV field had been usually assumed
to be the dominant grain charging mechanisms in the cold ISM \citep[e.g.,][]{DraineSutin1987,Draine2011book,
WeingartnerDraine2001}.

\citet{Ivlev+2015} showed that CRs strongly affect charging of dust grains in cold dense molecular clouds. They investigated
two additional mechanisms of dust charging: collection of suprathermal CR electrons and protons by grains, and photoelectric
emission from grains due H$_2$ fluorescence generated by CRs in the Lyman and Werner bands
\citep{PrasadTarafdar1983,Cecchi-PestelliniAiello1992}. While the former mechanism turns out to always be negligible, the CR-induced
photoemission was shown to dramatically modify the charge distribution for submicron grains for column densities of up to
$\sim3\times10^{22}$~cm$^{-2}$. The competition between the collection of electrons and ions, which primarily produce singly
charged negative grains, and the CR-induced photoemission, which results in positive charging, significantly broadens the charge
distribution.

\section{Galactic cosmic rays in circumstellar discs}
\label{protostars}



The ionisation rate critically shapes protoplanetary disc properties from beginning to end, influencing disc/planet formation, evolution and lifetimes. Protoplanetary discs are dense and relatively cold making any potential source of ionisation very important.  The degree of ionisation in the collapsing core affects the gas-field coupling and thus formation time and extent of the initial disc  \citep{Padovani+2014}. 
Throughout the disc lifetime, ionisation dictates disc chemistry and thereby regulates the efficiency of angular momentum transport 
via the magneto-rotational instability \citep[MRI,][]{BalbusHawley1991,Gammie1996}. Magneto-centrifugally launched winds \citep{BlandfordPayne1982} may also play a role in removing angular momentum from discs, thus allowing mass accretion onto the central star. These winds can also only be efficient if sufficient ionisation exists below the disc surface layers.
Meanwhile, reduced ionisation near the disc midplane enables planetesimal formation to commence \citep{Gressel+2012}. 

While far ultraviolet (FUV) and X-ray photons dominate the ionisation of the disc surface layers \citep{PerezBecker+2011,Glassgold+2012},
CRs, with their ability to reach surface densities of $\Sigma > 10^2$\,g\,cm\mm~\citep{UmebayashiNakano1981,Padovani+2018a} potentially play a starring role in disc chemistry and dynamics.
External sub-GeV CRs are likely screened by the stellar magnetosphere  \citep[or {\it T-Tauriosphere},][]{Cleeves+2013b}, reducing the influence of Galactic CRs at $\Sigma \lesssim 10^2$\,g\,cm\mm. However, CRs produced locally by stellar flares (see Sect.~\ref{cr_lum}), the accretion or jet shocks of the young star itself are, by definition, not excluded and may significantly enhance the disc ionisation \citep{RodgersLee+2017,Offner+2019,RodgersLee+2019}.

\subsection{Ionisation at high column densities}\label{ionisation_HD}

A comprehensive model of ionisation at high densities, relevant for the inner regions of collapsing clouds and circumstellar
discs, has been recently developed by \citet{Padovani+2018a}. The choice of the proper transport regime of CR protons (see
Sect.~\ref{propagation}) as well as of accurate models for the generation and transport of secondary CRs (see
Sect.~\ref{generation}) is crucial for computing ionisation in these environments. The authors calculated dependencies for
$\zeta_{\rm H_2}(N)$, representing several characteristic energy spectra of CRs. Apart from an extreme (and also poorly
constrained) case of ionisation due to enhanced flux of stellar protons, the obtained dependencies can be considered fairly universal and applicable to any relevant environment. The principal limitation of these results is that they cannot
be generally used to describe ionisation in regions dominated by MHD turbulence, which may lead to diffusive CR transport (see Sect.~\ref{section_D}).

\citet{Padovani+2018a} adopted an analytical expression for the interstellar spectra of CR
electrons and protons, described by Eq.~(\ref{model_j}) (spectra of heavier nuclei, with the corresponding abundances,
are described by the same energy dependence). 

\begin{figure}[!h]
\begin{center}
\resizebox{10cm}{!}{\includegraphics{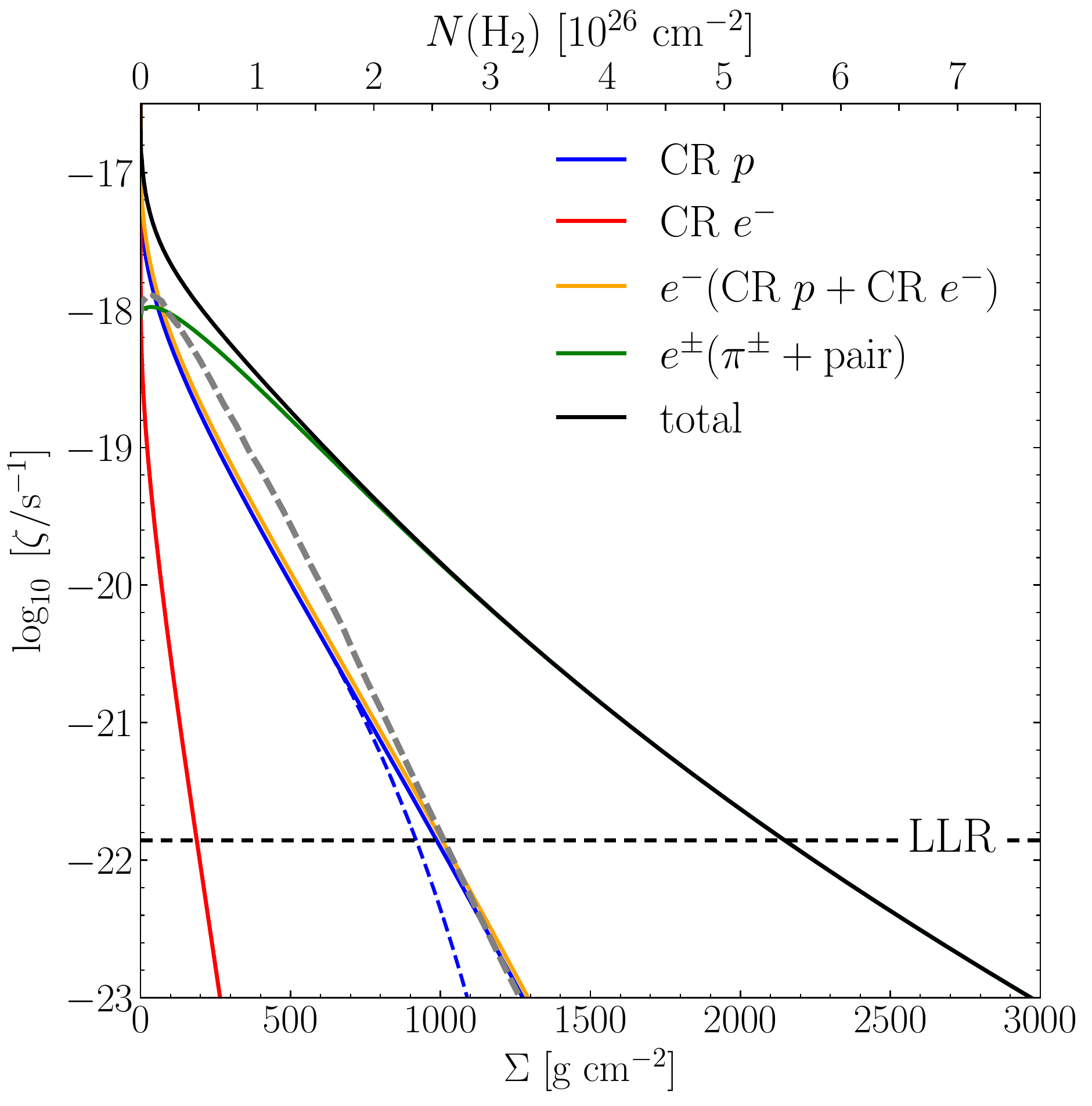}}
\caption{Production rate of molecular hydrogen ions, $\zeta_{\rm H_2}$, plotted vs.
the surface density $\Sigma$ (bottom scale) and the column density $N$ (top scale). The
black solid line shows the total rate; partial contributions to $\zeta_{\rm H_2}$ include ionisation due to primary CR
protons and electrons (blue and red solid lines, respectively), ionisation due to secondary electrons created by primary CRs
(orange solid line), and ionisation due to electrons and positrons created by charged pion decay and pair production (green
solid line). The horizontal dashed line at $1.4\times10^{-22}$~s$^{-1}$ indicates the total ionisation rate set by
long-lived radioactive nuclei (LLR). For comparison, we also plot $\zeta_{\rm H_2}(N)$ obtained by
\citet{UmebayashiNakano1981} (grey dashed line). The total rate of the electron production (due to CR ionisation of heavier gas
species) is approximately $1.11\zeta_{\rm H_2}$.
Figure from \citet{Padovani+2018a}.} \label{crion}
\end{center}
\end{figure}

By summing the contributions due to primary CRs (protons, heavier nuclei, and electrons) and secondary CR species
(electron-positron pairs and photons), 
one can derive the total production rate of molecular hydrogen ions,
H$_2^+$. Figure~\ref{crion} presents the total ionisation rate and partial contributions from various species. For
convenience, $\zeta_{\rm H_2}$ is plotted versus both the column density, $N$, and the surface density, $\Sigma$; numerically,
the relation between $N\:[\rm cm^{-2}]$ and $\Sigma\:[\rm g~cm^{-2}]$ is given by $N=2.55\times10^{23}\Sigma$, which
corresponds to an ISM elemental abundance with mean molecular weight of 2.35.

Figure~\ref{crion} shows that for $\Sigma$ below the transition surface density, $\Sigma_{\rm tr}\sim130$~g~cm$^{-2}$,
ionisation is mainly due to CR protons (and their secondary electrons), while at higher surface densities the
contribution of electron-positron pairs produced by photon decay becomes progressively dominant. At
$\Sigma\gtrsim600$~g~cm$^{-2}$, pairs fully determine the ionisation -- their contribution is about a factor of 10 larger
than that of CR protons. Hence, for $\Sigma\lesssim\Sigma_{\rm tr}$ the relevant quantity is the {\it effective} surface
density seen by CRs moving along magnetic field lines; depending on the magnetic field configuration \citep[see,
e.g.,][]{Padovani+2013}, the latter is generally larger or much larger than the line-of-sight surface density. Otherwise, if
$\Sigma\gtrsim\Sigma_{\rm tr}$, the ionisation is no longer affected by the magnetic field and therefore is controlled by
the {\it line of sight}, rather than the effective column density.

These results 
are substantially different from commonly adopted calculations by
\citet{UmebayashiNakano1981}. The latter found the total ionisation rate decreases exponentially with a characteristic attenuation
scale of about 115~g~cm$^{-2}$ for $100~\mathrm{g~cm}^{-2}\lesssim\Sigma\lesssim500~\mathrm{g~cm}^{-2}$ and about
96~g~cm$^{-2}$ at larger surface densities. Conversely, \citet{Padovani+2018a} find a characteristic scale that continuously
increases with surface density, from $\sim112$~g~cm$^{-2}$ to $\sim285$~g~cm$^{-2}$ in the range
$100~\mathrm{g~cm}^{-2}\lesssim\Sigma\lesssim2100~\mathrm{g~cm}^{-2}$, within an error lower than 10\%. This difference is
primarily due to the more elaborate transport models for primary CR protons and of secondary CR photons employed by
\citet{Padovani+2018a} (see their paper for detailed discussion).

\subsubsection{Discs around T-Tauri stars} Low-energy Galactic CRs may be prevented from penetrating an extended heliosphere (or
{\it T-Tauriosphere}) surrounding a young star due to its more powerful wind and increased magnetic stellar activity \citep{Cleeves+2013}. Unfortunately, little can be said about the extent and
shape of this region of CR exclusion other than scaling up the properties of the Sun's heliosphere. \citet{Cleeves+2013}
suggested that the T-Tauriosphere may well surround the entire disc. However, the energies of CR particles mostly
responsible for the ionisation at column densities above $\sim 100$~g~cm$^{-2}$ exceed a few GeV. The effect of the
modulation by the stellar wind at these energies is uncertain. For T-Tauri (TT) stars, \citet{Cleeves+2013} estimated values
for the modulation potential, $\phi$, at a distance of 1~AU in the range $\phi=4.8-18$~GeV, leading to a reduction of the CR flux at $E=10$~GeV by a factor of $\sim 6$ and $100$, respectively ($\phi$ is an unknown function of distance, which
could be determined from detailed magnetospheric models). On the other hand, the presence of a young magnetically active star may lead to increased production of SCRs (see Sect.\,\ref{cr_lum}).

\citet{Padovani+2018a} applied their model of CR propagation to estimate the ionisation produced at a distance of 1~AU from
the protostar by two different input proton spectra: a spectrum of Galactic protons modulated by TT stellar winds
\citep{Usoskin+2005,Cleeves+2013} and an enhanced flux of stellar protons generated by flares in an active TT star
\citep{Feigelson+2002,Rab+2017}.  The result is significantly different than that found in prior work. 

\begin{figure}[!h]
\begin{center}
\resizebox{10cm}{!}{\includegraphics{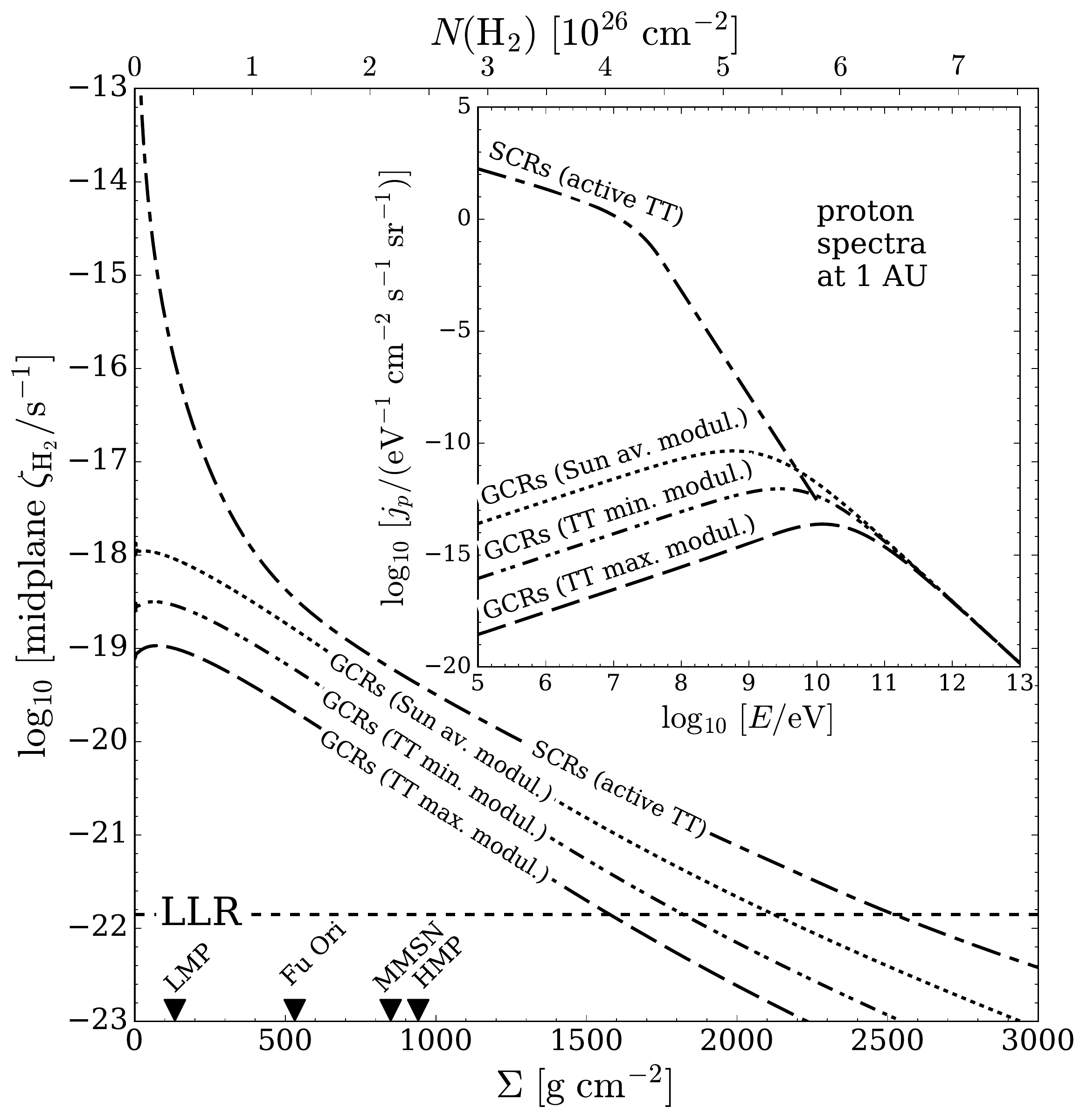}}
\caption{Mid-plane CR ionisation rate $\zeta_{\rm H_2}$ vs. the surface density $\Sigma$
(bottom scale) and column density $N$ (top scale) plotted for several proton spectra (at a distance of 1~AU from the central
star), as shown in the inset: Galactic CRs with solar mean modulation (dotted line); minimum and maximum modulation by a TT
wind (dash-dot-dotted and long dashed line, respectively); and stellar CRs from an active TT star (dash-dotted line). The
horizontal dashed line shows the ionisation rate set by LLR. For reference, the black triangles on the horizontal axis
indicate the characteristic values of the mid-plane disc density ($\Sigma_{\rm disc}/2$) at 1~AU for low- and high-mass
protostars (LMP and HMP) as well as for FU Orionis stars (FU Ori) and the minimum-mass solar nebula (MMSN).
Figure from \citet{Padovani+2018a}.}
\label{TTauri}
\end{center}
\end{figure}

({\em i}\/) Figure~\ref{TTauri} shows the CR ionisation rate 
for maximum and minimum
modulation by a TT stellar wind at 1~AU, corresponding to $\phi=18$~GeV and $\phi=4.8$~GeV, respectively, and labelled by
``GCRs (max. modul.)'' and ``GCRs (min. modul.)''. For completeness, the figure also shows the ionisation rate for the
Galactic CR flux modulated by an average solar wind ($\phi=1$~GeV) labelled ``GCRs (Sun. av. modul.)''. Compared to
\citet{Cleeves+2013}, the result for the minimum modulation model is larger by a factor of $\sim30$ at
$\Sigma\lesssim100$~g~cm$^{-2}$, while above $\Sigma\sim1200$~g~cm$^{-2}$ it decreases much more abruptly. The difference
is even more dramatic for the maximum modulation model. \citet{Padovani+2018a} predict an ionisation rate that is larger by a
factor of $\sim260$ at $\Sigma\lesssim100$~g~cm$^{-2}$ and decreases faster above $\Sigma\sim1400$~g~cm$^{-2}$.
The discrepancy at low surface densities is large because \citet{Padovani+2018a} included the process of electron-positron
pair creation by photon decay and also adopted the relativistic behaviour of the ionisation cross section for protons. The
faster decrease of their results at high surface densities is caused by losses due to heavy elements in the medium (see
Sect.~\ref{energylosses}). It is noteworthy that the ionisation rate for the minimum and maximum modulation is almost
entirely due to relativistic protons, propagating diffusively (see Sect.~\ref{propagationp}).

({\em ii}\/) For the proton flux generated in a TT flare, labelled in Fig.~\ref{TTauri} by ``SCRs (active TT)'', results
by \citet{Padovani+2018a} for up to $\Sigma\sim300$~g~cm$^{-2}$ agree with those obtained by \citet{Rab+2017} within 5\%. At
higher surface densities the ionisation rate 
decreases slowly, since electron-positron
pairs increase the ionisation by orders of magnitude. It is important to remark that it is still unclear what fraction of CRs
generated in a flare event can be channeled into the disc through magnetic field lines, without crossing the turbulent zone,
and what part may follow open field lines perpendicular to the disc. 

\subsection{Impact of cosmic rays on the dynamical and chemical evolution of discs}\label{feedback}

Exactly how disc ionisation varies with distance and vertical scale height 
in a protostellar disc 
is critical to the onset of the MRI, which requires sufficient ionisation in order for the gas to couple with the magnetic field. The MRI can be described by two dimensionless numbers: the magnetic Reynolds number, $Re$, quantifies the coupling between the ionised gas and magnetic field, while the Ambipolar diffusion number, $Am$, quantifies the coupling between the ions and neutrals: 
\begin{equation}
Re \equiv \frac{c_s h}{\eta} \approx \left( \frac{n_i/n_n}{10^{-13}} \right) \left( \frac{T}{100~ {\rm K}} \right)^{1/2} \left( \frac{r}{{\rm au}}\right)^{3/2}, ~~~~~ Am \equiv \frac{n_i\beta_{in}}{\Omega}.
\end{equation}
Here $c_s$ is the local sound speed, $h = c_s/\Omega$ is the scale height, $\Omega$ is the Keplerian orbital frequency and $\eta$ is the magnetic diffusivity, $n_i$ is the number density of charged species, and $\beta_{in} \sim 2 \times 10^{-9}$\,cm$^3$\,s\m is the collisional rate coefficient for singly charged species to distribute their momentum to neutrals. The degree of ionisation enters through the magnetic diffusivity, which can be defined as \citep{BlaesBalbus1994}: 
\begin{equation}
\eta = 234 \left( \frac{T}{{\rm K}} \right)^{1/2} \frac{n_n}{n_i} \,{\rm ~cm}^2\,{\rm s}^{-1}.
\end{equation}
High-resolution, global disc simulations suggest the minimum value of $Re$ for fully active MRI is $\sim 3000$ \citep{Flock+2012}. However, hydrodynamic simulations including ambipolar diffusion suggest there is no minimum value for $Am$ for MRI \citep{BaiStone2011}, which underscores the importance of $Re$ and hence local ionisation in disc dynamics.
 
Many studies that resolve MRI turbulence directly  assume a uniform value for the CR ionisation rate \citep{Simon+2018,Bai+2019}, but care should be taken because recent calculations show that the CR spectrum including a variety of energy loss mechanisms predict that the CR ionisation rate is far from being constant especially at the high column densities, typical of discs (see Sect.~\ref{ionisation_HD} and Fig.~\ref{crion}).
 If CR-produced ionisation exceeds thermal ionisation (inner $<1$\,au) and/or ionisation produced by radionuclides (disc midplane), CRs will regulate the cycle of MRI activation and accretion  \citep{Offner+2019}.  However, the coupling between the mean magnetic field and the CRs is also mediated by the degree of turbulence, which in turn is dictated by the MRI activity \citep{RodgersLee+2017,Bai+2019} -- circumscribing a thoroughly non-linear problem.

 CRs impinging on the disc are more likely to arise locally (see Sect.~\ref{localcosmicrays}) than from external sources as mentioned above. If the CRs are produced by the stellar accretion shock then the nature and properties of the accretion flow are important to the disc evolution. While protostellar accretion rates are expected to be high ($\dot M > 10^{-6}~{\rm M}_\odot$\,yr\m) at early times, CRs accelerated in the accretion shock will likely undergo severe energy losses as they travel through the dense accretion flow to the disc \citep{Offner+2019}. For accretion rates of $\sim 10^{-7}$--$10^{-9}~{\rm M}_\odot$\,yr\m, the CR flux accelerated at the shock is lower but likely less attenuated \citep{GachesOffner2018,Offner+2019}. The relationship between the CRs accelerated at the shock, energy losses in the accretion flow and the MRI, which in turn dictates the rate of accreting gas, may form a feedback loop that mediates accretion in time and produces luminosity variation \citep{Offner+2019}. 
 
 Older stellar sources also produce CRs at a lower level through stellar activity (as our Sun does).
Direct modelling of particle trajectories in the inner 0.5\,au of a TT disc
 suggests the local ionisation rate is spatially  inhomogeneous and depends on the magnetic field morphology and assumed degree of turbulence \citep{Fraschetti+2018}. However, it is worth noting that these simulations use a simplified model for the protoplanetary disc. The disc is assumed to be infinitely thin and massive. This implies that the SCRs suffer catastrophic energy losses when they penetrate the disc, which prevents radial transport of the SCRs in the disc. The complex relationship between CRs, gas chemistry and the magnetic field encompass a challenging, multi-scale, multi-physics problem that requires further study to explore the details of the propagation and impact of CRs on disc evolution.


\section{Locally accelerated cosmic rays}\label{localcosmicrays}
Recent observations have shown the presence of an 
unexpected extremely high ionisation rate in protostellar 
environments (e.g., $4\times10^{-14}$~s$^{-1}$ in
OMC 2~FIR-4, see \citealt{Ceccarelli+2014,Fontani+2017,Favre+2018}) as well as
the detection of synchrotron emission 
(the fingerprint of the presence of relativistic 
electrons) 
in the shocks at the
position of the knots that develop along
protostellar jets
\citep[e.g.][]{Beltran+2016,Rodriguez-Kamenetzky+2017,Osorio+2017,Sanna+2019}. 
Both signatures cannot be explained by
interstellar CRs since their flux is strongly attenuated
at the high densities typical of protostellar environments.
Arguments based on the energetics of the system justify the possibility of particles accelerating inside protostellar systems, namely, to generate local CRs.

The luminosity of an accretion shock on the surface of a protostar is
\be\label{Lgrav}
L_{\rm accr}=\frac{GM\dot{M}}{R_{\rm sh}}\,,
\ee
where $G$ is the gravitational constant, $M$ is the protostellar mass, $\dot{M}$ is the accretion rate,
and $R_{\rm sh}$ is the shock radius. If we consider the gravitational collapse of an early Class~0 protostar with $M=0.1~M_{\odot}$,
$\dot{M}=10^{-5}~M_{\odot}$~yr$^{-1}$ \citep[e.g.][]{Shu+1987,Belloche+2002}, $R_{\rm sh}=2\times10^{-2}$~AU \citep{MasunagaInutsuka2000}, then
$L_{\rm accr}=3\times10^{34}$~erg~s$^{-1}$. The luminosity of the interstellar CRs impinging on a molecular cloud core can be estimated by
\be\label{Lcr}
L_{\rm CR}=R_{\rm core}^{2}v_{\rm A}\epsilon_{\rm CR}\,,
\ee
where $R_{\rm core}$ is the core radius, $v_{\rm A}$ is the Alfv\'en speed in the surrounding medium, supposed to be the warm neutral medium,
and $\epsilon_{\rm CR}$ is the energy density of interstellar CRs based on the latest 
Voyager observations~\citep{Cummings+2016,Stone+2019}. 
Here we adopt $R_{\rm core}=0.1$~pc, \mbox{$v_{\rm A}=9.3\times10^{5}$~cm~s$^{-1}$} (based on $n_{\rm H}=0.5$~cm$^{-3}$ and $B=3~\mu$G; \citealt{Ferriere2001}), 
and $\epsilon_{\rm CR}=1.3\times10^{-12}$~erg~cm$^{-3}$, 
then $L_{\rm CR}=1.2\times10^{29}$~erg~s$^{-1}\ll L_{\rm accr}$. 
Since studies of propagating interstellar CRs
show that their flux is strongly
attenuated at high column densities
(see Sect.~\ref{ionisation_HD}),
$\epsilon_{\rm CR}$ 
at the protostellar surface
is much lower than its interstellar value, so that
$L_{\rm CR}\ll L_{\rm accr}$ close to the protostar. 
Thus, if a small fraction of the gravitational energy can be used to produce local CRs, they could easily dominate over the interstellar CR flux.
The gravitational energy available for a massive protostar to generate CRs is even higher since $\dot{M}$ could reach $10^{-3}~M_{\odot}$~yr$^{-1}$. 
Thus, in principle, it is possible to also observe $\gamma$ emission ~\citep{Araudo+2007,Bosch-Ramon+2010,Munar-Adrover+2011}.
\\

If CRs are produced locally by protostellar sources they must be accelerated efficiently enough to reach the kinetic energies needed to ionise the surrounding matter and explain non-thermal radio emission from jets. 
The three main acceleration mechanisms are \citep{Marcowith16}:
\begin{enumerate}
\item Stochastic Fermi acceleration (SFA): SFA occurs because, on average, 
CRs at a speed $v$ interact with scattering centres moving at a speed $U$, where $v \gg U$,
more often through head-on collisions than through rear-on collisions. For each head-on interaction CRs undergo an energy boost, while they decelerate in rear-on collisions. The averaged relative energy gain is $\langle \Delta E/E \rangle \propto (v/U)^2$. Under the conditions that prevail in the ISM or near protostars, $U$ is close to the local Alfv\'en speed (unless the plasma parameter $\beta$ is large), SFA is relevant for moderately relativistic particles so is potentially important for particles involved in the ionisation process.
\item First-order Fermi acceleration, also known as diffusive shock acceleration (DSA): DSA is produced because shock waves carry the scattering waves and then impose a bulk motion. At each shock crossing particles start to interact via head-on collisions and then systematically gain energy. On average (over a Fermi cycle, e.g., up-down-up stream) CRs gain $\langle \Delta E/E \rangle \propto (v/U)$. This process is of particular interest because beside being more efficient it also produces power-law solutions that only depend on one parameter: the shock compression ratio (at least in the linear stage). Shocks are ubiquitous in the ISM and near protostars, but may be collisional and not expected to accelerate particles.  
Conversely, collisionless shocks require both sufficiently low density and high magnetisation. 

\item Magnetic reconnection: There is no unique way to accelerate particles in magnetic reconnection, i.e., the process by which the topology of magnetic field lines is rearranged and magnetic energy is converted
into heat, plasma bulk motion and CRs. Instead, particles can gain kinetic energy via at least seven ways during such events: in thermal exhausts, in contracting plasmoid, in colliding plasmoid, by the reconnecting electric field, by Fermi first order acceleration in converging reconnection flows, by magnetic drifts, and by turbulence generated by tearing instabilities during the reconnection process. Such processes are expected to occur in corona above protostellar accretion discs, at the interface between the young stellar magnetosphere and its disc or within the protostellar jet. 

\end{enumerate}

In the following we focus on DSA as the primary mechanism that can efficiently accelerate thermal particles and {\it transform} them into CRs.

\subsection{Timescales and emerging fluxes at the shock surface}\label{timescales}
DSA is a well-studied process that is invoked 
for supernova remnants 
as the main mechanism explaining the acceleration of high-energy CRs. Unlike supernova remnants, the gas associated with protostars is not fully 
ionised and the friction between ions and neutral particles can strongly quench DSA. However, if neutrals and ions are coupled, namely if they move coherently, then the waves generated by ions are weakly damped and particles can be efficiently accelerated~\citep{Drury+1996,Padovani+2016}.

The maximum energy reached by a thermal particle crossing a shock surface can be 
obtained by comparing a series of timescales.
In particular, the acceleration must take
place before particles lose energy 
due to collisions with neutrals, diffuse
towards the protostar or in the transverse 
direction, and the shock disappears.
In this section we summarise the basic 
equations to compute the timescales. For a detailed review of the methods, we refer the reader to
\cite{Drury+1996} and \cite{Padovani+2015,Padovani+2016}.

The acceleration timescale, $t_{\rm acc}$, is given by
\be\label{tacc}
t_{\rm acc}=2.9(\gamma-1)\frac{r[1+r(k_{d}/k_{u})^\sigma]}{(r-1)k_{u}^\sigma} U_5^{-2} B_{-5}^{-1}~\mathrm{yr}\,,
\ee
where $U_5$ and $B_{-5}$ are the upstream flow velocity in the shock reference frame in units
of $100$~km~s$^{-1}$ and the upstream magnetic field strength in units of $10$~$\mu$G, respectively.\ Furthermore, $k_{u,d}$ is the upstream and 
downstream diffusion coefficient that is normalised to the Bohm coefficient
\be\label{diffcoeff}
k_{u}=\left(\frac{\kappa_{\rm B}}{\kappa_{u}}\right)^\sigma=\left(\frac{\gamma\beta^{2}m_pc^{3}}{3eB\kappa_{u}}\right)^\sigma
\ee
where $e$ is the elementary charge, $\gamma$ is the Lorentz factor,  $\beta=\gamma^{-1}\sqrt{\gamma^{2}-1}$, $m_p$ is the proton mass, and $c$ is the light speed. 
For a perpendicular shock $k_{u}=rk_{d}$ and $\sigma=1$, while for a parallel\footnote{A parallel and perpendicular shock is when the shock normal is parallel and perpendicular, respectively, to the ambient magnetic field.} shock $k_{u}=k_{d}$ and $\sigma=-1$.
Here, $r$ is the shock compression ratio defined by
\be\label{compressionratio}
r=\frac{\left(\gamma_{\rm ad}+1\right)M_{s}^{2}}{\left(\gamma_{\rm ad}-1\right)M_{s}^{2}+2}\,,
\ee
where $M_{s}=U/c_{s}$ is the sonic Mach number,
\be\label{cs} 
c_{s}=9.1[\gamma_{\rm ad}(1+x)T_{4}]^{0.5}~\mathrm{km~s^{-1}}
\ee
is the sound speed, $x=n_i/(n_n+n_i)$ is the ionisation fraction,
$T_{4}$ is the upstream temperature in units of $10^{4}$~K, and $\gamma_{\rm ad}$ is the
adiabatic index.

The general equation for the collisional energy loss timescale is given by
\be\label{tloss}
t_{\rm loss}=10\frac{\gamma-1}{\beta} n_{6}^{-1} L_{-16} ~\rm{yr} \ ,
\ee
where $n_{6}$ is the volume density in units of $10^{6}$~cm$^{-3}$ and $L_{-16}$ is the energy loss function in units of $10^{-16}$~eV~cm$^{2}$ (see Sect.~\ref{energylosses}). 
We can evaluate the mean loss timescale by averaging it over the particle's up- and downstream residence times \citep{Parizot+2006} 
\be\label{tloss1}
\langle t_{\rm loss}\rangle= \left(\frac{t_{{\rm loss},u}^{-1}+r t_{{\rm loss}, d}^{-1}}{1+r}\right)^{-1}\ .
\ee
The up- and downstream loss timescales differ in the density (downstream is a factor $r$ higher) and in the Coulomb component of the energy loss function, which depends on temperature\footnote{The relation between up- and downstream temperatures is given by the classic Rankine-Hugoniot condition.} \citep{MannheimSchlickeiser1994}.

The upstream diffusion timescale, $t_{{\rm diff},u}$, is obtained by assuming that the diffusion length in the upstream medium must be a fraction $\epsilon<1$ of the distance between the central source and the shock, $R$, which is $\kappa_{u}/U=\epsilon R$. The corresponding timescale is given by
\be\label{tescu}
t_{{\rm diff},u}=22.9\epsilon\frac{k_{u}^{\sigma}}{\gamma\beta^{2}}B_{-5}R_2^2~\mathrm{yr}\,,
\ee
where $R_2$ is the shock radius in unit of 100~AU.
In the case of jet shocks, particles may also escape in the downstream medium in the transverse direction. The corresponding
timescale\footnote{The factor of 4 in the denominator accounts for the fact that the 
downstream diffusion in the transverse direction is in two dimensions.}, $t_{{\rm esc},d}=R_{\perp}^{2}/(4\kappa_{d})$, can be rewritten as
\be\label{tescd}
t_{{\rm diff},d}=5.7\frac{\mathscr{C}}{\gamma\beta^{2}}B_{-5}R_{\perp,2}^{2}~\mathrm{yr}\,,
\ee
where $\mathscr{C}=r^{2}$ or 1 for a perpendicular or a parallel shock, respectively, and $R_{\perp,2}$ is the
transverse jet size in units of 100~AU.

Finally, the dynamical timescale, $t_{\rm dyn}$, is
computed as a function of the shock radius and velocity,
\be\label{tdyn}
t_{\rm dyn}=4.7R_2U_2^{-1}~\mathrm{yr}\,.
\ee
In order to effectively accelerate the
thermal particles, the flow has to be 
super-Alfv\'enic and supersonic, that is $U>\max(v_{\rm A},c_{s})$,
where the Alfv\'en speed is defined by
\be
v_{\rm A}=2.2\times10^{-2}n_6^{-0.5}B_{-5}~\mathrm{km~s^{-1}}\,.
\ee
Once this condition is satisfied, the maximum energy reached by a thermal proton, $E_{{\rm max
},p}$, imposes 
\be\label{Emaxcond}
t_{\rm acc}=\min(t_{\rm loss},t_{{\rm diff},u},t_{{\rm diff},d},t_{\rm dyn})\,.
\ee
%
%

The distribution per unit volume and energy
of the shock-accelerated protons is given by
\be\label{Nshp}
\mathcal{N}_p(E)=4\pi p^{2}f(p) \frac{\ud p}{\ud E}\,,
\ee
where $f(p)$ is the momentum distribution at the shock surface.
In the so-called test-particle regime, 
the latter is described by a power-law momentum function 
\be\label{fp}
f(p)=f_{0}\left(\frac{p}{p_{\rm inj}}\right)^{-q}\,,
\ee
with $q=3r/(r-1)$. The normalisation constant, $f_0$, is given by
\be\label{f0}
f_0=\frac{3}{4\pi}\frac{n\widetilde P}{\mathscr{I}}\left(\frac{U}{c}\right)^2(m_pc)^{q-3}p_{\rm inj}^{-q}\,,
\ee
where
\be\label{I2}
{\mathscr I}=\int_{\widetilde{p}_{\rm inj}}^{\widetilde{p}_{\rm max}}\frac{p^{\,4-q}}{\sqrt{p^{\,2}+1}}\ud p\,,
\ee
$\widetilde{P}$ is the fraction of the ram pressure, $nm_pU^2$, which is transferred to 
the
accelerated thermal protons \citep{BerezhkoEllison1999},
and $\widetilde{p}_{k}=p_{k}/(m_p c)$ is the normalised momentum.
Subscripts $k={\rm inj, max}$ refer to the injection momentum of a proton able to cross the shock that begins accelerating and the maximum momentum reached by an accelerated proton
related to $E_{{\rm max},p}$, respectively.

The process of electron injection at a
shock surface is poorly understood. However,
the distribution per unit volume and energy
of the shock-accelerated electrons, $\mathcal{N}_e$, can be estimated
following the approach by~\cite{BerezhkoKsenofontov2000} who
find that at relativistic energies the
energy distributions of electrons and
protons are related by
\be\label{NeNp}
\frac{\mathcal N_e}{\mathcal N_p}=\left(\frac{m_e}{m_p}\right)^{(q-3)/2}\,.
\ee
For the case of electrons one also 
has to consider the synchrotron timescale,
given by 
\be\label{tsyn}
t_{\rm syn}=2.7\times10^{11}%
\frac{\gamma-1}{\gamma^2}B_{-5}^{-2}~\mathrm{yr}\,,
\ee
to be compared with 
the acceleration timescale (Eq.~\ref{tacc}).%
At energies larger than $E^*$, where the condition
$t_{\rm syn}(E^*)<t_{\rm dyn}$ is fulfilled, 
the slope of the electron 
distribution, 
$s$, is modified from
$\mathcal{N}_e(E)\propto E^s$ to 
$\mathcal{N}_e(E)\propto E^{s-1}$ \citep{BlumenthalGould1970}.
Finally, the local flux of shock-accelerated CRs
is related to the distribution 
per unit volume and energy by
\be
j_k=\frac{\beta_kc}{4\pi}\mathcal{N}_k\,,
\ee
where $k=p,e$.

\subsection{Jets and protostars}\label{localjets}
\begin{figure}[!htb]
\begin{center}
\resizebox{.6\hsize}{!}{\includegraphics{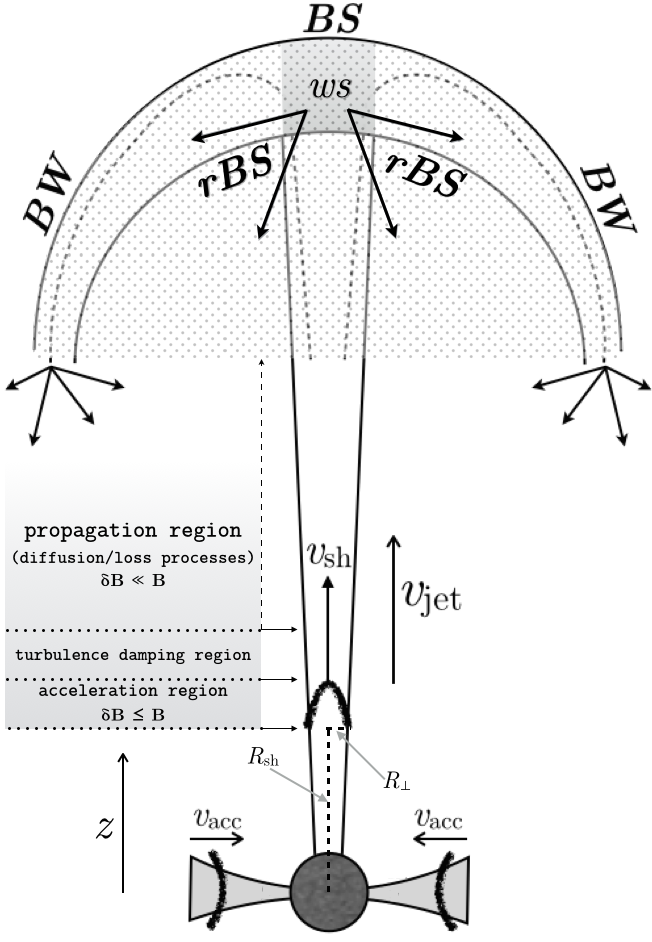}}
\caption{Sketch of the protostar configuration.
Accretion flow, jet, and shock velocities 
($v_{\rm acc}$, $v_{\rm jet}$, and $v_{\rm sh}$, respectively) are in the observer reference frame.
Shock and transverse radii are labelled  $R_{\rm sh}$ and $R_{\perp}$, respectively.
Bow shock, reverse bow shock, bow wings, and working surface are labelled  {\em BS}, {\em rBS}, {\em BW}, and {\em ws}, respectively.
The shaded areas show the regions where CR acceleration takes place and where 
turbulence is damped and
particle propagation occurs. 
The dot-filled region corresponds to the hot spot region.
Figure from \citet{Padovani+2016}.}
\label{sketch}
\end{center}
\end{figure}

Shocks located along protostellar jets at the position of the knots and on the surface of a protostar produced by infalling material have been investigated as possible acceleration sites of CRs~\citep{Padovani+2015,Padovani+2016}.
Figure~\ref{sketch} outlines the configuration of a protostar used for the modelling. Shocks in accretion flows
are assumed to be stationary, while 
shocks inside jets move more slowly than the flow
so that the upstream region is close to the protostar. 
The reverse bow shock (also known as Mach disc) 
and the bow shock, when observed,
usually
move slowly or are stationary~\citep{Caratti+2009}.
After passing through the jet shock, the gas flow spreads out until it yields a bow shock and a reverse bow shock.
In the intershock region, known as the working surface, the pressure gradient forces perpendicular to the jet are so high that the gas is 
ejected radially and  propagates into the surrounding region, called the hot spot region~\cite[e.g.][]{StahlerPalla2005book}, 
through the bow wings.

Table~\ref{paramspace}
lists the range of the main parameters that determine the timescales described in Sect.~\ref{timescales} 
estimated from observations \citep[e.g.][]{Raga+2002,Nisini+2005,HartiganMorse2007,Raga+2011,Agra-Amboage+2011,LeFloch+2012,Gomez-Ruiz+2012,Frank+2014,Maurri+2014} or from numerical simulations \citep[e.g.][]{MasunagaInutsuka2000,Tesileanu2009,Tesileanu2012}.
Ionisation fractions~\citep{Strelnitskii1984} and
shock velocities~\citep{Ceccarelli+2000} in collapsing envelopes are so small that the CR acceleration is quenched; the magnetic field strength is also high enough to produce a sub-Alfv\'enic shock.
This means that accretion flows are not likely CR acceleration
sites.
Conversely, shocks along the jets and on the protostellar surface can accelerate enough CRs to explain observations
~\citep{Padovani+2016}.

We note that the range for the flow velocity in the shock reference frame in Table~\ref{paramspace} refers to low-mass protostellar jets while, for high-mass sources, $U$ can reach values up to $\sim1000$~km~s$^{-1}$. In the latter case
the locally accelerated particles can reach up to
$\sim1-10$~TeV energies;
their $\gamma$ emission could be
observed with the next generation of ground-based telescopes such as the Cherenkov Telescope Array (CTA).

\begin{table}[!ht]
\setlength{\tabcolsep}{3pt}
\caption{Ranges of values of the parameters described in the text: upstream flow velocity in the shock reference frame ($U$),
temperature ($T$), total hydrogen density ($n_{\rm H}$), ionisation fraction ($x$), and magnetic field strength ($B$).}
\begin{center}
\begin{tabular}{cccccc}
\hline\hline
site$^{*}$ & $U$ & $T$ & $n_{\rm H}$ & $x$ & $B$\\
     & $\mathrm{[km~s^{-1}]}$ & $\mathrm{[K]}$ & $\mathrm{[cm^{-3}]}$ & & $\mathrm{[G]}$\\ 
\hline
${\mathscr E}$ & $1-10$ & $50-100$ & $10^{7}-10^{8}$ & $\lesssim10^{-6}$ & $10^{-3}-10^{-1}$\\
${\mathscr J}$ & $40-160$ & $10^{4}-10^{6}$ & $10^{3}-10^{7}$ & $0.01-0.9$ & $5\times10^{-5}-10^{-3}$\\
${\mathscr P}$ & 260 & $9.4\times10^{5}$ & $1.9\times10^{12}$ & $0.01-0.9$ & $1-10^{3}$\\
\hline
\end{tabular}\\
$^{*}{\mathscr E}$ = envelope;
${\mathscr J}$ = jet; 
${\mathscr P}$ = protostellar surface.
\end{center}
\label{paramspace}
\end{table}
\normalsize

 \subsection{Observational arguments for the luminosity and spectrum of stellar CRs}
\label{cr_lum}

To estimate the ionising effect of SCRs in protoplanetary discs originating from the central young star, a luminosity and spectrum for the SCRs must be adopted. The only available direct observations of SCRs from a star are from the Sun. Based on purely energetic arguments it is likely that the luminosity of SCRs from young stars is much larger than for the Sun.

There have been estimates of the expected luminosity and SCR spectrum for a young star using scaling law arguments relating to the Sun and X-ray observations of young stars \citep{Feigelson+2002}. These estimates do not consider the exact acceleration mechanism but generally assume that the SCRs originate from stellar flares. Other specific locations for accelerating particles in the accretion and jet shocks are discussed in Sect.~\ref{localcosmicrays}. The idea is to take the observed relationship between X-ray luminosity and proton fluence for the Sun to estimate the proton fluence for a young stellar object (YSO) using the known X-ray luminosity of YSOs.

\citet{Feigelson+2002} estimated that proton fluences produced by YSOs will be $10^5$ times greater than observed from the most powerful solar flares. This factor of $10^5$ combines a number of different factors. First, $Chandra$ X-ray observations of low-mass stars in the Orion Nebula Cluster (ONC) show increased X-ray luminosities in comparison to extreme solar events, which contributes a factor of $\sim30$. The ONC stellar flares also occur approximately $\sim300$ times more frequently than solar flares. This gives a total increase of $10^4$ resulting from stellar X-ray flare emission. Finally, a factor of $10$ represents the increased proton flux compared to the X-ray flux. This scaling has been used in \citet{Rab+2017} for instance where the slope of the stellar CR spectrum is assumed to be the same as for the Sun. A similar argument for the SCR luminosity is presented in \citet{RodgersLee+2017}, instead assuming a relationship between the power in the stellar wind and the power in the SCRs. Note that, while the flux of solar CRs is not constant, \citet{Feigelson+2002} argue that the flux of YSO CRs can be considered to be constant due to overlapping magnetic flare events.

 The last important detail of this scaling argument is the high-energy cut-off for the stellar CRs. It seems likely that a YSO will be able to accelerate  particles to higher energies than the Sun. As discussed in \citet{RodgersLee+2017}, the break in the solar spectrum at $\sim$20 MeV to a steeper power law has been seen to shift up to an order of magnitude higher in energy during solar flare events \citep{ackermann_2014long,ajello_2014long}. This is perhaps indicative of the type of behaviour expected from YSOs. The location of this high-energy cut-off in the stellar CRs spectrum is of interest because MeV protons have a very large energy loss rate per grammage meaning that they will lose their energy faster relative to GeV protons, which are minimally ionising. Additionally, if diffusive transport is assumed for CRs, from quasi-linear theory GeV protons will have larger diffusion coefficients than MeV protons meaning that GeV protons will spend less time in the very dense inner regions of protoplanetary discs. Therefore, the presence of GeV protons would mean it is much more feasible for stellar CRs to contribute ionisation further out in the disc.

 \subsection{Propagation models of stellar cosmic rays in protoplanetary discs}

To date, a number of groups have estimated the influence of stellar energy particles in protoplanetary discs  \citep{Rab+2017,RodgersLee+2017,Fraschetti+2018,RodgersLee+2019}. The main ingredients of these models, which have been discussed in previous sections in detail, are the following: the properties of the SCRs (power and assumed spectrum), the loss rate of the SCRs, the propagation of the SCRs (ballistic, free streaming or diffusive transport) and the assumed properties of the protoplanetary disc itself. Modelling the influence of CRs in protoplanetary discs began with \citet{TurnerDrake2009} who studied Galactic and stellar CRs with energies above 0.1\,GeV and considered ballistic transport without any potential modulation by the T-Tauriosphere at these energies. More recently the idea that Galactic CRs would be significantly modulated by the stellar wind and its entrained magnetic field at $\sim$GeV energies led to the idea that instead stellar CRs might contribute significantly to the ionisation rate in protoplanetary discs.
 
\citet{Rab+2017} used a scaled up solar-like energy spectrum for the stellar CRs, following \citet{Feigelson+2002}, composed of two power-law components, which result in a sharp cut-off in the spectrum at energies above 20\,MeV. They consider ballistic transport, i.e., ignore the effect of any possible magnetic fields, using CSDA 
 for the energy losses of the CRs. \citet{RodgersLee+2017} also focused on the influence of $\sim$GeV stellar CRs in diffusive transport, which includes the influence of turbulent magnetic fields, with a constant energy loss rate for the CRs. \citet{Fraschetti+2018} performed test particle simulations of stellar CRs propagating in the wind of a T Tauri star including catastrophic energy losses for the CRs, meaning that once particles crossed the plane of the geometrically thin disc they lost all their energy. These modelling approaches all have different strengths and weaknesses. \citet{Rab+2017} very importantly focus on the chemical signatures of stellar CRs, which is vital for direct comparisons with observations and also treat the energy losses of the stellar CRs the most accurately. On the other hand, it seems unlikely that stellar CRs travel ballistically in protoplanetary discs. Here, the diffusive transport treatment and test particle simulations of \citet{RodgersLee+2017} and \citet{Fraschetti+2018}, which include the influence of magnetic fields,
 seem more reasonable. Overall, the stellar CRs did not penetrate far into the disc in any of these models for different reasons. A more consolidated approach considering similar systems is necessary to understand the different results obtained by these models.
 
 It is non-trivial to compare these models, not simply because of the important differences mentioned above in the assumed properties, transport and energy losses of the stellar CRs, but because the stellar and disc properties considered are also different. \citet{RodgersLee+2019}, using the diffusive approximation, present a parameter study that varies the disc surface density profile and the total disc mass using ranges constrained by ALMA dust observations of large samples of discs \citep{tazzari_2017,Pascucci+2016}. For discs around solar-mass stars with a range of disc masses and surface density profiles, they find that stellar CRs produce larger ionisation rates than those expected from unmodulated Galactic CRs at $\sim 70$\,au close to the disc midplane. The $1/r$ profile expected in the absence of energy losses is recovered for many  combinations of disc parameters, namely,  for lower-mass discs combined with shallower surface density profiles.
 This indicates that such discs are not dense enough to significantly attenuate  stellar CRs in the limit of diffusive transport. In future work, it is important to investigate whether discs likely to exhibit high ionisation rates at large radii are observable with ALMA.

\subsection{Cosmic-ray feedback from embedded protostellar clusters}\label{clusters}

Enhanced CR fluxes may also arise locally from sources embedded within the cloud as a result of CRs produced by young forming stars. Models estimate that proto-clusters with $>10^3$ accreting members can produce a local CR flux that exceeds the typical Galactic value \citep{GachesOffner2018}. This CR feedback dissociates CO and warms the nearby molecular gas, raising the temperatures 10s of K \citep{Gaches+2019a}. Massive clouds forming  $10^6\,{\rm M}_\odot$ star clusters, such as the progenitors of globular clusters, may experience local CR fluxes that yield ionisation rates of $\zeta > 10^{-14}$\,s\m. Such high values are comparable to the observed CR ionisation rates inferred for extreme regions such as the Galactic centre. However, in the Galactic centre, the CRs responsible for ionising the diffuse gas are likely produced predominantly by supernovae (SNe) and other high energy events (see Sect.~\ref{cmz}).

CR feedback from forming star clusters thus shapes carbon and nitrogen chemistry from the inside out. Astrochemistry models suggest that CR feedback decreases $X_{\rm CO}$ with increasing ionisation rate \citep{Gaches+2019b}. As expected, the abundances of C and C$^+$ are also enhanced. The abundance of the high-density tracer NH$_3$ declines, while the abundance of HCO$^+$ increases \citep{Gaches+2019a}. These results underscore that accounting for both external and internal CR ionisation is important for accurately modelling the chemistry and temperatures of active star-forming regions.

\subsection{H{\sc ii} regions}\label{HII}
\HII regions are usually dominated by thermal emission \citep[e.g.,][]{WoodChurchwell1989, Kurtz2005, Sanchez-Monge2008, Sanchez-Monge2011, Hoare+2012, Purcell+2013, Wang+2018, Yang+2019}. However, observations with  facilities such as the Very Large Array (VLA) or the Giant Metrewave Radio Telescope (GMRT) disclosed the presence of non-thermal emission in a small number of sources. 
Usually this emission appears 
as spots surrounded by thermal emission 
\citep[e.g.,][]{Nandakumar+2016,Veena+2016}, contiguous to thermal emission
as in cometary \HII regions \citep[e.g.,][]{Mucke+2002},
or sometimes it appears  isolated \citep{Meng+2019}.
As for the synchrotron emission observed in jet knots,
non-thermal emission observed in \HII regions 
cannot have interstellar origin.
In fact, the interstellar CR electron
flux based on the most recent Voyager observations 
\citep{Cummings+2016,Stone+2019}
is too low to explain the observed flux density.
The same conclusion 
holds for the flux of secondary electrons created through ionisation processes by interstellar CRs,
since \HII regions are often embedded within molecular 
clouds.

A possible explanation for the origin of 
these relativistic electrons is local acceleration inside
the \HII region itself through DSA, as proposed by~\cite{Padovani+2019}. This
assumes that shocks are located at the position
where non-thermal emission is detected. 
This model has been successfully applied to an expanding
\HII region close to the Galactic centre, Sgr~B2(DS),
recently observed with the VLA~\citep{Meng+2019}. 

\section{Low-energy cosmic rays at 
different Galactic scales}\label{diffgalscales}

\subsection{Intermediate Galaxy (100 pc scales)}\label{intgalaxy}
Here we define intermediate scales as the typical scales where CR are injected in the ISM, i.e.,  of order 100 pc, which is
similar to the size of a SN remnant at the end of its lifetime or the sizes of superbubbles produced by massive star clusters. This scale is also comparable to the Galactic disc height. 
It is not immediately obvious that CRs should have any dynamical impact at these scales. Indeed, consider a simple calculation comparing the typical turbulence energy cascade at a scale $L$, $\tau_{c}=L/c_{s}$, where $c_{s}$ is the local sound speed and the CR (1D) diffusion time over the scale, $L$, with diffusion coefficient $D$, i.e., $\tau_{\rm diff}=L^2/2 D$. CRs will leak out if $\tau_{\rm diff} < \tau_{c}$, namely when \citep{Commercon+2019}
\be
D > \frac{L}{2}c_{s}\approx 1.5\times 10^{23} \left(\frac{L}{\rm pc}\right) \left(\frac{c_{s}}{\rm{km~s\m}}\right)~\rm{cm^2~s\m}\,.
\ee
If CRs diffuse with a diffusion coefficient typical of that inferred from direct CR measurements, i.e., $D \sim 10^{28}~\rm{cm^2~s\m}$, then no dynamical effects are expected at intermediate scales.\\

\subsection{Galaxy (kpc scales)}\label{galaxy}

\subsubsection{Dynamical impact of CRs}
One key aspect of CRs on scales of kiloparsecs is their dynamical impact via a global CR pressure gradient and the ability to impact the dynamical evolution of galaxies via outflows. Consequently, on galactic scales we are primarily interested in the CR component that carries most of the energy, which are protons at a few GeV. The integrated energy in CRs is approximately $\epsilon_\mathrm{CR}\sim0.5\,\mathrm{eV\,cm}^{-3}$ and thus comparable to the magnetic energy in the ISM, $\epsilon_\mathrm{mag}\sim0.25\,\mathrm{eV\,cm}^{-3}$ for a magnetic field strength of $3\,\mu\mathrm{G}$ \citep{Haverkorn2015, Beck2015}. This approximate energy equipartition makes CRs a potentially relevant energy reservoir and suggests that there is mutual interaction between CRs and the magnetic field. Treating CRs as test particles is therefore not justified. The typical gyroradii or scattering lengths of GeV protons with the magnetic field are significantly smaller than the scales under consideration, which allows for a fluid treatment of CRs. This high-energy fluid is then dynamically coupled to the MHD equation in a two-fluid approach with an effective adiabatic index and a total pressure that includes the CR contributions \citep{HanaszLesch2003, PfrommerEtAl2017}.

GeV protons are primarily produced in SN remnants \citep{Ackermann+2013}, which are among the strongest and most abundant shocks in the ISM. SNe in our galaxies are not uniformly distributed \citep{MillerScalo1979, Heiles1987, TammannLoefflerSchroeder1994}. In our Galaxy, 20\% of the SN are type~Ia, which explode not only in the vicinity of star-forming regions but also high above the Galactic disk. Their distribution can be described by a Gaussian with a scale height of $\sim300\,\mathrm{pc}$. The remaining 80\% are core collapse SNe, connected to massive stars above $\sim8\,\mathrm{M}_\odot$. The clustered formation of stars combined with the relatively short lives of massive stars results in clustered SNe. Approximately 60\% of core collapse SNe explode close to their birth places, following a vertical Gaussian distribution with a scale height of only $\sim100\,\mathrm{pc}$. The remaining 40\% are {\it walk-away} or runaway stars that escape the star cluster prior to explosion. The resulting vertical scale height is then comparable to the one of the type~Ia SNe, i.e., $300\,\mathrm{pc}$. The majority of the CRs are thus produced close to the disk with an effective scale height of $\sim150\,\mathrm{pc}$. The explosion sites extend approximately to a height of $1\,\mathrm{kpc}$. Consequently, on scales at $\sim100\,\mathrm{pc}$, as discussed in the previous section, the CR distribution is relatively smooth due to the fast diffusion of CRs as well as the distributed injection. In contrast, on kpc scales above the disk, the enhanced injection close to the disk and the resulting CR pressure gradient must be taken into account.

\subsubsection{CR clocks and Galactic CR scale height}

Under the assumption that the ISM in a galaxy has approximately uniform metallicity and that all particles in the ISM are equally likely to be accelerated to high-energies, we should expect a very similar composition for the CRs and the thermal gas. However, we notice an overabundance of light elements like Li, Be and B, which must be produced during CR transport through the medium \citep{AloisioBlasi2013, Stone+2013, Aguilar+2015, AguilarEtAl2015b, AguilarEtAl2016, AguilarEtAl2016b}. Spallation is the most likely process to produce these secondary particles. A very common method to estimate the ratio of primary to secondary particles is the B/C abundance 
ratio. This ratio depends on the particle energy, but to zeroth order we observe approximately B/C = $0.3$. 

To trigger spallation we need to know the grammage of CRs, which is the amount of material the CRs statistically pass through before colliding with a thermal gas particle,
\begin{equation}
\chi = \int\rho(l) dl.
\end{equation}
We can relate the grammage to the effective cross section for spallation of carbon to boron $\sigma_{\mathrm{C}\rightarrow\mathrm{B}}$ in an average ISM particle mass $m$ as
\begin{equation}
\chi\frac{\sigma_{\mathrm{C}\rightarrow\mathrm{B}}}{m} \approx \frac{\mathrm{B}}{\mathrm{C}}\sim 0.3,
\end{equation}
which yields a value of $\chi\sim 10~\mathrm{g\,cm}^{-2}$. At a solar radius the gas surface density is $\Sigma\sim10~\mathrm{M}_\odot\,\mathrm{pc}^{-2} = 2\times10^{-3}\,\mathrm{g\,cm}^{-2}$, i.e., the CR must cross the disc $\sim10^3$ times before interacting with the ISM. The associated lifetime of the CRs, which is effectively the travel or residence time in the Galaxy, is given by
\begin{equation}
t_\mathrm{res}=\chi\frac{h}{v\Sigma}\sim 10^6\,\mathrm{yr},
\end{equation}
assuming a gas scale height of  $h\sim100\,\mathrm{pc}$ and the speed of light for the CR velocity, $v=c$.

During this residence time the CRs travel a linear distance of $l=t_\mathrm{res}c\sim500\,\mathrm{kpc}$, which is much longer than galactic scales, which suggests the CRs must be confined to the galaxy. Unstable secondary CRs provide a valuable tool to estimate the escape time. The best measured unstable isotope is $^{10}$Be, which is also the longest lived. Its decay time is $t_\mathrm{decay}=1.39\,\mathrm{Myr}$, which is comparable to the residence time. The cross sections for the spallation of carbon into stable ($^{9}$Be) and unstable ($^{10}$Be) are similar, so $\sigma_{\mathrm{C}\rightarrow\mathrm{^{10}Be}}\approx\sigma_{\mathrm{C}\rightarrow\mathrm{^9Be}}$. Consequently, an initial abundance ratio of ($^{10}$Be/$^{9}$Be) of order unity declines over time by $t_\mathrm{decay}(^{10}\mathrm{Be})/t_\mathrm{esc}$. The observed ratio yields an escape time of $10$--$20~\mathrm{Myr}$, i.e., an order of magnitude larger than the residence time of CRs.

Assuming for simplicity a simple diffusive propagation model, the transport equation along the vertical dimension reads
\begin{equation}
\frac{\partial {\mathcal N}}{\partial t} = Q_0(p)\delta(z) + \frac{\partial}{\partial z}\left[D \frac{\partial {\mathcal N}}{\partial z}\right]
\label{eqn:NQ}
\end{equation}
for a CR density per unit energy, ${\mathcal N}$, a CR source function, $Q_0$ and  diffusion coefficient, $D$. The Dirac-$\delta$ distribution indicates the injection of CRs in the midplane of the galactic disc. Assuming steady state, Eq.~(\ref{eqn:NQ}) reduces to
\begin{equation}
Q_0(p)\delta(z) = D\frac{\partial^2 {\mathcal N}}{\partial z^2}.
\end{equation}
For $z>0$,
\begin{equation}
\frac{\partial {\mathcal N}}{\partial z} = \mathrm{const.} \Leftrightarrow {\mathcal N}(z) = {\mathcal N}_0\left(1-\frac{z}{H}\right),
\end{equation}
where $H$ is the size of the halo, which is poorly confined to approximately $5\,\mathrm{kpc}$. A total grammage of $\chi=\overline{\rho}t_\mathrm{esc} c$ is reached by moving though the average density of the total volume (disc plus halo), $\overline{\rho}=\mu m_p n h/H$. Here, $h=100\,\mathrm{pc}$ is the scale height of the gaseous galactic disc, $\mu=1.4$ is the mean molecular weight, $n=1\,\mathrm{cm}^{-3}$ is the average number density of the ISM. The resulting estimated diffusion coefficient is
\begin{align}
D &= \frac{H^2}{t_\mathrm{esc}}=\frac{\mu m_p n h H c}{\chi} \nonumber \\
&\sim 3\times10^{28} \left(\frac{H}{5\,\mathrm{kpc}}\right)\left(\frac{\chi}{10\,\mathrm{g\,cm}^{-2}}\right)^{-1}\mathrm{cm}^2\,\mathrm{s}^{-1}
\end{align}

Despite their simplicity, the derived estimates provide valuable features of Galactic CRs, which remain valid even under more complex assumptions. One result is that GeV CRs are distributed relatively smoothly in the ISM. Local changes in the CR energy density are smaller than the changes in the gas structures like the thermal and magnetic energy density. As a result, molecular clouds are located in an almost uniform sea of GeV CRs. Due to frequent scattering, the CR distribution is locally isotropic.

\subsubsection{Pressure gradient and dynamical impact}

Frequent SNe create a relatively smooth distribution of GeV CRs in the ISM on scales of 0.1--1~kpc. The resulting small CR pressure gradients provide a relatively small force acting on the ISM  compared to locally strongly varying thermal pressure gradients or gravitational forces in star-forming regions. As a result, the dynamical impact of CRs is most effective in the diffuse ISM and acts over longer timescales compared to the short dynamical time in the ISM (see also the previous section). In contrast to SN shocks, CRs provide an effective background force in the vertical direction that counteracts the gravitational attraction of the disk. Hydrodynamical simulations of the ISM including thermal and CR feedback confirm this simple picture of driving outflows. Including CRs increases the total amount of outflowing gas by a factor of a few up to an order of magnitude (Girichidis 2016, 2018, Kim \& Ostriker 2018, Mao \& Ostriker 2018). The outflows are overall smoother and colder compared to their thermally driven counterpart.

\subsubsection{Galactic outflows}
On even larger scales, namely for full galaxies, CRs appear to be dynamically important for driving galactic outflows. Idealised simulations using anisotropic diffusion and an isothermal equation of state show that CRs alone are able to drive outflows with mass-loading factors of order unity \citep{HanaszEtAl2013}. The efficiency of the outflows depend on the diffusion coefficient as well as the SN energy fraction that is converted into CRs. \citet{SalemBryan2014} investigate this aspect in isolated disk galaxies. Indirectly, CRs also influence the dynamics by changing the magnetic strength and morphology in the disk \citep{PakmorEtAl2016}. Anisotropic diffusion results in stronger magnetic fields, which in turn reduces the formation of dense molecular clouds and star formation \citep{GirichidisEtAl2016a, GirichidisEtAl2018a}. The details of CR-driven outflows like the mass and energy loading as well as the resulting heating in the ISM depend on the details of the transport \citep{Zweibel2013, Zweibel2017, RuszkowskiYangZweibel2017, FarberEtAl2018}. Currently developed models therefore couple CRs to the thermal gas as well as the magnetic fields in a more consistent way \citep{JiangOh2018, ThomasPfrommer2019} and investigate the CR interactions with spectrally resolved methods \citep{GirichidisEtAl2019}.

\subsection{The Central Molecular Zone}\label{cmz}
The CMZ is a region about 250~pc in radius near the centre of the Galaxy where a large amount of molecular gas resides \citep{MorrisSerabyn1996,Bally+2010}. This material faces a variety of extreme conditions, including high turbulence, pressure, temperature, CR flux and ionising radiation flux, compared to gas elsewhere in the Galactic disc \citep{Ao+2013,Ginsburg+2016,Henshaw+2016,Schmiedeke+2016,Krieger+2017}. Intriguingly, the observed star-formation rate is at least a factor of 10 below that predicted given the amount of dense gas available \citep{Longmore+2013,Kruijssen+2014,Barnes+2017,Kauffmann+2017}. The CMZ can serve as a template for Galactic nuclei in general and potentially for the conditions prevalent during the era of prolific star formation at redshifts of about 2--3 \citep{KruijssenLongmore2013}. At a distance of 8~kpc \citep{Gravity+2019}, the CMZ is the nearest such region that we can study, so understanding star formation there has a unique potential to improve our understanding of star formation in similar regions throughout the Universe.

As for the Galactic disc, molecular abundances (e.g., H$_3^+$, OH$^+$, H$_2$O$^+$, and H$_3$O$^+$) have been used to infer CR ionisation rates within the CMZ. Observations of H$_3^+$ absorption toward several different sight lines in the CMZ indicate $\zeta_{\rm H_2}\sim10^{-14}$~s$^{-1}$ across most of the region \citep{Oka+2019}. This is consistent with previous studies using H$_3^+$ \citep{Goto+2013} and oxygen bearing ions \citep{Indriolo+2015} and with results from more detailed chemical modelling \citep{LePetit+2016}. All of these results indicate ionisation rates much higher than found in the Galactic disc, again implying spatial variations in the underlying particle spectrum.

Another constraint on the CR ionisation rate in the CMZ can be inferred from the temperatures of molecular clouds. Ionisation of H$_2$ by CRs liberates electrons with about 30~eV of kinetic energy on average \citep{CravensDalgarno1978}, which then heat the gas through collisions. This means that any given ionisation rate sets a floor gas temperature due to CR-induced heating. By inferring gas temperatures in molecular clouds throughout the CMZ from molecular emission line observations, \citet{Ginsburg+2016} placed an upper limit of $\zeta_{\rm H_2}<10^{-14}$~s$^{-1}$ on the ionisation rate in the Galactic centre based on the floor gas temperature. Even with such a high ionisation rate, it is thought that turbulence, rather than CR ionisation, dominates gas heating in the CMZ. Still, at high ionisation rates CR heating will affect the gas temperature and thus the initial conditions in molecular clouds from which stars form \citep[e.g.][]{Clark+2013}.

Recent radio observations with the MeerKAT telescope show outflow bubbles that suggest a recent ($<$~few~Myr) energetic event in the CMZ \citep{Heywood+2019}, plausibly a burst of star formation \citep[e.g.][]{Krumholz+2017}, although feedback from the central supermassive black hole cannot be ruled out. Irrespective of its origin, it is possible that this event contributed to the high CR ionisation rate, especially given its large energy budget of $\sim7\times10^{52}$~ergs \citep{Heywood+2019}. These CRs may themselves be responsible for the Galactic wind emanating from the CMZ, in the process explaining the origin of the energetic non-thermal filaments throughout the CMZ, as well as their perpendicular orientation relative to the disc plane \citep{YusefzadehWardle2019}.

\subsection{Low-energy cosmic rays in superbubbles}\label{feedback}

Clustering of young massive stars in star-forming regions has a profound effect on the structure and 
 evolution of the ISM on 100 pc scales with a possibility of the break out of the thin HI Galactic disc in the form of {\it chimneys} \citep[] 
 []{MacLowMacCray1988,TomisakaIkeuchi1986,Kim+2017,El-Badry2019}. The massive star-forming regions are structured as they are observationally represented by  compact clusters of young massive stars as well as by highly substructured OB associations with lower stellar density than in the compact clusters \citep[see, e.g.,][]{Krumholz+2019,Ward+2019}. The winds of massive stars and their subsequent SNe create  large-scale superbubbles and supershells in the ISM  through their great momentum and energy release \citep[see, e.g.,][]{Heiles1979,MacLowMacCray1988,El-Badry2019}.  Such structures distribute the chemical elements freshly produced by massive stars and accelerate CRs. These components together with radiation and magnetic fields are essential components of stellar feedback inside star-forming molecular clouds  
 \citep[e.g.][]{Fierlinger+2016}.  Moreover, this stellar feedback may drive Galactic winds at kpc scales \citep[e.g.][]{Fielding+2017}. 
 The high mechanical power released by the winds of massive early type stars and SNe
 in young stellar clusters or OB associations can exceed 10$^{38}$ erg s$^{-1}$ over many million years. 
 This mechanical luminosity blows out superbubbles of different sizes surrounded by massive supershells, which in some cases may produce sequential star formation in galaxies. The  power of winds and SNe should produce  mechanical work on the surrounding cold ISM and provide gas and dust heating through advection and thermal conduction. On the other hand, \citet{Rosen+2014} studied the mechanical energy partition in some compact stellar clusters and found that a relatively small fraction of the injected energy goes into these channels. They suggested 
that turbulent mixing at the hot-cold interfaces removes the kinetic energy injected by the winds of clustered massive stars. A similar issue was discussed by \citet{Cooper+2004} for the superbubble DEM L192 located in the Large Magellanic Cloud. They found that a significant fraction of the mechanical power is not accounted for by the observed constituents. 

\begin{figure}[!h]
\begin{center}
\resizebox{.8\hsize}{!}{\includegraphics{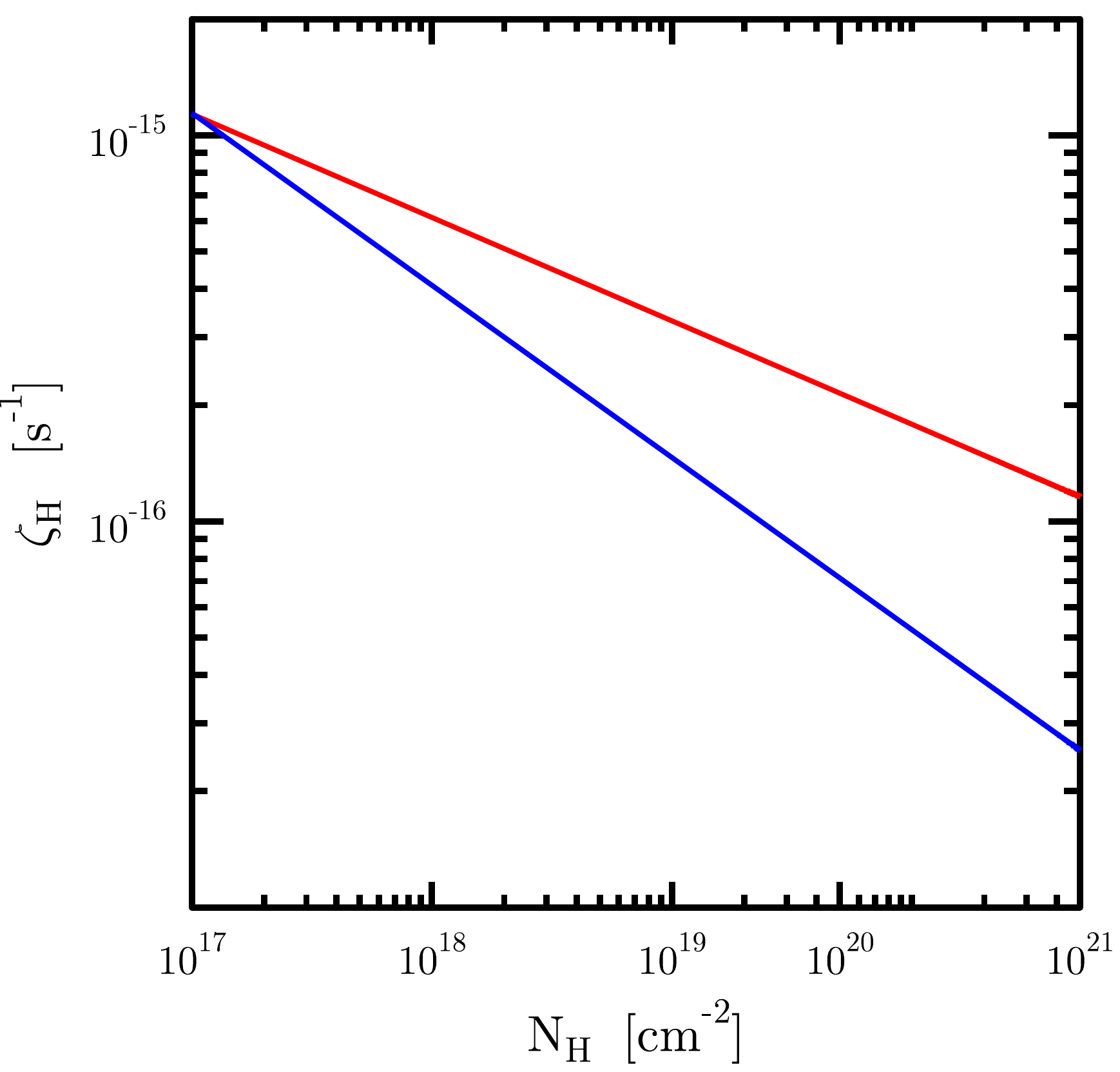}}
\caption{Hydrogen ionisation rate by  low-energy CRs in an HI supershell surrounding a superbubble. The CRs are accelerated in the superbubble and propagate diffusively into the supershell. The red curve shows the ionisation rate for the energy-dependent CR diffusion coefficient, while the blue curve shows the result for the energy independent turbulent-advection and diffusion regime.}
\label{SB_ion}
\end{center}
\end{figure}

CRs can be accelerated in superbubbles \citep[see for a review][]{Binns+2007,Bykov2014,Lingenfelter2018}. The efficiency of the kinetic power transfer to CRs in the models with particle acceleration by multiple shocks can exceed 20\% \citep{Bykov2001}. Therefore, CR pressure and the escaped CR energy fluxes should be accounted for in the energy balance of superbubbles \citep{ButtBykov2008}. Models of low-energy CRs accelerated in superbubbles by shocks and MHD turbulence predict time variable CR spectra at different evolutionary stages. The hydrogen in large-scale supershells surrounding the superbubbles is ionised by both the soft X-ray photons produced by the hot gas inside the superbubble and the low-energy CRs,  which can penetrate deep into the shell depending on the CR transport in the shell. Fig.~\ref{SB_ion} illustrates a model for the hydrogen ionisation rate by CRs in the HI supershell of mass 10$^5 ~M_{\odot}$, thickness 30 pc  and  number density of 10 cm$^{-3}$. The red curve in the figure shows the ionisation rate of the HI shell by CRs accelerated in the superbubble derived for the model of the diffusive propagation of CRs in the shell with the diffusion coefficient $D \propto p^{0.33}$ , which corresponds to a Kolmogorov spectrum of magnetic fluctuations. For large turbulent advection rms velocities in the subparsec scale or in the case of  multiple weak shocks in the supershell \citep[e.g.][]{BykovToptygin1987} the CR diffusion can be energy independent. This results in lower ionisation rates deep inside the HI shell, as illustrated by the blue curve in Fig.~\ref{SB_ion}.                    
   
\subsection{Extragalactic studies}\label{extra}
Observations of the molecules used to infer CR ionisation rates are difficult even within the Galaxy, so these techniques 
have only been applied to a small sample of other galaxies. OH$^+$ and H$_2$O$^+$
were observed with {\it Herschel}
toward the nuclei of NGC~4418, Arp~220, and Mrk~231,  suggesting ionisation rates on the order of about $10^{-13}$--$10^{-12}$~s$^{-1}$ \citep{Gonzalez-Alfonso+2013,Gonzalez-Alfonso+2018}. Like our own Galactic centre, these regions have rather high ionisation rates. OH$^+$ and H$_2$O$^+$ have also been observed in absorption toward the discs of a small sample of nearby galaxies, such as NGC~253, M82, NGC~4945, CenA, where much lower ionisation rates -- on the order of 10$^{-16}$~s$^{-1}$ -- were inferred \citep{Vandertak+2016}. These values are more in line with those determined for the Galactic disc and provide further evidence for different ionisation rates and particle fluxes in Galactic nuclei compared to discs.

Recently, ALMA has provided the opportunity to observe OH$^+$ and H$_2$O$^+$ in galaxies at higher redshift where the rotational transitions out of the ground states are shifted to frequencies where the Earth's atmosphere is transparent. Both molecules have been detected in a foreground absorber at $z=0.89$ toward the lensed quasar PKS~1830$-$211. The absorber is thought to be Milky-Way-like, and the two lensed images of the quasar provide two different sight lines through the disc of this galaxy, in which ionisation rates of about $10^{-15}$~s$^{-1}$ and $10^{-14}$~s$^{-1}$ are inferred \citep{Muller+2016}. Oxygen-bearing ions are now also being detected in lensed submillimetre bright galaxies at $z>2$ by ALMA, and while high ionisation rates have been estimated in these objects, the uncertainties in those analyses are rather large \citep{Indriolo+2018}.



\section{Summary and outlook}\label{conclusions}
In this review it has been shown that low-energy CRs represent a key element in several physical and chemical processes of the ISM, having a strong impact from the large scales of molecular clouds to the small scales of protostellar systems. 
During the last ten years, growing attention has been given to the study of the interaction of CRs with the ISM, and a significant progress has been made in understanding their transport regimes at different depths of a cloud and their interaction with magnetic fields. These studies demonstrated that the assumption of a constant CR ionisation rate, based on the fundamental studies of L. Spitzer in the '50s and used for many decades, is inaccurate. Indeed, CR ionisation rates span over at least $4$--$5$ orders of magnitude from diffuse clouds ($\zeta\sim10^{-15}$~s$^{-1}$) to circumstellar disc midplanes ($\zeta\lesssim10^{-19}$~s${-1}$). This is an important warning for modelers of
 astrochemical codes and non-ideal MHD simulations of collapsing clouds and MRI in circumstellar discs. The use of a constant $\zeta$ can produce unrealistic values of chemical abundances and heavily affect magnetic decoupling and hence the fragmentation of molecular clouds.
Besides, one must also consider the effect of the presence of local CR sources in single protostars and protostellar clusters that can locally enhance the CR flux, exceeding the typical Galactic value, with important feedback on the ionisation degree of the local medium, on the temperature, and ultimately on its chemical composition and dynamical evolution.

The study of low-energy CRs has opened a new line of research, allowing a strong synergy between theoretical models and observations. 
For instance, theory predicts that at the shock surface of high-mass protostellar jets should be possible to accelerate CRs up to tens of TeV from which should arise a spatially limited, but non-negligible, $\gamma$-ray emission. 
While current instruments such as H.E.S.S. and Fermi can hardly detect this localised emission because of their low resolution, it will be interesting to understand if the Cherenkov Telescope Array (CTA) will be able to distinguish it from the background Galactic $\gamma$-ray emission. This detection, due
to the local production of CR protons, that 
should be localised at the same spatial position of the synchrotron emission
caused by locally accelerated CR electrons, would represent a double
proof of the possibility of accelerating CRs in protostellar systems.
We also note that thanks to future instruments such as the Square Kilometre
Array (SKA) with its huge field of view, high sensitivity, resolution,
and survey speed, the detection of synchrotron sources would represent
the rule rather than the exception. SKA precursors such as ASKAP, MeerKAT, E-LOFAR, will be of great importance for the first characterisation of
synchrotron emission at different scales.


A new and growing frontier of astrophysics research explores the impact of CRs on exoplanetary atmospheres and habitability. 
CRs of both Galactic and stellar origin likely play a crucial role in the chemistry of planetary atmospheres and the development and characteristics of life. Exoplanet searches such as the
MEarth project have focused specifically on finding exoplanets around M-dwarf stars, which are considered prime candidates to host rocky planets \citep{NutzmanCharbonneau2008}. However, such stars experience strong stellar activity, and any planets within their relatively close-in habitable zones  ($<0.2$ au) will be bombarded by stellar CRs \citep{Griessmeier+2005}. 
The planet magnetosphere regulates the flux of CRs reaching the top of the planetary atmosphere \citep{Atri+2013,Griessmeier+2015}. Close-in planets, such as those in the M-dwarf habitable zone, however, are likely to be tidally locked to their host star, which leads to a smaller magnetic moment and weaker magnetosphere \citep{Griessmeier+2005,Griessmeier+2016}.  CRs drive chemical reactions in the planet atmosphere that may interfere with the detection of bio-signatures (chemical markers produced by biological organisms), for example, by enhancing the abundance of methane 
and reducing the presence of ozone \citep{Griessmeier+2016,Tabataba-Vakili+2016,Scheucher+2018}. 
 The planet atmospheric density, even more than the magnetosphere, dictates the particle and radiation flux at the surface \citep{Atri+2013}. 
 CR interactions in the atmosphere produce electromagnetic radiation as well as high-energy secondary particles, which are harmful for DNA-based life on the surface \citep{Dar1998}. 
 Models suggest that close-in, rocky planets with Earth-like atmospheres are sufficiently shielded that CR induced reduction of the ozone has a minor effect on the biological radiation dose at the surface \citep{Griessmeier+2016}. In contrast, the surfaces of planets with atmospheres 10\% of that of Earth may receive 200 times higher radiation doses, which is lethal to most Earth-like life \citep{Atri+2013}.
Future facilities such as ELT/HIRES, JWST, and ARIEL will
help to 
disentangle the products of CR-driven chemistry from biological signatures and assessing the impact of high-CR environments on the prevalence of life in the Universe.

\begin{acknowledgements}
M.P. acknowledges funding from the INAF PRIN-SKA 2017 program 1.05.01.88.04.
J.M.D.K. gratefully acknowledges funding from the German Research Foundation (DFG) in the form of an Emmy Noether Research Group (grant number KR4801/1-1) and a DFG Sachbeihilfe Grant (grant number KR4801/2-1), from the European Research Council (ERC) under the European Union's Horizon 2020 research and innovation programme via the ERC Starting Grant MUSTANG (grant agreement number 714907), and from Sonderforschungsbereich SFB 881 ``The Milky Way System'' (subproject B2) of the DFG.
S.S.R.O. acknowledges funding from NSF Career grant AST-1650486 and NASA ATP grant 80NSSC20K0507. P.G. acknowledges funding from the European Research Council under ERC-CoG grant CRAGSMAN-646955.
\end{acknowledgements}




%
%

\bibliographystyle{spbasic}      
\bibliography{shortenedbibliography}

%
%

\end{document}